\documentclass[fleqn,usenatbib]{mnras}

\usepackage{newtxtext,newtxmath}

\usepackage[T1]{fontenc}

\DeclareRobustCommand{\VAN}[3]{#2}
\let\VANthebibliography\thebibliography
\def\thebibliography{\DeclareRobustCommand{\VAN}[3]{##3}\VANthebibliography}

\usepackage{graphicx}	
\usepackage{amsmath}	
\usepackage{xspace}
\usepackage{pdflscape}
\usepackage{bigfoot}

\usepackage{scalerel}
\usepackage{tikz}
\usetikzlibrary{svg.path}

\definecolor{orcidlogocol}{HTML}{A6CE39}
\tikzset{
  orcidlogo/.pic={
    \fill[orcidlogocol] svg{M256,128c0,70.7-57.3,128-128,128C57.3,256,0,198.7,0,128C0,57.3,57.3,0,128,0C198.7,0,256,57.3,256,128z};
    \fill[white] svg{M86.3,186.2H70.9V79.1h15.4v48.4V186.2z}
                 svg{M108.9,79.1h41.6c39.6,0,57,28.3,57,53.6c0,27.5-21.5,53.6-56.8,53.6h-41.8V79.1z M124.3,172.4h24.5c34.9,0,42.9-26.5,42.9-39.7c0-21.5-13.7-39.7-43.7-39.7h-23.7V172.4z}
                 svg{M88.7,56.8c0,5.5-4.5,10.1-10.1,10.1c-5.6,0-10.1-4.6-10.1-10.1c0-5.6,4.5-10.1,10.1-10.1C84.2,46.7,88.7,51.3,88.7,56.8z};
  }
}

\newcommand\orcid[1]{\href{https://orcid.org/#1}{\mbox{\scalerel*{
\begin{tikzpicture}[yscale=-1,transform shape]
\pic{orcidlogo};
\end{tikzpicture}
}{|}}}}

\newcommand{\eve}{\texttt{EVEREST 2.0}\xspace}
\newcommand{\gaia}{\textit{Gaia}\xspace}
\newcommand{\ktwo}{\textit{K2}\xspace}
\newcommand{\tfaw}{\texttt{TFAW}\xspace}
\newcommand{\tls}{\texttt{TLS}\xspace}
\newcommand{\tri}{\texttt{TRICERATOPS}\xspace}
\newcommand{\vespa}{\texttt{vespa}\xspace}

\newcommand{\msun}{$M_{\odot}$\xspace}
\newcommand{\rsun}{$R_{\odot}$\xspace}
\newcommand{\mstar}{\ensuremath{M_{\rm s}}\xspace}
\newcommand{\rstar}{\ensuremath{R_{\rm s}}\xspace} 
 
\newcommand{\feh}{\ensuremath{[\mbox{Fe}/\mbox{H}]}\xspace}
\newcommand{\teff}{\ensuremath{T_{\mathrm{eff}}}\xspace}  
\newcommand{\logg}{\ensuremath{\log g}\xspace}

\newcommand{\mearth}{$M_\oplus$\xspace}
\newcommand{\rearth}{$R_\oplus$\xspace}

\title[TFAW survey II]{TFAW survey II: 6 Newly Validated Planets and 13 Planet Candidates from \ktwo}

\author[del Ser et al.]{D.~del Ser\orcid{0000-0001-6776-3211},$^{1,2}$\thanks{E-mail: danieldelser@icc.ub.edu}
O.~Fors\orcid{0000-0002-4227-9308},$^{2}$
M.~del Alc{\'a}zar\orcid{0000-0003-1070-6579},$^{1,2}$
V.~Dyachenko\orcid{0000-0002-4973-270X}$^{3}$
E.~P.~Horch\orcid{},$^{4}$
\newauthor
A.~Tokovinin\orcid{0000-0002-2084-0782},$^{5}$
C.~Ziegler\orcid{0000-0002-0619-7639},$^{6}$
G.~T.~van Belle\orcid{0000-0002-8552-158X},$^{7}$
C.~A.~Clark\orcid{0000-0002-2361-5812},$^{7,8}$
and Z.~D.~Hartman\orcid{0000-0003-4236-6927}$^{9}$
\\
$^{1}$Observatori Fabra, Reial Acad{\`e}mia de Ci{\`e}ncies i Arts de Barcelona, Rambla dels Estudis, 115, E-08002 Barcelona, Spain\\
$^{2}$Dept. de F{\'i}sica Qu{\`a}ntica i Astrof{\'i}sica, Institut de Ci{\`e}ncies del Cosmos (ICCUB), Universitat de Barcelona, IEEC-UB, Mart\'{\i} i Franqu{\`e}s 1, E-08028 Barcelona, Spain\\
$^{3}$Special Astrophysical Observatory, 369167 Nizhnij Arkhyz, Russia\\
$^{4}$Department of Physics, Southern Connecticut State University, 501 Crescent Street, New Haven, CT 06515, USA\\
$^{5}$Cerro Tololo Inter-American Observatory, NSF's NOIRLab, Casilla 603, La Serena, Chile\\
$^{6}$Department of Physics, Engineering and Astronomy, Stephen F. Austin State University, TX 75962, USA\\
$^{7}$Lowell Observatory, 1400 W. Mars Hill Road, Flagstaff, AZ 86001, USA\\
$^{8}$Northern Arizona University, 527 S. Beaver Street, Flagstaff, AZ 86011, USA\\
$^{9}$Gemini Observatory/NSF's NOIRLab, 670 N. A'ohoku Place, Hilo, HI, 96720, USA\\
}

\date{Accepted XXX. Received YYY; in original form ZZZ}

\pubyear{2022}

\begin{document}
\label{firstpage}
\pagerange{\pageref{firstpage}--\pageref{lastpage}}

\maketitle

\begin{abstract}
Searching for Earth-sized planets in data from Kepler's extended mission (\ktwo) is a niche that still remains to be fully exploited. The \tfaw survey is an ongoing project that aims to re-analyze all light curves in \ktwo C1-C8 and C12-C18 campaigns with a wavelet-based detrending and denoising method, and the period search algorithm \tls to search for new transit candidates not detected in previous works. We have analyzed a first subset of 24 candidate planetary systems around relatively faint host stars ($10.9 < K_{\rm p} < 15.4$) to allow for follow-up speckle imaging observations. Using \vespa and \tri, we statistically validate six candidates orbiting four unique host stars by obtaining false-positive probabilities smaller than 1$\%$ with both methods. We also present 13 vetted planet candidates that might benefit from other, more precise follow-up observations. All of these planets are sub-Neptune-sized, with two validated planets and three candidates with sub-Earth sizes, and have orbital periods between 0.81 and 23.98 days. Some interesting systems include two ultra-short-period planets, three multi-planetary systems, three sub-Neptunes that appear to be within the small planet Radius Gap, and two validated and one candidate sub-Earths (EPIC 210706310, EPIC 210768568, and EPIC 246078343) orbiting metal-poor stars.
\end{abstract}




\section{Introduction}

The \ktwo mission \citep{Howell2014}, represented a way to continue \emph{Kepler}'s observations after the failure of the spacecraft reaction wheels. This mode, which became fully operational in May 2014, led to a series of 19 sequential campaigns each of which observed a set of independent target fields distributed along the ecliptic plane during $\sim$80 days.

Given the degraded photometric precision of the \ktwo light curves compared to those from the original \emph{Kepler} one, improvements in the data analysis have played a key role in increasing the number of detected planet candidates in \ktwo light curves. The first example was the series of pixel decorrelation and detrending algorithms \citep{Deming2015,Lund2015,Vanderburg2014} which culminated in the \eve pipeline \citep{Luger2018}. These have provided the best photometric precision for \ktwo light curves and can return photometric precisions very similar to the ones from the original \emph{Kepler} mission to K$_{\rm p}$=15 mag. Most planet searches in \ktwo campaigns used these \eve-corrected light curves, yielding an appreciable fraction of the currently confirmed planets and candidates \citep{Mayo2018,Kovacs2020,Zink2020,Adams2021,deLeon2021, Castro-Gonzalez2021,Zink2021,Christiansen2022}. The development of new transit search tools has also helped increase the number of planets detected. For example, \citet{Heller2019} was especially sensitive to Earth-sized planets thanks to the use of the Transit Least Squares (\texttt{TLS}) algorithm \citep{Hippke2019} as a new transit detection tool, which was designed and optimized to detect smaller planets. The definition of robust vetting and statistical validation procedures \citep{Morton2012,Morton2015,Kruse2019,Heller2019,Giacalone2020,Giacalone2021}, have also allowed to improve the characterization of false-positive signals originating from background stars, non-associated blended eclipsing binaries, or non-associated stars with transiting planets. All this has led to the admirable current \ktwo mission legacy of 537 confirmed planets exclusively discovered from \ktwo observations, and 969 candidates yet to be confirmed.

The current goal of the \tfaw survey \citep{delSer2018} is to search for new exoplanet candidates previously missed by former studies by further improving the photometric precision of the \eve-corrected light curves. The survey makes use of \tfaw, a novel wavelet-based detrending and denoising algorithm developed by \cite{delSer2018}, the \eve \citep{Luger2018} processed \ktwo light curves, and the \tls \citep{Hippke2019} transit search algorithm. As shown in \citet{delSer2020}, \tfaw delivers both better photometric precision and planet characterization than any detrending method applied to \ktwo light curves. The increased photometric precision achieved with \tfaw, especially for faint \ktwo magnitudes, together with \tls improved capabilities to detect small planets, enable us to detect new, Earth-sized, and smaller planets orbiting G-, K- and M-type stars. As an example of this, \citet{delSer2020} reported the discovery of two new statistically validated Earth-sized planets, K2-327 b, and K2-328 b, orbiting an M-type and a K-type star, respectively.

In this work, we present a first sample of 27 new planetary candidates detected by the \tfaw survey with new speckle imaging follow-up observations. In Section \ref{sec:Data}, we describe the observations and ancillary data used in this work, consisting of \ktwo \eve-corrected light curves, stellar host characterization, archival high-resolution images, speckle imaging follow-up observations, and \gaia eDR3 \citep{Gaia2021} photometry and astrometry. In Section \ref{sec:methods}, we briefly describe the \tfaw algorithm and the transit search method, we present our vetting method, the MCMC-based transit modeling, the mass-radius estimation, and resonance analysis, our validation approach, and the candidate disposition procedure. In Section \ref{sec:results}, we present and characterize our final validated, candidate, and false-positive sample, and discuss some of the most interesting systems found in this work.

\section{Data and Observations}
\label{sec:Data}

\subsection{\ktwo photometry}
\label{subsec:k2phot}

The \tfaw survey focuses on \ktwo campaigns C1 to C8, and C12 to C18. We exclude campaigns C9, used to study gravitational microlensing events, and C10 and C11, which were separated into sub-campaigns. We download $\sim$300000 \eve long cadence target light curves recorded before 4 Jan 2019 available at the MAST archive\footnote{\url{https://archive.stsci.edu/hlsps/everest/v2/bundles/}}. Given the characteristics of the wavelet transform used by \tfaw (for more details on the algorithm see \citet{delSer2018} and \citet{delSer2020}), for campaigns C1 to C8 we use 3072 epochs while, for campaigns C12 to C18, we use 2432. Also, \tfaw was designed as a general detrending and denoising tool, and not specifically to analyze \ktwo data. To deal with intrapixel and interpixel variations, we use the Pixel Level Decorrelation (PLD) \citep{Deming2015}, and single co-trending basis vector (CBV) corrected fluxes provided by the \eve pipeline. We also retrieve the available \ktwo Target Pixel Files (TPF) and the \eve photometric apertures of each target. While most of the 27 systems presented in this work were observed in a single \ktwo campaign, three (EPIC 211436876, EPIC 246078343, and EPIC 246220667) were observed in two separate campaigns.

The \ktwo targets studied in this work together with their corresponding observing campaigns are listed in Table~\ref{tab:obs}.

\begin{table}
    \caption{Summary of the \ktwo targets and campaigns, and speckle imaging follow-up facilities used in this work.}
    \begin{tabular}{ccccc}
\hline
EPIC & Campaign & BTA & SOAR & LDT \\
\hline
205979483 & 3 & & x & \\
206461841 & 3 & & x & \\
210418253 & 4 & x & & \\
210706310 & 4 & x & & \\
210708830 & 4 & x & & \\
210768568 & 4 & x & x & \\
210945680 & 4 & x & & \\
210967369 & 4 & & x & \\
211436876 & 5/18 & x & & \\
218701083 & 7 & & x & \\
220356827 & 8 & x & x & x \\
220471100 & 8 & x & & \\
246022853 & 12 & & x & \\
246048459 & 12 & & x & \\
246078343 & 12/19 & & & x \\
246163416 & 12 & & x & \\
246220667 & 12/19 & & & x \\
247223703 & 13 & x & & \\
247422570 & 13 & x & & \\
247560727 & 13 & x & & \\
247744801 & 13 & x & & \\
247874191 & 13 & x & & \\
211572480 & 18 & x & & \\
211705502 & 18 & x & & \\ 
\hline
\end{tabular}
    \label{tab:obs}
\end{table}

\subsection{Stellar characterization}
\label{subsec:stellar}

Robust stellar parameters are critical to ensure unbiased planetary characterization. When available, we update the EPIC catalog data \citep{Huber2017} setting the host stellar parameters of our targets to the ones derived by \citet{Hardegree2020}. They were obtained using a combination of Pan-STARRS DR2 photometry \citep{Flewelling2020}, \gaia data, and spectroscopic parameters from the Large Sky Area Multi-Object Fibre Spectroscopic Telescope (LAMOST, \citet{Cui_2012}) DR5 spectra. \citet{deLeon2021} find that these parameters and the ones obtained with the \texttt{isochrones} package \citep{Morton2015b} using 2MASS \citep{Skrutskie2006} photometry and \gaia parallaxes and extinctions are in good agreement within 1$-\sigma$. For seven of our targets, we also compare their listed stellar parameters with the ones from the \verb|GALAH+ DR3| \texttt{K2-HERMES} survey \citep{Buder2021}. For all of them, except for the metallicities of EPIC 206461841, EPIC 210706310, and EPIC 210967369, the \texttt{K2-HERMES} parameters are in good agreement with the \citet{Hardegree2020} ones. For EPIC 206461841 and EPIC 210768568, for which there are no derived \citet{Hardegree2020} stellar parameters, we use the most recent values from the TESS Input Catalog (TIC) version 8.2 \citep{Paegert2021}. In the case of EPIC 211572480 and EPIC 211705502 (see full discussion in Section \ref{subsec:false}) where, neither \citet{Hardegree2020} or EPIC data is available, we do not report stellar information given the astrometry from \gaia (see Section \ref{subsec:gaia}). The stellar limb darkening coefficients are obtained from the tabulated values in \citet{Claret2018} using the available \teff, \logg, and \feh. Distances to our candidate host stars are obtained from \gaia data \citep{Bailer-Jones2021}. A summary of the stellar parameters of our targets is listed in Table \ref{tab:stars}.

\subsection{Speckle imaging}
\label{subsec:speckle}

High angular resolution imaging of our targets has been made using speckle instruments at three telescopes as listed in Table \ref{tab:obs}.

The speckle observations at the 6-m Large Alt-Azimuthal Telescope (BTA) of the Special Astrophysical Observatory of the Russian Academy of Sciences (SAO RAS) were obtained in October and December 2021 using its digital speckle interferometer based on EMCCD detectors \citep{Maksimov2009}. 10 of our targets were observed using the 550/20, 700/50, and 800/100 nm filters, three with the 550/20, and 700/50 nm ones, and one target using only the 550/20 mm filter. Most (73\%) of the observations were done under good weather conditions, while the remaining ones were done under low SNR conditions. The calibration methods for the speckle images are listed in \cite{Mitrofanova2020}. Positional parameters and magnitude differences were determined using the method described in \cite{Balega2002} and \cite{Pluzhnik2005}. One companion was detected at sub-arcsecond separation (see Table~\ref{tab:speckle_cont}).

The 4.3-m Lowell Discovery Telescope (LDT) speckle observations were obtained in August and September of 2021 using the Quad-camera Wavefront-sensing Six-channel Speckle Interferometer (QWSSI) \citep{Clark2020}. Depending on brightness, one thousand to several thousand speckle frames were taken and subsequently analyzed according to methods detailed in e.g. \citet{Horch2015}. None of the Lowell observations revealed companions, so detection limit curves were constructed from the reconstructed images in each case. These were used to rule out stellar companions with separations and magnitudes that would have been detectable by QWSSI. For these observations, only four of the six wavelength channels were available for use, and of those, the reconstructed images with the highest signal-to-noise were those taken at 880 nm. Thus, only these were used for the final detection limit curves.

Nineteen EPIC targets from this program have been observed by the High-Resolution speckle camera at the 4.1-m Southern Astrophysical Research Telescope (SOAR) in Chile. The instrument and data processing are described in \citet{Tokovinin2018}. The observations were carried out in October-November 2021 (2021.75 to 2021.80) in the $I$ filter (880/140\,nm) using the UNC partner time. Three companions at sub-arcsecond separations were detected (see Table~\ref{tab:speckle_cont}). The resolution limits were from $0.07"$ to $0.1"$ and the typical contrast limit at $1"$ separation was around 4 mag.

\subsection{Archival imaging}
\label{subsec:archival}

Following a similar approach as the one in \citet{deLeon2021}, we downloaded Palomar Observatory Sky Survey (POSS-I) images taken in the 1950s from the Space Telescope Science Institute (STScI) Digitized Survey (DSS)\footnote{\url{https://archive.stsci.edu/cgi-bin/dss_form}} for our targets and compare them to Pan-STARRS DR2\footnote{\url{https://ps1images.stsci.edu/cgi-bin/ps1cutouts}} (taken between 2012, and 2014) cutouts, and with the \ktwo TPFs. We do this to study the possibility of a chance alignment of our targets with a foreground or background star; especially in the cases of stars with relatively high proper motions ($\ge$50 mas $yr^{-1}$) or with low galactic latitudes (as is the case for targets in campaigns C7 and C13). 

\begin{landscape}
\begin{table}
    \caption{Summary of stellar parameters. (a) astrometric goodness of fit; (b) astrometric excess noise significance; (c) Renormalised Unit Weight Error.}
    \scriptsize
\begin{tabular}{ccccccccccccc}
\hline
EPIC &          \rstar [\rsun] &          \mstar [\msun] &             \teff [K] &             \logg [cgs] &                \feh [dex] &  $K_p$ [mag] &  $\verb|GOF_AL|^a$ &  $\verb|D|^b$ & $\verb|RUWE|^c$ & d [pc] & pm [mas/yr] & notes \\
\hline
205979483 &  0.814$^{+ 0.054}_{- 0.051}$ & 0.945$^{+ 0.425}_{- 0.288}$ & 5414 $\pm 138$ &  4.595$^{+ 0.150}_{- 0.150}$ & -0.028 $\pm  0.235$ & 12.77 & 7.19 & 15.1 & 1.41 & 278.08 & 3.57 & $\star\S$ \\
206461841 & 0.746$\pm0.047$ & 0.800$\pm0.093$ & 4893$\pm119$ & 4.596$\pm0.090$ &0.040$\pm$0.048** &10.89 & 2.10 & 24.1 & 1.11 & 100.19 & 118.51 & $\dagger$\\
210418253 & 1.129$^{+ 0.073}_{- 0.067}$ & 1.325$^{+ 0.590}_{- 0.411}$ & 5296$\pm 138$ & 4.455$^{+ 0.150}_{- 0.150}$ & 0.110$\pm 0.235$ & 12.15 & 1.41 & 5.86 & 1.06 & 217.98 & 34.07 & \\
210706310 & 0.954$^{+ 0.055}_{- 0.052}$ & 0.709$^{+ 0.304}_{- 0.219}$ & 5941$\pm 138$ & 4.328$^{+ 0.150}_{- 0.150}$ & -0.252 $\pm$0.081** & 12.29 & 1.33 & 7.07 & 1.05 & 274.92 & 62.46 & \\
210708830 & 0.760$^{+ 0.018}_{- 0.018}$ & 1.067$^{+ 0.203}_{- 0.173}$ & 5342 $\pm 45$ & 4.704$^{0.074}_{-0.074}$ & 0.015$\pm  0.043$ & 13.26 & 1.35 & 0.62 & 1.05 & 261.71 & 3.78 & \\
210768568 & 1.375$\pm0.068$ & 1.018$\pm0.131$ & 5711$\pm105$ & 4.1693$\pm0.074$ &0.1415$\pm0.0152$ &11.94 & 2.66 & 14.1 & 1.09 & 295.87 & 62.34 & $\dagger$ \\
210945680 & 1.059$^{+ 0.018}_{- 0.017}$ & 0.994$^{+ 0.082}_{- 0.075}$ & 5969$\pm 20$ & 4.386$^{+ 0.031}_{- 0.031}$ & 0.115$\pm  0.017$ & 11.32 & -0.14 & 19.6 & 0.99 & 226.52 & 24.37 & \\
210967369 & 0.953$^{+0.058}_{-0.055}$ & 0.845$^{+0.372}_{-0.252}$ & 5534 $\pm 138$ & 4.411 $\pm 0.150$ & 0.320$\pm$0.071** & 12.40 & 5.44 & 0.08 & 1.28 & 266.98 & 29.01 & \\
211436876 & 1.057$^{+ 0.022}_{- 0.020}$ & 0.992$^{+ 0.068}_{- 0.063}$ & 5992$\pm 14$ & 4.386$^{+0.023}_{-0.023}$ & -0.095$\pm  0.012$ & 12.30 & -2.72 & 2.36 & 0.88 & 370.04 & 14.73 & \\
218701083 & 1.476$^{+ 0.093}_{- 0.086}$ & 1.198$^{+ 0.521}_{- 0.359}$ & 6262$\pm 138$ & 4.178$^{+ 0.150}_{- 0.150}$ &-0.184 $\pm 0.235$ & 12.49 & -1.43 & 0.85 & 0.93 & 544.18 & 8.99 & \\
220356827 & 1.270$^{+ 0.086}_{- 0.084}$ & 0.982$^{+ 0.437}_{- 0.298}$ & 5986$\pm 138$ & 4.222$^{+ 0.150}_{- 0.150}$ & 0.030$\pm 0.235$ & 12.58 & -0.84 & 2.68 & 0.97 & 504.52 & 1.95 & \\
220471100 & 0.960$^{+ 0.035}_{- 0.035}$ & 1.365$^{+ 0.234}_{- 0.197}$ & 5197$\pm 37$ & 4.609$^{+ 0.061}_{- 0.061}$ & 0.108$\pm 0.035$ & 14.21 & 21.02 & 21.1 & 1.83 & 547.04 & 21.99 & $\S$\\
246022853 & 1.114$^{+ 0.162}_{- 0.141}$ & 0.883$^{+ 0.470}_{- 0.315}$ & 5909$\pm 138$ & 4.287$^{+ 0.150}_{- 0.150}$ &-0.147$\pm 0.235$ & 11.48 & 32.43 & 297 & 3.12 & 466.31 & 30.48 & $\star\S$ \\
246048459 & 0.645$^{+ 0.047}_{- 0.044}$ & 0.769$^{+ 0.341}_{- 0.239}$ & 4514$\pm 138$ & 4.703$^{+ 0.150}_{- 0.150}$ &-0.368$\pm 0.235$ & 11.60 & 5.22 & 0.00 & 1.07 & 83.97 & 14.66 & \\
246078343 & 0.700$^{+ 0.055}_{- 0.048}$ & 0.808$^{+0.364}_{-0.256}$ & 4116$\pm 138$ & 4.656$^{+ 0.150}_{- 0.150}$ &-0.205$\pm 0.235$ & 14.57 & 2.92 & 0.97 & 1.16 & 292.52 & 3.37 & \\
246163416 & 0.515$^{+ 0.026}_{- 0.025}$ & 0.512$^{+ 0.058}_{- 0.055}$ & 3734$\pm 138$ & 4.724$^{+ 0.062}_{- 0.068}$ &-0.101$\pm 0.235$ & 13.48 & 24.62 & 131 & 2.44 & 85.52 & 199.01 & $\star\S$ \\
246220667 & 0.732$^{+ 0.055}_{- 0.052}$ & 0.814$^{+ 0.363}_{- 0.251}$ & 4343$\pm 138$ & 4.621$^{+ 0.150}_{- 0.150}$ &-0.102$\pm 0.235$ & 13.96 & 0.29 & 0.00 & 1.01 & 255.88 & 3.86 & \\
247223703 & 0.741$^{+ 0.056}_{- 0.051}$ & 0.861$^{+ 0.394}_{- 0.264}$ & 4434$\pm 138$ & 4.631$^{+ 0.150}_{- 0.150}$ &-0.087$\pm 0.235$ & 14.28 & 1.90 & 1.07 & 1.07 & 257.94 & 31.16 & \\
247422570 & 0.977$^{+ 0.066}_{- 0.063}$ & 0.893$^{+ 0.406}_{- 0.275}$ & 5590$\pm 138$ & 4.412$^{+ 0.150}_{- 0.150}$ & 0.014$\pm 0.235$ &15.11 & 1.21 & 0.00 & 1.05 & 668.62 & 1.46 & \\
247560727 & 0.779$^{+ 0.059}_{- 0.054}$ & 0.693$^{+ 0.301}_{- 0.212}$ & 5634$\pm 138$ & 4.494$^{+ 0.150}_{- 0.150}$ &-0.130$\pm 0.235$ & 15.43 & -0.83 & 0.00 & 0.96 & 680.62 & 4.92 & \\
247744801 & 0.975$^{+ 0.065}_{- 0.059}$ & 1.027$^{+ 0.451}_{- 0.317}$ & 5214$\pm 138$ & 4.466$^{+ 0.150}_{- 0.150}$ & 0.028$\pm 0.235$ & 13.83 & -0.71 & 0.00 & 0.97 & 368.81 & 36.72 & \\
247874191 & 1.290$^{+ 0.085}_{- 0.080}$ & 1.053$^{+ 0.468}_{- 0.315}$ & 5998$\pm 138$ & 4.241$^{+ 0.150}_{- 0.150}$ &-0.163$\pm 0.235$ & 14.54 & 0.67 & 0.00 & 1.02 & 865.32 & 4.67 & \\
211572480 & -- & -- & -- & -- & -- & 14.10 & 174.37 & 1530 & 12.25 & 499.84 & 8.33 & $\star\S$\\
211705502 & -- & -- & -- & -- & -- & 13.21 & 30.94 & 57.8 & 2.42 & 774.16 & 6.02 & $\S$\\
\hline
\multicolumn{13}{l}{$\star$: detected companion in Speckle data, $\S$: probable binary from \gaia data, $\dagger$: data from TIC catalogue \citep{Paegert2021}, **: \feh from GALAH+ DR3 survey \citep{Buder2021}.}
\end{tabular}

    \label{tab:stars}
\end{table}
\end{landscape}

\subsection{Gaia eDR3 photometry and astrometry}
\label{subsec:gaia}

We use \gaia eDR3 to search for neighboring stars close to our targets. We do this to minimize the chances of biasing our planetary candidates' characterization due to the presence of unresolved stars within the \eve photometric aperture \citep{Evans2016}. Resolved \gaia detections are plotted in our \ktwo TPF Validation Images (see Figure \ref{fig:gaia_cont}) and checked during our vetting and validation procedure (see Sections \ref{subsec:vetting} and \ref{subsec:validation}). We also check for indirect evidence of potential contamination from unresolved stars using the available \gaia data for our targets. First, we use \gaia Astrometric Goodness of Fit of the astrometric solution for the source in the Along-Scan direction (\verb|GOF_AL|) and the Astrometric Excess Noise significance (\verb|D|) to determine which of our targets could be poorly-resolved binaries \citep{Evans2018,Gandhi2022}. \citet{Evans2018} manually set \verb|D|>5 and \verb|GOF_AL|>20 to match the boundary between confirmed binaries and confirmed singles. Given that no star in our candidate sample is too bright or has a very high proper motion, we do not expect any large offset of these parameters to be related to difficulties in modeling saturated or fast-moving stars. Additionally, we use the Renormalised Unit Weight Error (\verb|RUWE|) provided by \gaia eDR3 as an extra parameter to identify binary systems from astrometric deviations \citep{Penoyre2022}. \gaia sources with \verb|RUWE| values significantly greater than one (i.e. significant deviations from the single-body model fit) can be candidate binary systems. We use a rather restrictive value of \verb|RUWE|>1.4 as our threshold to determine which of our targets might be unresolved binaries. We choose this value from our analysis of EPIC 205979483 (see Section \ref{subsec:speckle}) which has \verb|D|=15.1, \verb|GOF_AL|=7.19, and \verb|RUWE|=1.41. Although \verb|GOF_AL| is smaller than its proposed threshold value, \verb|D| exceeds it. In addition, we also detect a very faint object separated 0.5751$\arcsec$ from our target using SOAR speckle imaging data confirming the binary/contaminated nature of this target. We present these three parameters for each of the targets in our sample in Table \ref{tab:stars}. A full discussion on these parameters and their implications on the candidate dispositions is presented in Section \ref{sec:results}. 

\section{Methods}
\label{sec:methods}

\subsection{\tfaw and \tls}
\label{subsec:tfaw}

\tfaw \citep{delSer2018} is a wavelet-based algorithm that is able to denoise and reconstruct the input signal without any \textit{a priori} feature assumption or modify its astrophysical properties. It combines the Stationary Wavelet Transform (hereafter SWT) potential to characterize and denoise the input signal with the detrending and systematic removal capabilities of \texttt{TFA} \citep{Kovacs2005}.

The \tfaw detrending and denoising algorithm can be summarized as follows (see \citet{delSer2018} for a complete description): 1) as with \texttt{TFA}, a template of reference stars is used to create an initial filter to remove trends and systematics from the target light curve, 2) using the detrended light curve, the noise-free underlying signal is estimated by means of the SWT decomposition levels and its corresponding power spectrum. 3) outliers are removed based on the previous SWT signal estimation and the high-frequency noise contribution is removed from the target light curve using the SWT decomposition level/s with the highest frequency resolution/s, 4) a search for significant periodicities is run over the denoised signal, 5) if a significant period is found, the detrended and denoised light curve is phase folded and the underlying signal (i.e. the astrophysical signal) is estimated using the SWT, and 6) the final noise-free signal is iteratively denoised and reconstructed.

As shown in \cite{delSer2020}, \texttt{TFAW} delivers both better photometric precision and planet characterization than any previous detrending method applied to \ktwo light curves. In order to increase the  transit detection potential of the algorithm, we make use of \tls during the \tfaw period search step. \tls makes use of the stellar limb-darkening parameters of the target star and includes the effects of planetary ingress and egress in the search for transit-like features. This leads to an increase in the detection efficiency compared to the commonly used \texttt{BLS} \citep{Kovacs2002} and is particularly suited for the detection of small planets. The combination of \tls and \tfaw can yield detection efficiencies for \ktwo data $\sim$8.5$\times$ higher for \texttt{TFAW}-corrected light curves than for \texttt{EVEREST 2.0} ones, specially for faint magnitudes \citep{delSer2020}.

\subsection{Vetting procedure}
\label{subsec:vetting}

We follow a transit search, vetting, and False-Positive Probability (FPP) approach similar to the one detailed in \citet{Heller2019}. A candidate period is considered to be significant if its peak in the \texttt{TLS} power spectrum during \tfaw period search step (see Section \ref{subsec:tfaw}) has a Signal Detection Efficiency (SDE$_{\rm TLS}$) above 9.0 (i.e. false-positive rate $<$10$^{-4}$ \citep{Hippke2019}). Any target light curve that matches these criteria undergoes the full \tfaw iterative denoising and signal reconstruction. Following the recommendation in \citet{Luger2018}, and to avoid any over-fitting of the transit signal by the PLD correction, we mask the candidate transits and recompute their \eve light curves prior to rerunning the full \tfaw correction.

Our vetting procedure consists of the following steps: 1) we visually inspect all \tfaw-corrected light curves and keep those that have transit-like features. 2) we compare the \tls periodograms for the original \eve and the \tfaw light curves to verify that we have not introduced any systematic signature in the data during the \tfaw analysis. We also compare our results with the available \ktwo pipeline and \texttt{K2SFF} \citep{Vanderburg2014} light curves, and with PLD-corrected light curves obtained from \ktwo TPFs using the \texttt{lightkurve} \citep{Lightkurve2018} package. The latter is done with extra care if a nearby star is contaminating the \eve aperture. In this case, we check how the transit feature is affected for different aperture sizes and positions. 3) We iteratively run \tls to search for extra transiting signals in the light curve. 4) We also rule out that no other light curve in the same CCD module presents transit-like features with similar periods and transit epochs as the candidates. We also check for any systematic bias by plotting the overall distribution of periods in the CCD module and comparing them to our candidate period. 5) Using \tls output, we check that all transiting signals have good signal-to-noise ratios (SNR) (long period candidates should have SNR$>$10) and that the average depth of the odd/even transits agree within $<$3$\sigma$, and secondary eclipses at half an orbital phase after the candidate transit are not present at the $>$3$\sigma$ level. We visually inspect the transits positions in the light curves and require that they are at least 0.5 days away from the beginning or end of any gaps in their light curves to avoid false positives, especially in the case of long period candidates. 6) we cross-match our candidates with the most up-to-date (March 2022) lists of confirmed or candidate exoplanets from the NASA Exoplanet Archive \footnote{\url{https://exoplanetarchive.ipac.caltech.edu}} or in the Vizier database \citep{Adams2016,Barros2016,Crossfield2016,Vanderburg2016,Crossfield2018,Hirano2018,Livingston2018,Mayo2018,Dattilo2019,Kruse2019,Castro-Gonzalez2020,Kovacs2020,Zink2020,Adams2021,Castro-Gonzalez2021,deLeon2021,Zink2021,Christiansen2022}. 7) we run \texttt{EDI-Vetter Unplugged} \footnote{\url{https://github.com/jonzink/EDI_Vetter_unplugged}}, a simplified version of \texttt{EDI-Vetter} \citep{Zink2020}, that uses the output from \tls to identify false-positive transit-like signals using a battery of tests: transit outliers, individual transit, even/odd transit, secondary transit, phase coverage, period and transit duration limits, period alias, and flux contamination checks. 8) finally, we use high-resolution imaging and \gaia photometry and astrometry (see Sections \ref{subsec:archival} and \ref{subsec:gaia}) to evaluate contamination from other stellar sources.

\subsection{Centroid testing}
\label{subsec:centroid}

The centroid test (i.e. measuring the changes in the position of the centroid of a target star during the transit) is an excellent tool to discern between bona fide planetary candidates and background transiting sources \citep{Batalha2010,Bryson2013} for \textit{Kepler} light curves. After the failure of the second reaction wheel of \textit{Kepler} primary mission in 2013, the \ktwo mission relied on the two remaining reaction wheels to balance against the radiation pressure of the Sun. In this way, \ktwo was able to reduce the pointing drifts and achieve a photometric precision close to the one for the original \textit{Kepler} mission \citep{Vanderburg2014}. However, as the spacecraft continuously and slightly rotated out of position and then was readjusted to its original pointing, this resulted in increased correlated noise in the \ktwo light curves on time scales typical of planetary transit durations. Although some algorithms such as \eve (which makes use of the PLD technique) were able to correct this effect, some previously validated planets have been found to be background eclipsing binaries (BEBs) near or within the photometric aperture.
As a final vetting tool in our procedure, we use \texttt{vetting} \citep{Hedges2021}, a Python-based implementation of the centroid test that takes into account the \ktwo motion. It makes use of the \ktwo TPF information, the transit period, $T_0$, and duration to return two distributions of centroids (in transit and out of transit), and a p-value corresponding to the likelihood that they are both drawn from the same, underlying distribution. We also pass our transit depths to the code to get the distance to which a companion can be ruled out. We added a modification to the code in order to account for the aperture size used by \eve as it is usually larger than the one used by the standard \textit{Kepler} pipeline. We use the same threshold for the p-value as \citet{Christiansen2022} to separate between false positives and possible planetary candidates. Only those candidates with $p>0.05$ are considered vetted planetary candidates.

\subsection{Stellar blending}
\label{subsec:blending}

The aperture radius of the \eve pipeline is usually $\sim$4 pixels in radius. Given \ktwo's relatively large pixel size (3.98$\arcsec$), it leads to the possibility of other objects being present within the photometric aperture. This flux contamination leads to a decrease in the observed transit depth, and, as a consequence, to biased planetary characterization \citep{Daemgen2009}. As explained in Section \ref{subsec:vetting}, in those cases where the contaminating object is far enough away from the target star, we recompute the light curve modifying the aperture position and size, and studying if there is any change in the transit depth. However, in some cases, the object is within a couple of pixels from the target, making it impossible to deblend their flux contributions. For these cases, we quantify the photometric contamination by computing the dilution factor \citep{Daemgen2009,Livingston2018} as $\gamma = 1 + 10^{0.4\Delta m}$, where $\Delta m$ denotes the difference between the magnitude of the fainter contaminating star and the brighter target star in a given photometric band (i.e. the formula assumes the brighter component to be the variable component). The relationship between the observed transit depth ($\delta '$) and the true transit depth ($\delta$) is then given by $\delta ' = \gamma^{-1} \delta$. Following the notation in \citet{Castro-Gonzalez2021} and \citet{deLeon2021}, we compute the dilution factors $\gamma_{pri}$ and $\gamma_{sec}$ considering that the transiting signal comes from the target (primary) star or from the nearby (secondary) star with transit depths $\delta_{pri}$, and $\delta_{sec}$ respectively. Faint eclipsing binaries, when blended, can have their eclipses diluted to depths similar to planetary transit ones. Assuming that their hypothetical eclipses can not be greater than 100\% (i.e. $\delta_{sec}\le1$), then if $\delta '>\gamma_{sec}^{-1}$, the observed depth $\delta '$ is too deep to be caused by the fainter neighboring star. We compare these results to the nearby star tests done by \tri to decide the final dispositions of those targets with contaminating/blended sources.

\subsection{Transit parameters modelling}
\label{subsec:MCMC}

To model the transit light curves, we use the probabilistic Keplerian Orbit model provided by the \texttt{exoplanet} package \citep{Foreman-Mackey2021}, and a quadratic limb darkening law as parameterized by \citet{Kipping2013} (implemented in \textsf{exoplanet}). As explained in Section \ref{subsec:stellar}, the limb darkening coefficients are obtained from the tabulated values in \citet{Claret2018}. We include a Gaussian Process (GP) model (implemented using \textsf{celerite2} \citep{Foreman-Mackey2017, Foreman-Mackey2018} consisting on a Matérn 3/2 kernel plus a jitter or "white" noise term to generalize the likelihood function in order to consider correlated noise, non-periodic variations and to minimize the bias of the inferred parameters. In the case of ultra-short-period (USP) candidates, following \citet{Adams2016}, we use super-sampling (7 points for 4$\le$period$\sim$24hr) to fit the transits given the few observations per transit for very short transit durations.

We assume circular orbits (i.e. eccentricity=0) and fit the following five transit parameters: the transit epoch, $T_0$, the orbital period, $P$, the semi-major axis of the orbit, $a$, the planetary radius, $R_{p}$, and the inclination of the orbit, $i$. We also include as free parameters the stellar radius, the logarithm of the Gaussian errors, a constant light curve baseline, and the quadratic limb darkening coefficients.

We use the MCMC sampler provided by \textit{PyMC3} \citep{Salvatier2016} to explore the posterior probability distribution. We optimize the model parameters to find the maximum a posteriori (MAP) parameters as a starting point for the MCMC sampler. We consider normal distributions of the priors for all free parameters with the exception of the stellar radius which is bounded by its cataloged uncertainties. We give wide enough bounds to let the chains explore the parameter space without getting close to the bound limit. We run the sampler with 100 walkers, 10,000 iterations with a burn-in phase of 2,000 iterations to ensure that each walker runs for more than 50 auto-correlation times for each parameter and that the mean acceptance fraction is between 0.25 and 0.5 \citep{Bernardo1996,Foreman-Mackey2013}. We also inspect the MCMC chains and posterior distributions as well as the final fitted model to ensure they are well-behaved. In Table \ref{tab:planet} we report the 50\% quantiles as the best-fit parameters and their upper and lower errors computed from the 25\% and 75\% quantiles, respectively. The transit light curves and their best-fit transit model are show in Figures \ref{fig:singplanets} and \ref{fig:multiplanets}.

\subsection{False Positive Probabilities and Validation}
\label{subsec:validation}
Statistical validation, i.e. the statistical confirmation that a transiting signal arises from a planet and not from an astrophysical false positive, is a challenging issue. Several planetary transit validation methods have been developed in the literature over the years \citep{Morton2012,Lissauer2014,Diaz2014,Morton2015,Torres2015,Giacalone2020,Giacalone2021,Armstrong2021} based in different techniques like Bayesian methods or machine learning. 
\vespa \citep{Morton2012,Morton2015} has been largely used to validate planets from the \textit{Kepler} and \ktwo missions (e.g., \cite{Livingston2018,Heller2019,Dattilo2019,Castro-Gonzalez2020,deLeon2021}). Using the stellar and photometric properties of the host star, \vespa generates a synthetic sample of stars around the target by means of the \texttt{isochrones} \footnote{\url{https://isochrones.readthedocs.io/en/latest/}} package \citep{Morton2015b}. Then, \vespa calculates the probabilities of the transiting signal being caused by different scenarios: non-associated blended eclipsing binaries, eclipsing binaries, hierarchical triples, non-associated stars with transiting planets, and lastly, the transiting planet scenario around the target star. Planetary candidates with False Positive Probabilities (FPPs) lower than 1\% are considered to be validated planets.

However, \citet{Armstrong2021} find concerning discrepancies with \vespa and caution against using only one method to validate planetary candidates. The use of independent methods is desirable to reduce the risk of model-dependent biases that could impact several exoplanet research fields and follow-up observations. To minimize the risk of misclassifying our planet candidates, we quantify their FPPs by combining the results from \vespa with those from \tri \citep{Giacalone2020,Giacalone2021}. \tri is a Bayesian tool that vets and validates planet candidates by calculating the probabilities for a set of transit-like scenarios using the target light curve, the photometric aperture, the stellar properties of the host star, and current models of planet occurrence and stellar multiplicities. It also computes the probability that the observed transit comes from a resolved, nearby star (denoted nearby false positive probability or NFPP). A planetary candidate is considered to be validated if they have FPP $<0.015$ and NFPP $<10^{-3}$.  

We supply both software with the \tfaw phase folded light curves of our candidates, their celestial coordinates, and the stellar parameters and photometric data of their host star. We also compute a limiting aperture radius obtained from the \texttt{EVEREST 2.0} information for each star and include the speckle imaging contrast curves (see Section \ref{subsec:speckle}) as additional constraints. In the particular case of \vespa, following \citet{Mayo2018}, we also include the \texttt{secthresh} value, computed using the 3-$\sigma$ deviation of the out-of-transit phase-folded light curve. This way, we consider the fact that no secondary transit is detected at any phase.

In the case of multi-planetary candidate systems, and given that neither \vespa nor \tri consider multiplicity, we apply a correction factor for the computed FPPs to account for the low probability of multiple false-positive signals \citep{Lissauer2011}. \citet{Lissauer2012} introduce correction factors derived from \textit{Kepler} data of $\sim$25, and $\sim$50, for systems with two and three or more planets respectively. Given the different Galactic environments and observational constraints of the \ktwo mission, \citet{Castro-Gonzalez2020} computed very similar correction factors of $\sim$28, and $\sim$40, based on candidates from several \ktwo campaigns.

\subsection{Mass-radius estimation and Multi planet resonance analysis}
\label{subsec:resonance}

In those stellar systems in which more than one transiting planet candidate is found, low-order mean motion resonances are estimated using a Python-based analytical tool \verb|analytical-resonance-widths| \footnote{\url{https://github.com/katvolk/analytical-resonance-widths}}. The algorithm originally uses the \citet{Lissauer2011b} mass-radius relationship, based on fitting a power-law relation to Earth and Saturn only, to estimate the masses of a given multi-planetary system. In our case, we use the Python-based \verb|mrexo|\footnote{\url{https://github.com/shbhuk/mrexo}} tool for non-parametric fitting and analysis of the mass-radius relationship for exoplanets. The code allows to choose between the mass-radius relationship obtained from the M-dwarf sample data set of \citet{Kanodia2019}, and the one obtained using the complete \textit{Kepler} exoplanet sample of \citet{Ning2018}. However, two effects have to be taken into account in order to estimate the masses of planets with $R_p\lesssim$1.2\rearth and to avoid biased results: first, the small amount of Earth-sized planets with a measured mass around FGK dwarf stars, and second, the M-dwarf dataset is strongly affected by the presence of the TRAPPIST-1 planets \citep{Gillon2017}. Thus, for planets with $R_p\lesssim$1.2\rearth, we estimate their masses with the widely used program \texttt{FORECASTER}\footnote{\url{https://github.com/chenjj2/forecaster}} \citep{Chen2016}. It uses a broken power-law to fit the mass-radius relationship across a wide range of planetary masses and radii, to take into account the potential differences in the physical mechanisms responsible for the planetary formation. To estimate the mass of each of our candidate planets, we select the corresponding sample, and algorithm depending on the cataloged spectral types of their host stars (see Section \ref{subsec:stellar}) and their MCMC best-fit planetary radius (see Section \ref{subsec:MCMC}).

\subsection{Candidate dispositions}
\label{subsec:disposition}

Following the vetting and validation procedure described in the previous sections, we assign the final dispositions of each of our candidates. First, those objects with \verb|D|>5, \verb|GOF_AL|>20 and \verb|RUWE|>1.4 (see Section \ref{subsec:gaia}) are designated as false positives (FP). If any combination of two of these parameters is above the previous limits, we also consider the target as a FP. Regardless of their values, if a contaminating object is found in the speckle imaging data, we also consider the candidate as a FP.

If a nearby star is found within the \eve aperture that cannot be established as a potential nearby eclipsing binary (using \gaia astrometric parameters), the candidate is designated as a planet candidate (PC). In the case that the contaminating star is far enough to recompute a new \eve aperture minimizing the parasitic flux, the light curve is recomputed to obtain the undiluted depth and the true radius of the planet candidate.

We also adopt an upper limit of $R_{p}<8R_{\oplus}$ similar to previous works \citep{Mayo2018,Giacalone2020,deLeon2021} to denote possible FPs that can be of brown dwarf or low-mass star origin. Following \citet{Kipping2014} we also check that the MCMC-derived stellar densities are consistent with the ones obtained from the cataloged values. The agreement between these two values is indicative of the transit coming from a planet and not from another astrophysical source.  

Finally, we use the FPPs computed by \vespa and \tri to assign the final disposition of the remaining candidates. Those planets with 1\%<FPP$_{\vespa}$ and FPP$_{\tri}$<99\% are designated as PC while those with FPP$_{\vespa}$ and FPP$_{\tri}$<1\% are designated as validated planets (VP). The final dispositions of each of our candidates and their FPPs are listed in Table \ref{tab:planet}.

    \begin{landscape}
        \begin{table}
        \centering
        \caption{Candidate MCMC posterior transit parameters, FPPs and dispositions.}
        \label{tab:planet}
        \scriptsize
\renewcommand{\arraystretch}{1.5}
\begin{tabular}{llllllllllll}
\hline
EPIC      & $T_0$ (BJD-2454833)             & \multicolumn{1}{l}{$P$ (days)} & \multicolumn{1}{l}{$a$ ($AU$)} & \multicolumn{1}{l}{$R_p (R_{\oplus})$} & \multicolumn{1}{l}{$i (^{\circ})$} & $S/S_{\oplus}$ & \multicolumn{1}{l}{SDE$_{\rm TLS}$} & FPP$_{\rm VESPA}$ & FPP$_{\rm TRICERATOPS}$ & notes & \multicolumn{1}{l}{Disposition} \\ \hline
205979483 & 2145.1578$^{+0.0016}_{-0.0015}$ & 12.4292$^{+0.0005}_{-0.0007}$  & 0.0974$^{+0.0027}_{-0.0031}$   & 1.1126$^{+0.0450}_{-0.0456}$           & 90.000$^{+0.3968}_{-0.3960}$       & 53.836         & 17.909                        & -                   & -                         &       & FP/CC                           \\
206461841 & 2149.6596$^{+0.0022}_{-0.0024}$ & 10.4404$^{+0.0007}_{-0.0011}$  & 0.0828$^{+0.0034}_{-0.0035}$   & 0.6739$^{+0.0394}_{-0.0370}$           & 89.9971$^{+0.5356}_{-0.5338}$      & 41.745         & 14.324                        & 0.3081              & 0.1168                    &       & PC                           \\
210418253 & 2233.8056$^{+0.0017}_{-0.0027}$ & 23.9683$^{+0.0044}_{-0.0013}$  & 0.1791$\pm0.0046$              & 1.6119$^{+0.1413}_{-0.1343}$           & 91.0247$^{+0.1476}_{-1.8806}$      & 28.046         & 14.439                        & 0.3004              & 0.2533                    &       & PC                              \\
210706310 & 2229.7095$\pm0.0015$            & 5.1718$\pm0.0002$              & 0.0510$^{+0.0024}_{-0.0029}$   & 0.8891$^{+0.0529}_{-0.0443}$           & 90.0307$^{+1.5873}_{-1.5942}$      & 391.097        & 15.612                        & 0.00239             & 0.1527                    &       & PC                              \\
210708830 & 2231.1694$\pm0.0015$            & 5.7408$\pm0.0002$              & 0.0627$^{+0.0026}_{-0.0029}$   & 1.0668$^{+0.0573}_{-0.0444}$           & 89.9948$^{+0.9284}_{-0.9156}$      & 107.349        & 12.503                        & 0.8392              & 0.2033                    &       & PC                              \\
210768568 & 2231.8355$^{+0.0019}_{-0.0018}$ & 3.2141$\pm0.0002$              & 0.0511$^{+0.0021}_{-0.0028}$   & 0.9898$^{+0.0498}_{-0.0486}$           & 90.0008$^{+2.2467}_{-2.5589}$      & 619.038        & 13.63                         & 0.0016              & 0.015                     &       & VP                              \\
210945680 & 2242.5226$^{+0.0019}_{-0.0013}$ & 20.5949$^{+0.0022}_{-0.0010}$  & 0.1477$^{+0.0047}_{-0.0046}$   & 1.5097$^{+0.0741}_{-0.0790}$           & 89.0790$^{+0.2482}_{-0.1533}$      & 58.550         & 17.533                        & 0.05459             & 0.116                     &   $\star$    & PC                           \\
210967369 & 2229.4873$^{+0.0028}_{-0.0030}$ & 7.1149$\pm0.0006$              & 0.0589$^{+0.0021}_{-0.0024}$   & 0.9835$^{+0.0483}_{-0.0499}$           & 90.0013$^{+0.8089}_{-0.8110}$      & 220.293        & 13.245                        & 0.8165              & 0.266                     &       & PC                              \\
218701083 & 2471.8892$\pm0.0030$            & 5.0521$\pm0.0004$              & 0.0529$^{+0.0019}_{-0.0022}$   & 2.1388$^{+0.0898}_{-0.0924}$           & 89.9891$^{+1.5234}_{-1.5050}$      & 1073.998       & 11.81                         & 0                   & 0.3271                    &       & PC                              \\
220356827 & 2563.1692$^{+0.0038}_{-0.0043}$ & 4.7535$\pm0.0005$              & 0.0559$^{+0.0034}_{-0.0039}$   & 1.5147$^{+0.1477}_{-0.1000}$           & 89.9707$^{+2.4107}_{-2.4000}$      & 594.594        & 12.743                        & 0.9834              & 0.4558                    &       & FP?                            \\
220471100 & 2560.0112$\pm0.0016$            & 7.3274$^{+0.0002}_{-0.0003}$   & 0.0760$^{+0.0023}_{-0.0026}$   & 2.1225$^{+0.0731}_{-0.0667}$           & 89.9968$^{+0.6323}_{-0.6336}$      & 104.428        & 26.11                         & -                   & -                         &       & FP/CC                           \\
246022853 & 2911.3949$^{+0.0036}_{-0.0042}$ & 10.7228$\pm0.0011$             & 0.0879$^{+0.035}_{-0.0037}$    & 1.2998$^{+0.0689}_{-0.0690}$           & 89.9955$^{+0.7779}_{-0.7816}$      & 175.686        & 10.433                        & -                   & -                         &       & FP/CC                           \\
246048459 & 2905.6563$^{+0.003}_{-0.0023}$  & 2.0507$\pm0.0001$              & 0.0258$^{+0.0023}_{-0.0025}$   & 0.4064$^{+0.0302}_{-0.0266}$           & 90.0112$^{+2.4869}_{-2.5015}$      & 232.818        & 10.592                        & 0.3407              & 0.9836                    &       & FP                              \\
246078343 & 2905.6924$^{+0.0017}_{-0.0021}$ & 0.8094$\pm0.00003$             & 0.0117$^{+0.0009}_{-0.0012}$   & 0.7599$^{+0.0758}_{-0.0496}$           & 89.9518$^{+6.0250}_{-6.1473}$      & 921.755        & 17.088                        & 6$\times$10$^{-4}$               & 0.009                         &       & VP                              \\
246078343 & 2906.1673$^{+0.0023}_{-0.0024}$ & 5.3301$\pm0.0003$              & 0.0426$^{+0.0016}_{-0.0019}$   & 1.2327$^{+0.0565}_{-0.0593}$           & 89.9985$^{+0.8243}_{-0.8117}$      & 69.529         & 14.329                        & 2$\times$10$^{-4}$             & 0.007                    &   $\dagger$    & VP                              \\
246163416 & 2905.8554$\pm0.0007$            & 0.8768$\pm0.00002$             & 0.0034$\pm0.0001$              & 8.4683$^{+4.3149}_{-2.8121}$           & 140.52$^{+2.2073}_{-2.1373}$       & 4001.692       & 32.173                        & -                   & -                         &       & FP/CC                           \\
246220667 & 2907.2563$^{+0.0010}_{-0.0012}$ & 6.6690$\pm0.0002$              & 0.0552$^{+0.0017}_{-0.0023}$   & 1.9287$^{+0.0630}_{-0.0728}$           & 89.9973$^{+0.6830}_{-0.6794}$      & 56.130         & 23.962                        & 0.0097            & 0.0067                     &       & VP                              \\
246220667 & 2909.0993$\pm0.0013$            & 4.3696$\pm0.0001$              & 0.0487$^{+0.0027}_{-0.0031}$   & 1.2190$^{+0.1004}_{-0.0731}$           & 90.0045$^{+1.3925}_{-1.3991}$      & 72.113         & 21.1445                       & 0.001              & 0.005                    &       & VP                              \\
247223703 & 2989.9313$^{+0.0092}_{-0.0057}$ & 3.1764$\pm0.0005$              & 0.0377$^{+0.0048}_{-0.0044}$   & 0.9532$^{+0.0796}_{-0.0734}$           & 89.9893$^{+2.1050}_{-2.1043}$      & 133.976        & 14.586                        & 0.5567              & 0.0943                    &       & PC                              \\
247422570 & 2990.1456$^{+0.0043}_{-0.0028}$ & 5.9382$^{+0.0006}_{-0.0004}$   & 0.0586$^{+0.0032}_{-0.0033}$   & 2.1160$^{+0.1050}_{-0.1052}$           & 89.9980$^{+1.3080}_{-1.3214}$      & 243.518        & 19.11                         & 0                   & 0.0036                    &       & VP                              \\
247560727 & 2989.6718$^{+0.0026}_{-0.0019}$ & 3.3733$\pm0.0002$              & 0.0279$^{+0.0013}_{-0.0016}$   & 1.5839$\pm0.0781$                      & 89.9954$^{+1.7526}_{-1.7522}$      & 109.437        & 21.932                        & 0.0013             & 0.0135                     &       & PC/CC                           \\
247560727 & 2993.7254$^{+0.0017}_{-0.0018}$ & 8.4356$^{+0.001}_{-0.0006}$    & 0.0708$^{+0.0051}_{-0.0056}$   & 2.9192$^{+0.4080}_{-0.3350}$           & 87.5945$^{+0.3104}_{-0.3697}$      & 704.732        & 14.6306                       & 0.03            & 0.028                    &       & PC/CC                           \\
247744801 & 2994.9350$^{+0.0016}_{-0.0018}$ & 12.5050$^{+0.0007}_{-0.0006}$  & 0.1034$^{+0.0033}_{-0.0037}$   & 1.6730$^{+0.0825}_{-0.0772}$           & 89.9995$^{+0.5221}_{-0.5273}$      & 58.958         & 20.705                        & 0.2907              & 0.2576                    &       & PC                              \\
247874191 & 2989.5529$^{+0.0040}_{-0.0034}$ & 7.6240$^{+0.0007}_{-0.001}$    & 0.0775$^{+0.0036}_{-0.0040}$   & 2.4551$^{+0.1655}_{-0.1378}$           & 90.0133$^{+1.6828}_{-1.7013}$      & 321.730        & 15.903                        & 0.0069              & 0.0485                    &       & PC                              \\
211436876 & 3419.1996$^{+0.0101}_{-0.0072}$ & 1.1524$^{+0.0003}_{-0.0004}$   & 0.0125$^{+0.0009}_{-0.0011}$   & 0.6746$^{+0.0454}_{-0.0392}$           & 89.9962$^{+7.9422}_{-7.9753}$      & 8270.029       & 10.786                         & 0.4624              & 0.1054                    &       & PC         \\  
211572480 & 3421.2436                       & 6.2043                         & 0.0661                         & 1.8527                                 & -              & -              & 15.46                         & -                   & -                         &       & FP                              \\
211705502 & 3418.7764                       & 2.5819                         & 0.0368                         & 1.9422                                 & -             & -              & 19.917                        & -                   & -                         &   *    & FP                              \\
\hline
\multicolumn{12}{l}{$\star$: listed in \citet{Zink2021}; $\dagger$: listed in \citet{Dattilo2019}}; *: listed in \citet{Castro-Gonzalez2021}
\end{tabular}
        \end{table}
    \end{landscape}

\section{Results}
\label{sec:results}

\begin{figure*}
            \includegraphics[clip,trim={0 0 0 0},width=0.8\textwidth]{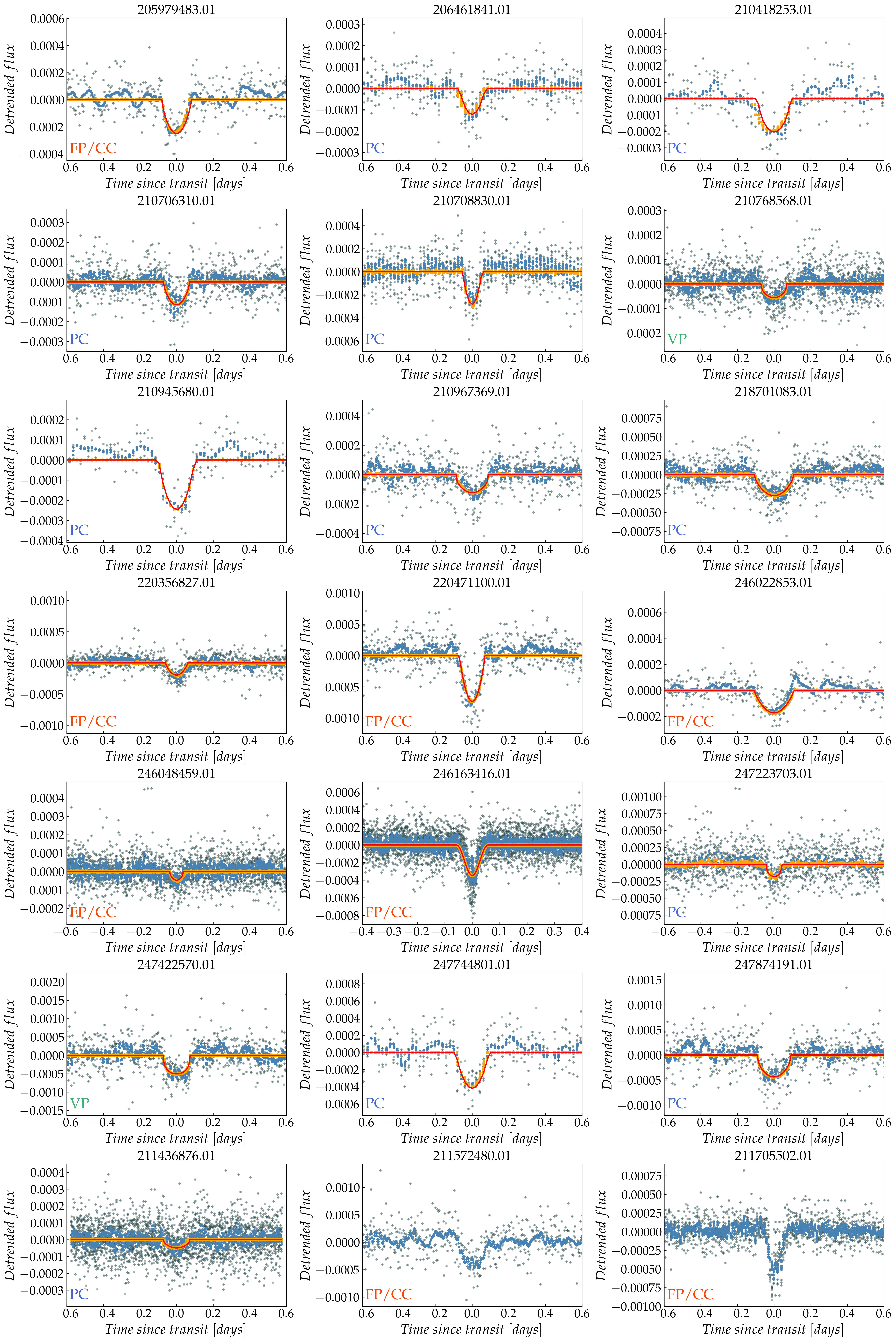}
            \caption{\eve (gray points), \tfaw-corrected (blue points), and \tfaw+GP corrected (orange points) light curves and superposed MCMC best-fit transit model (red line) for all single planet candidates in this work. Final dispositions in the lower left corner (VP=validated planet; PC=planet candidate; FP/CC=false positive/contaminated candidate).}
            \label{fig:singplanets}
        \end{figure*}

\begin{figure*}
            \includegraphics[clip,trim={0 0 0 0},width=0.8\textwidth]{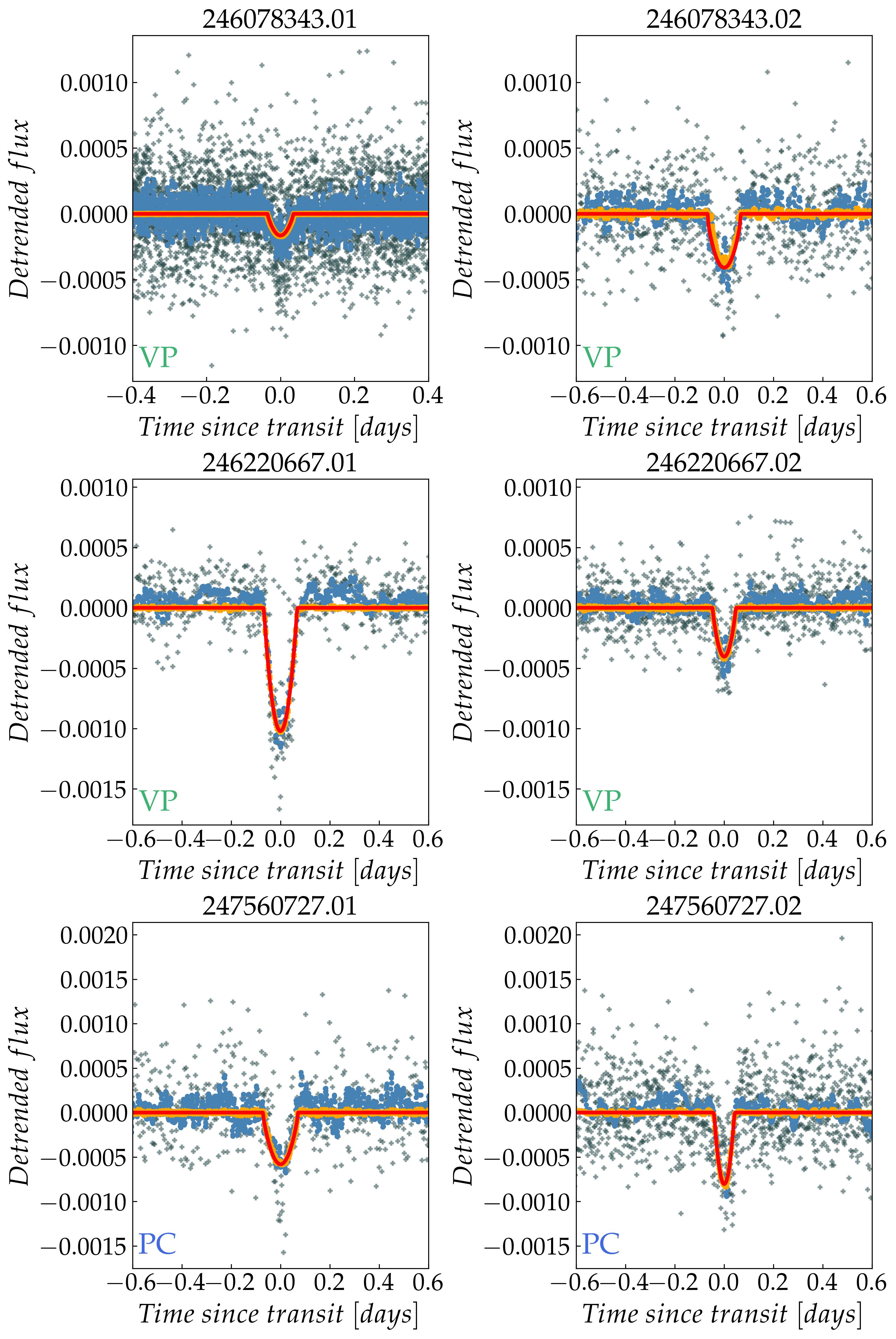}
            \caption{\eve (gray points), \tfaw-corrected (blue points), and \tfaw+GP corrected (orange points) light curves and superposed MCMC best-fit transit model (red line) for all multi planetary candidates in this work. Final dispositions in the lower left corner (VP=validated planet; PC=planet candidate).}
            \label{fig:multiplanets}
        \end{figure*}

Following the vetting and validation procedure described in the previous section, we consider as statistically validated planets those candidates that have passed all the above-mentioned criteria (i.e. having passed all the vetting tests, with no evidence of stellar companions from speckle imaging and \gaia photometry, and with FPP$_{\vespa}$ and FPP$_{\tri}$<1\%. From a total sample of 27 candidates in 24 systems (see Table \ref{tab:planet}), we statistically validate six planets in four different stellar systems: a highly-irradiated Earth (EPIC 210768568.01), a sub-Neptune (EPIC 247422570.01) orbiting a G4 star, a two-planet system (EPIC 246078343) consisting of a Super-Earth (EPIC 246078343.02) and a USP planet (EPIC 246078343.01) with a similar structure to Mercury's interior. Also, a Super-Earth (EPIC 246220667.01) and a sub-Neptune (EPIC 246220667.02) pair orbiting close to their 3:2 mean resonance motion around a K5 star. All, except EPIC 246078343.02 (listed in \citet{Dattilo2019}), are new detections missed by previous works. We do a more extended description of these validated systems in Section \ref{subsec:validated}. Out of the remaining systems, we present 13 new planet candidates. We highlight EPIC 247560727 (see Section \ref{subsubsec:247560727}), a multi-planetary candidate system consisting of a Super-Earth and sub-Neptune pair in a 5:2 resonant orbit, and EPIC 21436876.01 (see Section \ref{subsubsec:211436876}) a very-short period sub-Earth around a G2 star. The phase folded light curves with their MCMC best-fit transit models are shown in Figures \ref{fig:singplanets} and \ref{fig:multiplanets}. The stellar properties of our host star sample are represented in Figure \ref{fig:st_props}.

\subsection{Characteristics of our host star sample}
\label{subsec:host_stars}

\begin{figure*}
            \centering
            \includegraphics[clip,width=\columnwidth]{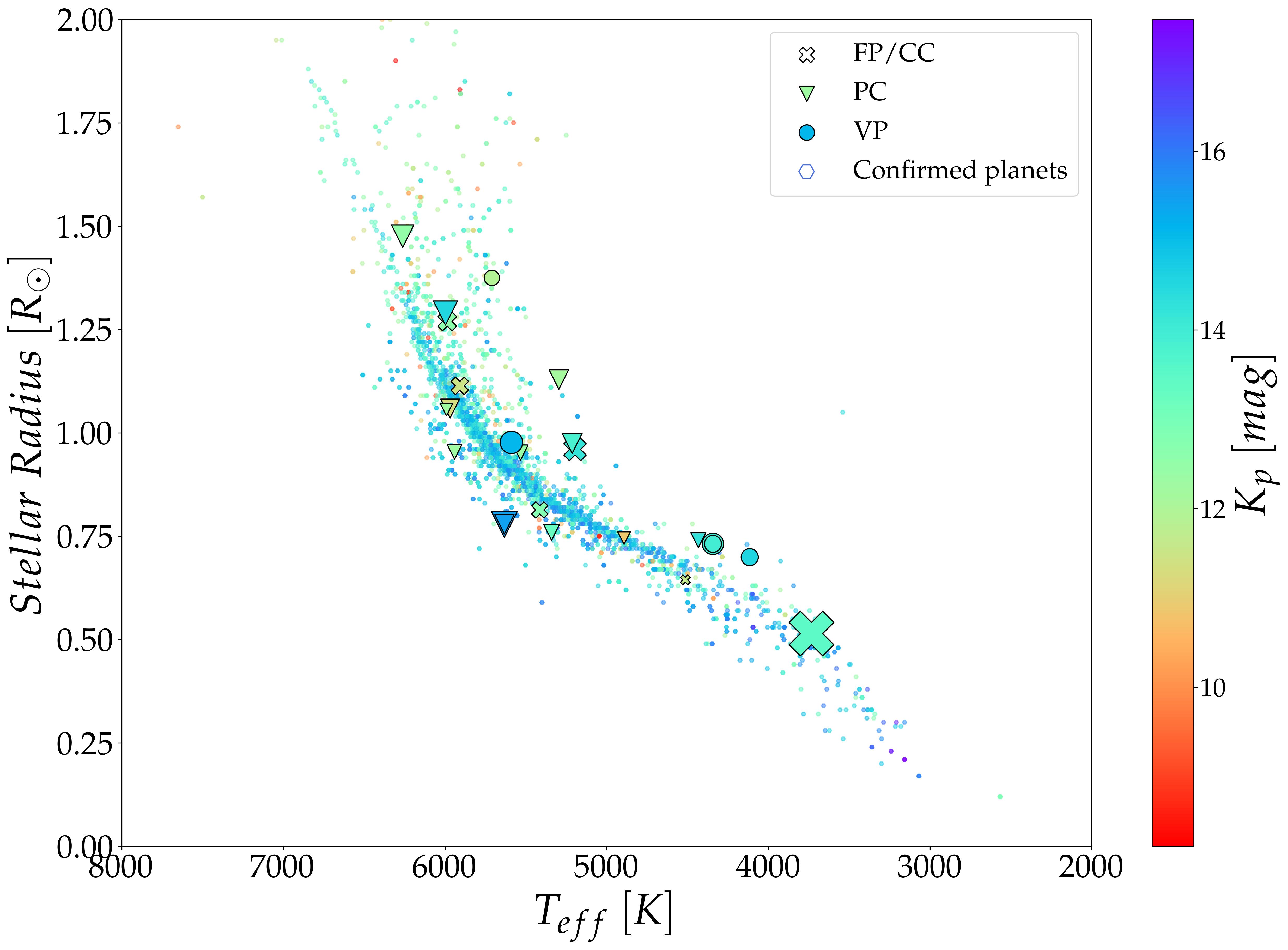}
            \caption{Distribution of the host stars of our validated (circles), candidate (triangles), and false positive (crosses) sample versus the stellar properties of the hosts stars of know planets from the \textit{NASA} Exoplanet Archive (hexagons). The sizes of the markers from our sample are scaled to the MCMC best-fit planetary radius.}
            \label{fig:st_props}
        \end{figure*}

\begin{figure*}
            \centering
            \includegraphics[clip,width=\columnwidth]{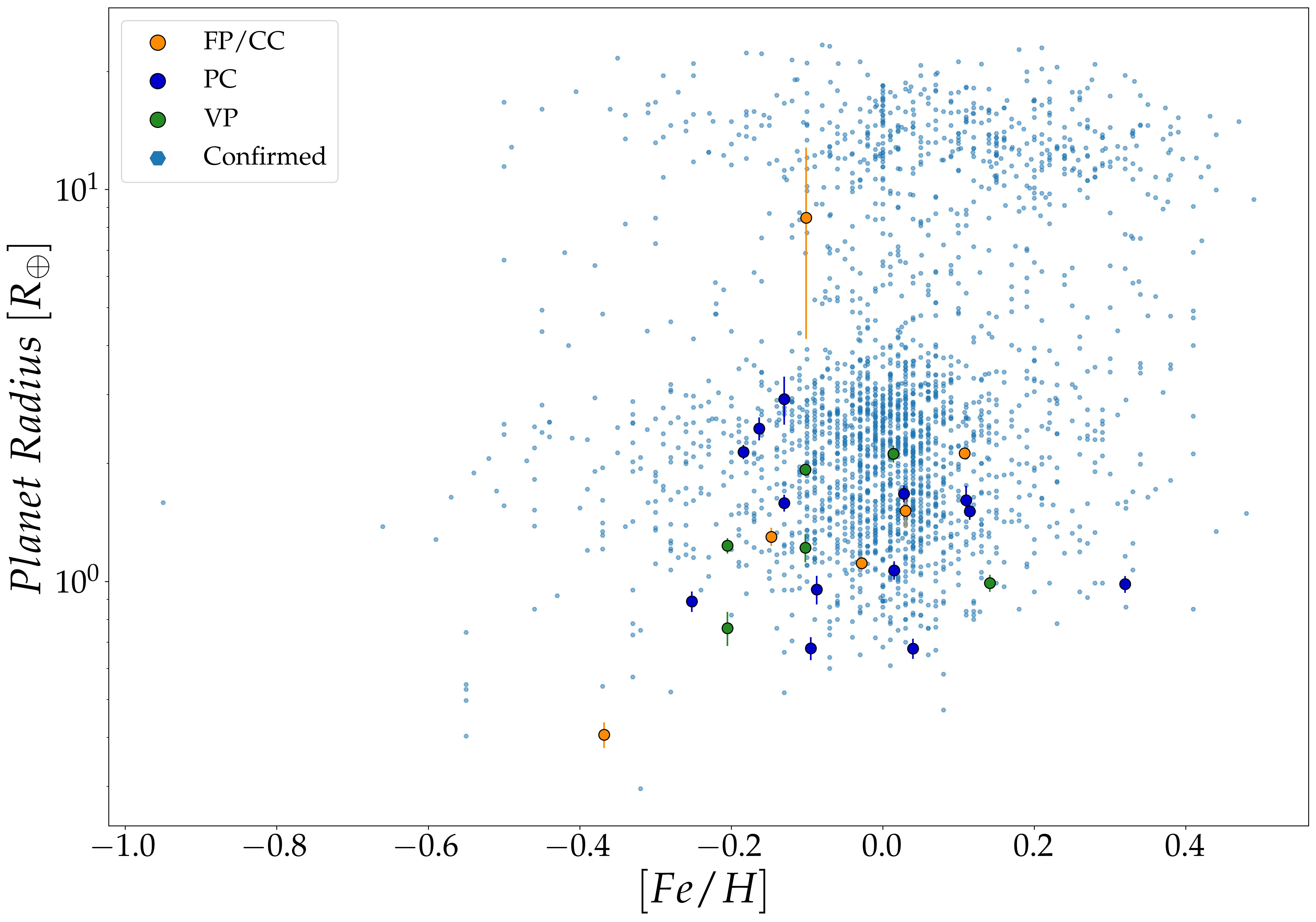}
            \caption{Planetary radius and host star metallicity distribution for our validated (green points), candidate (dark blue points), and false positive (orange points) sample versus the same distribution for confirmed planets with measured radius (blue hexagons) from the \textit{NASA} Exoplanet Archive.}
            \label{fig:rprad_vs_met}
        \end{figure*}

Our candidate host star sample (see Table \ref{tab:stars}) has a median magnitude of $K_p$=13.3, that is, $\sim$0.7 magnitudes fainter than the median $K_p$ magnitude for the \ktwo confirmed planets host stars ($K_p$=12.5). They comprise a small fraction of the \tfaw survey sample \citep{delAlcazar2021} ($\sim$10\%) and have been selected in part, for being bright enough to have good contrast in speckle imaging detection limit curves. Regarding their spectral types, 10 of our targets are G-type stars, six are K-type stars, three are F-type stars, one is an M-type star, and four of them are missing their spectral classification. Most of our validated and candidate planets are located in less populated areas of the confirmed planet host stellar radius vs \teff diagram (see Figure \ref{fig:st_props}). In addition, the sub-Earth planetary candidate EPIC 210706310.01 (see Section \ref{subsubsec:210706310} for a detailed discussion) seems to orbit a metal-poor host star (\feh=-0.402$\pm$0.235$[dex]$, \citet{Hardegree2020}; \feh=-0.463428$^{+0.35536}_{-0.230723}[dex]$, \citet{Anders2022}; \feh=-0.252370$\pm$0.081465$[dex]$, \citet{Buder2021}) (see Figure \ref{fig:rprad_vs_met}).

\subsection{Characteristics of our planetary sample}
\label{subsec:planet_sample}

\subsubsection{Planet period distribution}
\label{subsubsec:planet_period}

Figure \ref{fig:pl_rad} shows the orbital period distribution of our validated, candidate, and false-positive sample. Most of our candidates lie in the $P=\sim$3-10 day range. Given the length of the \ktwo observing campaigns these values do not differ from the typical distribution of the confirmed and candidate \ktwo sample. Two of our planet candidates (EPIC 210418253.01 and EPIC 210945680.01) have periods larger than 20 days. We remark that in the case of EPIC 210945680.01 (which appears listed as a planet candidate in \citet{Zink2021}), although it fails the centroid test (see Figure \ref{fig:centsingle}) we leave it as a planet candidate given that the \gaia astrometric parameters are below the thresholds defined in Sect. \ref{subsec:gaia}, and that we do not detect contaminating sources from BTA speckle observations (but future observations might help in the characterization of this candidate). 
Although the occurrence of Sub-Neptune planets, as a function of period, changes at $\sim$10 days \citep{Winn2018}, USP planets can be defined by the criteria of having a period shorter than $\sim$1 day \citep{Adams2016,Winn2018}. The occurrence rate of USP planets is dependent on the spectral type of the host star, being highest in M-type (1.1\% ± 0.4\%) and lowest in F-type (0.15\% ± 0.05\%) \citep{Winn2018}. The origin of the USP population is still not clear, with different formation scenarios proposed (see \citet{Uzsoy2021} and references within). All of the USP planets known so far are either hot Jupiters or apparently rocky planets \citep{Hamer2020, Uzsoy2021}. One of our validated planets (EPIC 246078343.01), and one planet candidate (EPIC 211436876.01) have periods ($P$=0.8094$\pm$0.00003 days, and $P$=1.1524$^{+0.0003}_{-0.0004}$ days, respectively) that allow us to characterize them as USP planets. For a more detailed discussion on our USP sample see Sections \ref{subsubsec:246078343} and \ref{subsubsec:211436876}.

\subsubsection{Planet radius distribution}
\label{subsubsec:planet_radius}

\begin{figure*}
            \begin{tabular}{ll}
                \includegraphics[clip,width=\columnwidth]{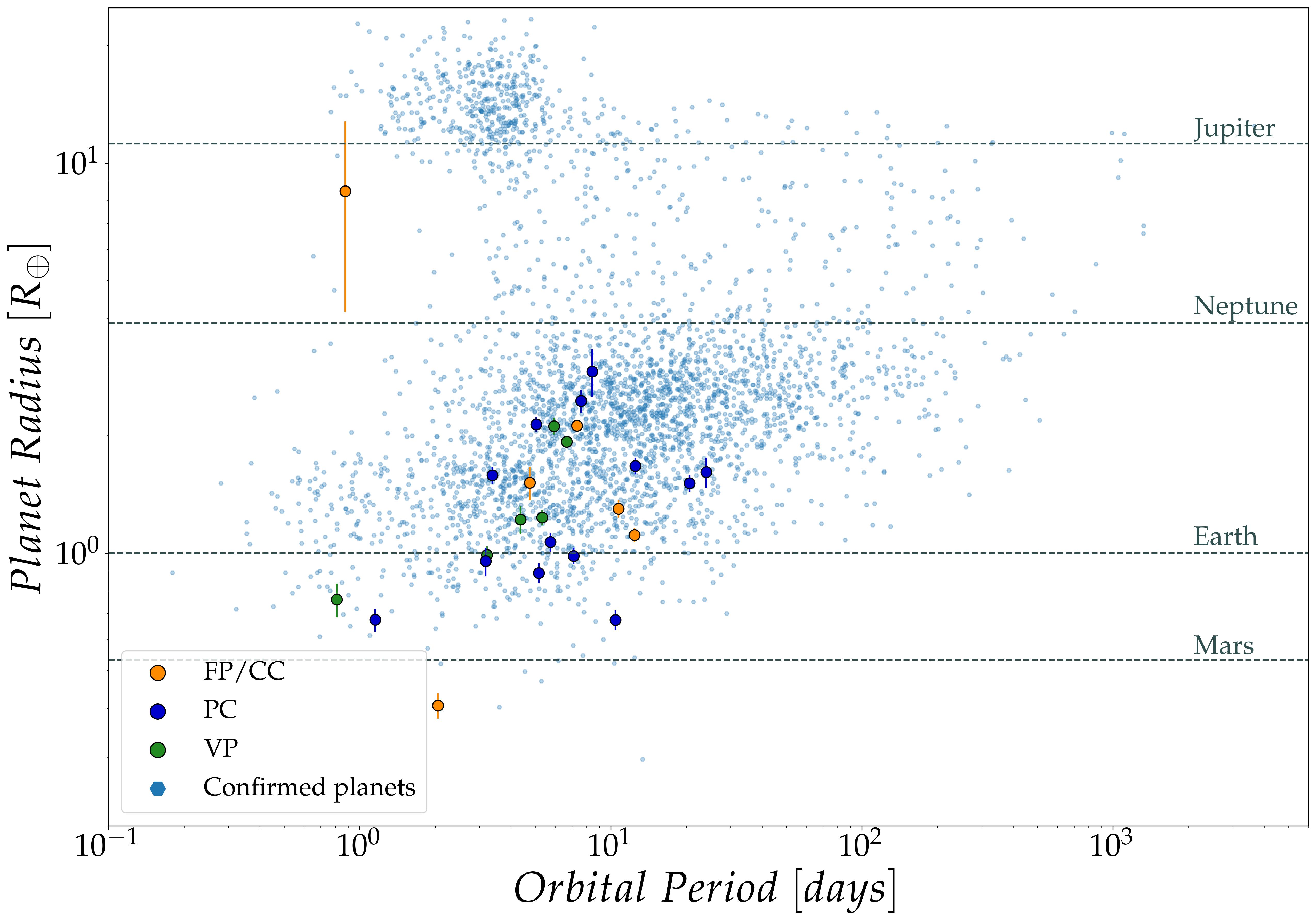} &
                \includegraphics[clip,width=\columnwidth]{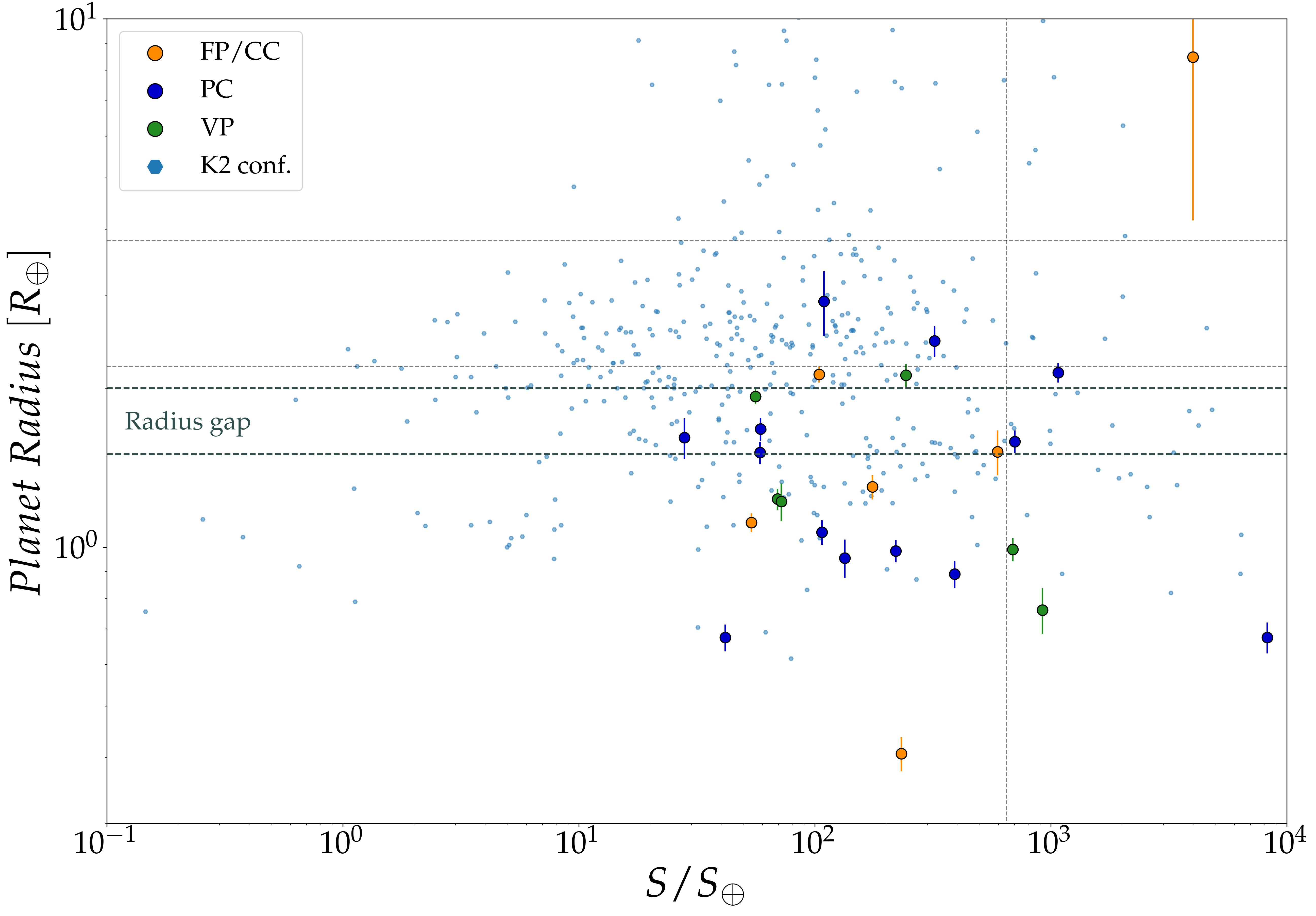}\\
            \end{tabular}
            \caption{\textbf{Left:} Planet radius as a function of the orbital period for our validated planets (green points), planet candidates (dark blue points), and false positive (orange points) sample versus the distribution of confirmed planets (blue hexagons) from \textit{NASA} Exoplanet Archive. \textbf{Right:} Planet radius as a function of the stellar insolation (same notation as left plot). Dark dashed lines denote the approximate location of the radius gap. The region enclosed by the light dashed lines at the right of the plot denotes the Hot Neptune desert.}
            \label{fig:pl_rad}
        \end{figure*}

Using a planet radius distribution similar to the one from \citet{Borucki2011} our sample of validated and candidate planets (see Figure \ref{fig:pl_rad}) is comprised of three sub-Earth planets ($R_p$<0.8\rearth), seven Earths (0.8\rearth$\le$ $R_p$<1.25\rearth), four super-Earths (1.25\rearth$\le$ $R_p$<2\rearth), and four sub-Neptunes ($R_p$<4\rearth). 

Planets with radii <2\rearth are most likely rocky planets; however, the internal nature of planets with 2\rearth$<R_{p}<$4\rearth is still a matter of debate. The two most accepted scenarios are that they might either be planets with a rocky core and a gaseous envelope or water worlds \citep{Zeng2019}. This bimodality of the distribution of small planets is separated by an observed scarcity of planets with radii 1.5\rearth$<R_p<$2\rearth known as the Radius Gap \citep{Fulton2017}. Several scenarios for the Radius Gap origin have been postulated \citep{Owen2013,Owen2017,Venturini2017,Ginzburg2018,Zeng2019}. In addition, the Radius Gap seems to depend on the stellar host type \citep{Fulton2017,Zeng2017,McDonald2019}, and metallicity, and evolves (as well as the whole planetary radius distribution) on a long timescale of giga-years \citep{Chen2022,Petigura2022}. Thus, planets within the Radius Gap can serve as valuable probes to analyze the processes that lead to planet formation, atmosphere loss, and evolution \citep{Petigura2020}. Four of our planet candidates (EPIC 210418253.01, EPIC 210945680.01, EPIC 247744801.01\footnote{This candidate is affected by the presence of a nearby ($\sim$7.7$\arcsec$), fainter ($G$=17.51 mag) star (\verb|GOF_AL|=1.79, \verb|D|=1.59, $RUWE$=1.06)}, and EPIC 247560727.01), and one validated planet (EPIC 246220667.02) lie within the Radius Gap based on our analysis. Also, eight out of our sample of 18 validated and candidate planets have radii smaller than that of the Radius Gap. This points towards the improved detection of smaller planets by the combination of the \tfaw corrected light curves and \tls \citep{delSer2020} in contrast with previous works \citep{Castro-Gonzalez2021}.

\begin{figure*}
            \centering
            \includegraphics[clip,width=\columnwidth]{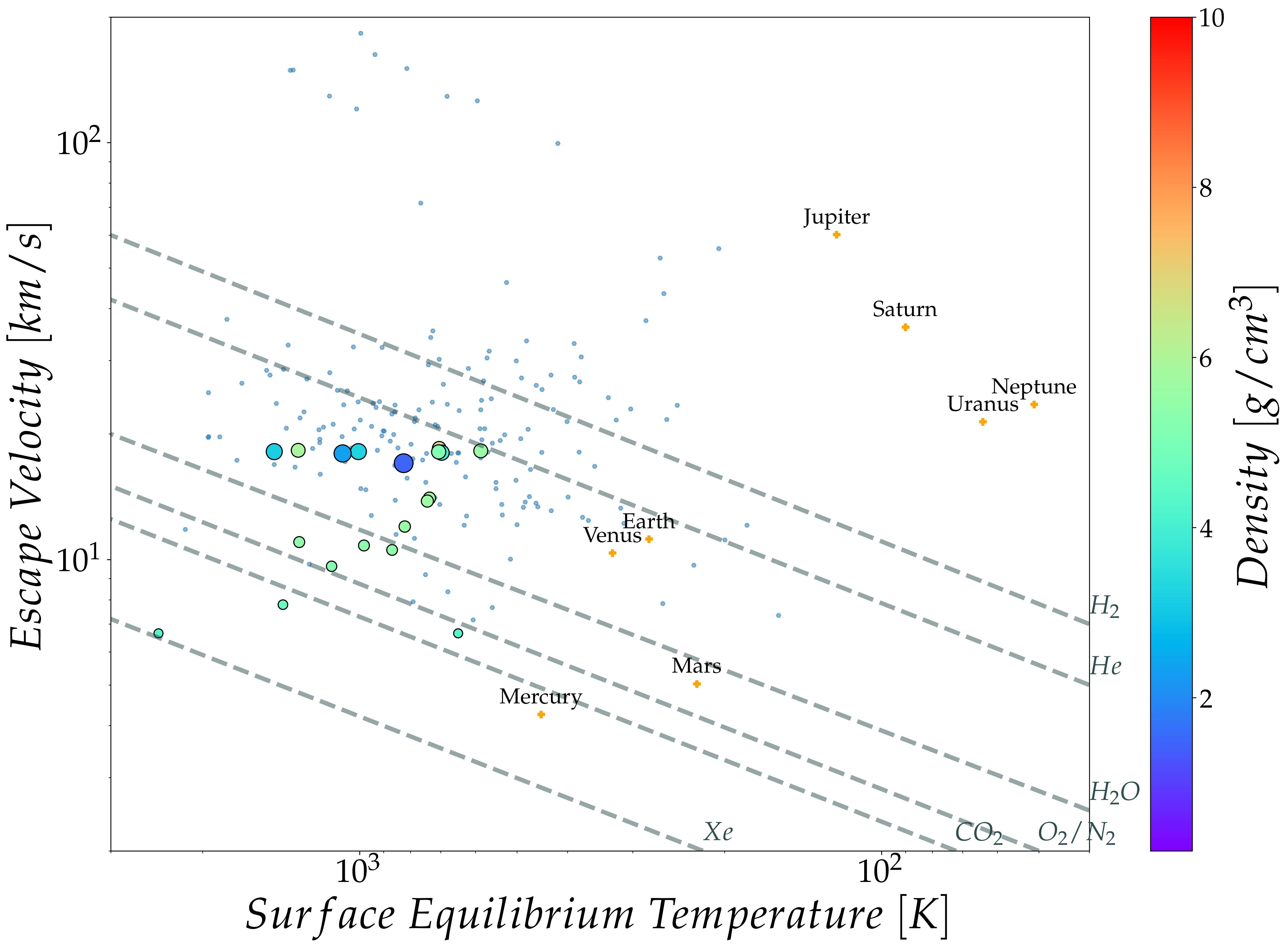}
            \caption{Atmospheric escape velocities versus surface equilibrium temperature for our sample of validated and candidate planets with $R_p$<4\rearth (black-edged points) versus. Their face colors correspond to their bulk densities computed from the estimated masses computed as per Section \ref{subsec:resonance}. For comparison purposes, Solar System bodies (orange crosses), and the confirmed planet (blue hexagons) from the \textit{NASA} Exoplanet Archive is also plotted. Dashed lines represent the threshold velocities of the atmospheric components labeled at each line.}
            \label{fig:vesc_vs_teq}
        \end{figure*}

According to the photoevaporation-driven mass-loss model, the planet’s atmosphere is heated, stripped off, and driven out by the host star high energy radiation, leaving the rocky cores \citep{Owen2013,Owen2017}. Planets with thicker H/He envelopes may still keep part of it after the first 100 Myr of the host star's lifetime when the high energy radiation shuts down \citep{Ribas2005}. The remaining atmosphere can significantly inflate the planet's radii and place them in the $R_p$>2\rearth part of the observed radii distribution. 
We analyze whether our 4 sub-Neptune candidates (i.e. with 2\rearth$<R_{p}<$4\rearth) can keep an atmospheric envelope over the first billion-year of their host star. Using the equations from \citet{Zeng2019}, we can estimate the atmospheric components that our candidate planets can hold. These estimates are obtained following the correlation that the escape velocities and the atmospheric composition of Solar System bodies have with the atmospheric escape. We use the masses estimated using the procedure explained in Section \ref{subsec:resonance} to derive both the escape velocities ($v_{esc}=\sqrt{2GM_pR^{-1}_p}$), and planet bulk densities ($\rho=M_p/(4/3 \pi R_p^3)$). We compute the surface equilibrium temperatures of our planet sample using the stellar radii, and \teff listed in Table \ref{tab:stars}, the MCMC best-fit value for the semi-major axis of the planetary orbit, and we assume a bolometric albedo \textbf{$A_B=0.3$} similar to that of the Earth and Neptune. In Figure \ref{fig:vesc_vs_teq} we show the escape velocities of our $R_p$<4\rearth validated and candidate planet sample as a function of their surface equilibrium temperatures. We find a clear differentiation between our Earth- and sub-Earth-sized planets, and our sub-Neptune sample. The first group seems to be rocky worlds consisting primarily of Mg–silicate–rock and (Fe,Ni)-metal \citep{Zeng2019}, having similar bulk densities to those of Earth and Venus. Validated planet, EPIC 246078343.01, and planet candidate EPIC 211436876.01 would be rocky planets with a composition similar to that of Mercury. Our sub-Neptune sample lies within a region with escape velocities of $\sim$20$km/s$, and equilibrium temperatures between 500 and 1500 K. Inside this region they are susceptible to the escape of $H_2$ and He and, except for the presence of an internal reservoir, they would not be able to retain their primordial H/He atmospheres during the first Myrs. \citet{Zeng2019} infer that the He escape threshold is the boundary separating the populations of puffy hot Saturns and smaller planets. More interestingly, our four planet candidates and the one validated planet lying in the Radius Gap, correspond to the 5 planets closer to the He boundary in Figure \ref{fig:vesc_vs_teq}. Given their estimated densities, all would be rocky planets, except for EPIC 247560727.02 which would probably be a water world given its estimated bulk density and equilibrium temperature \citep{Zeng2019}.

\begin{figure*}
            \centering
            \begin{tabular}{ccc}
                \includegraphics[width=0.6\columnwidth]{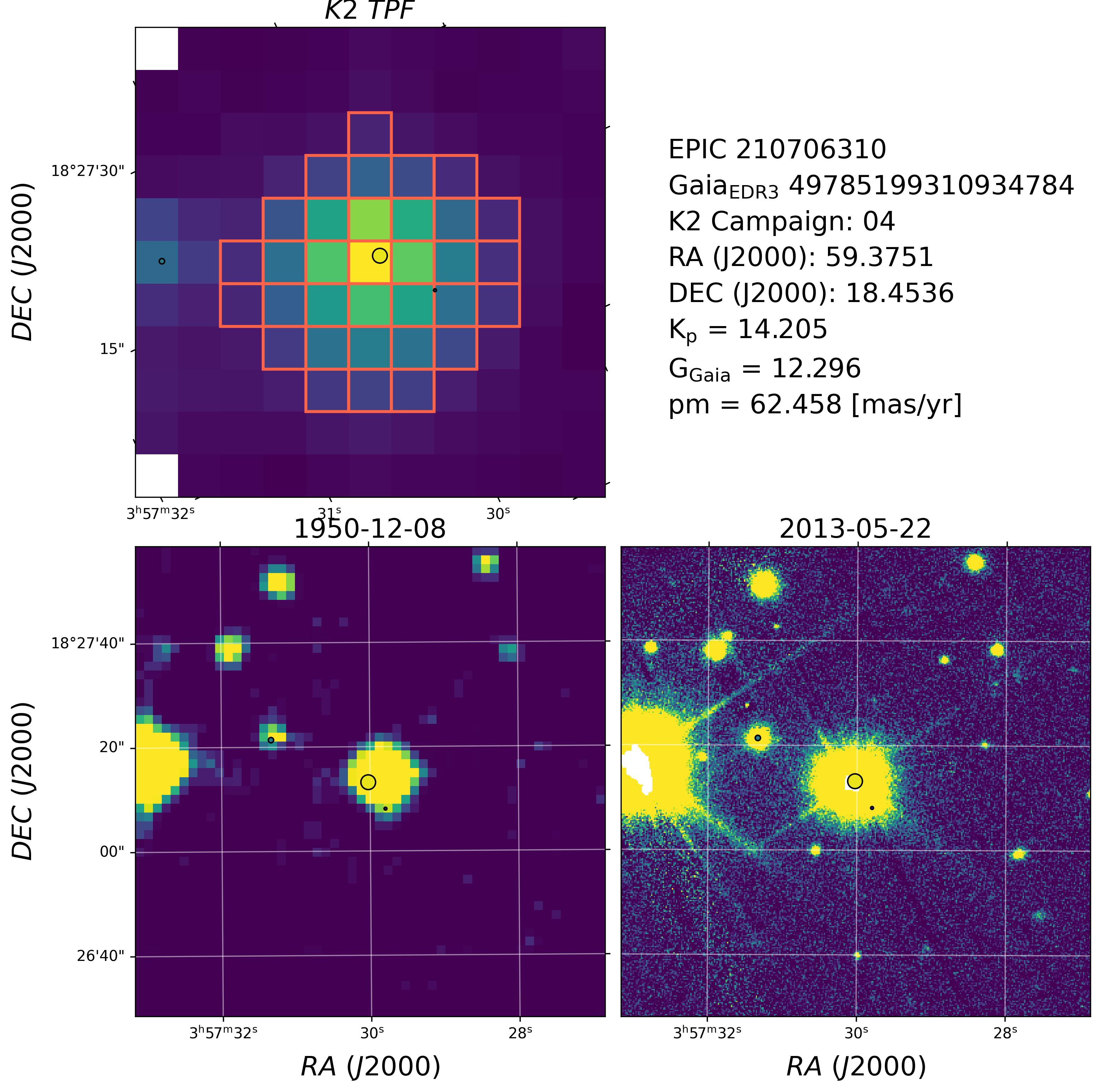}
                &
                \includegraphics[width=0.6\columnwidth]{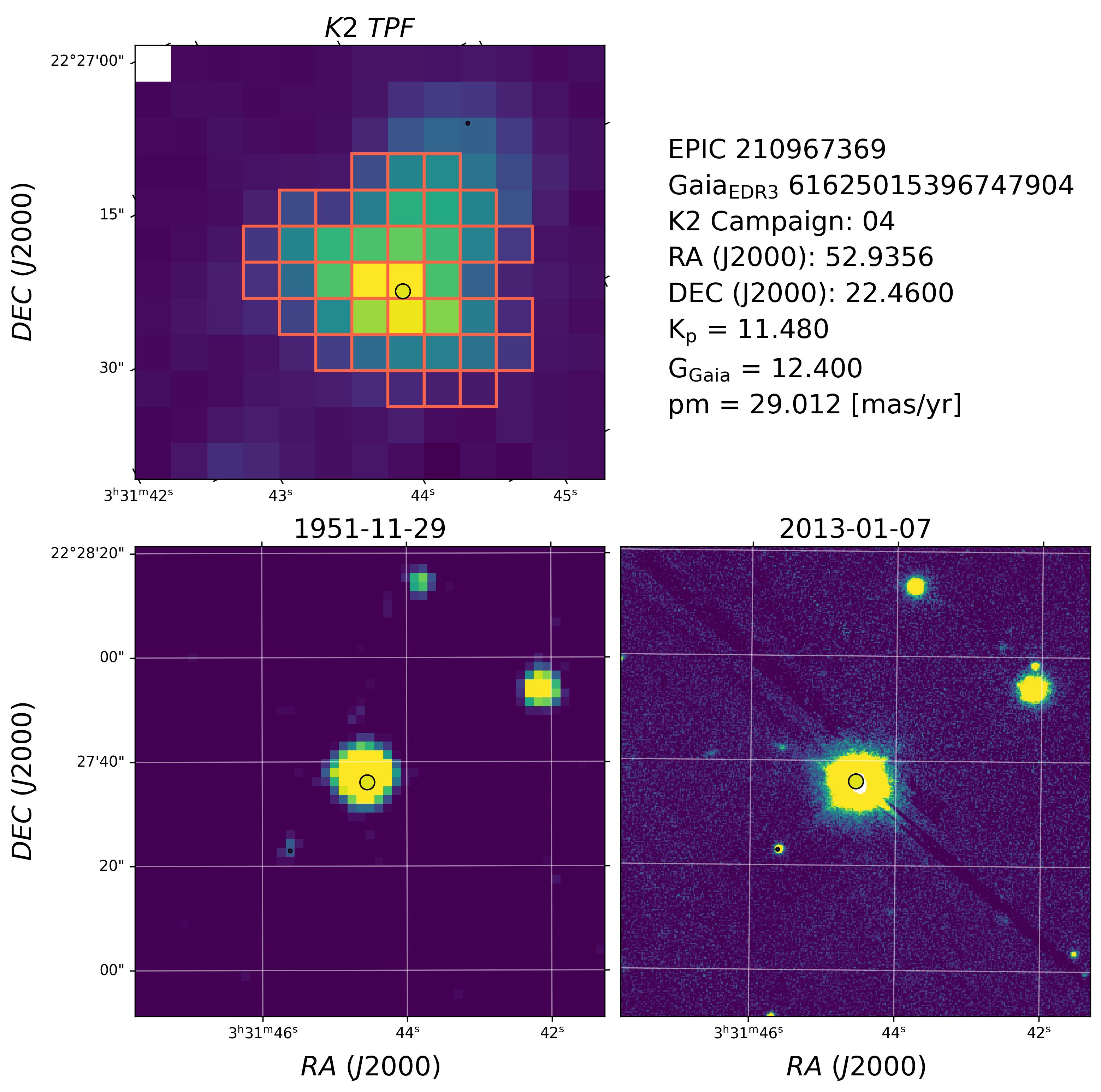}
                &
                \includegraphics[width=0.6\columnwidth]{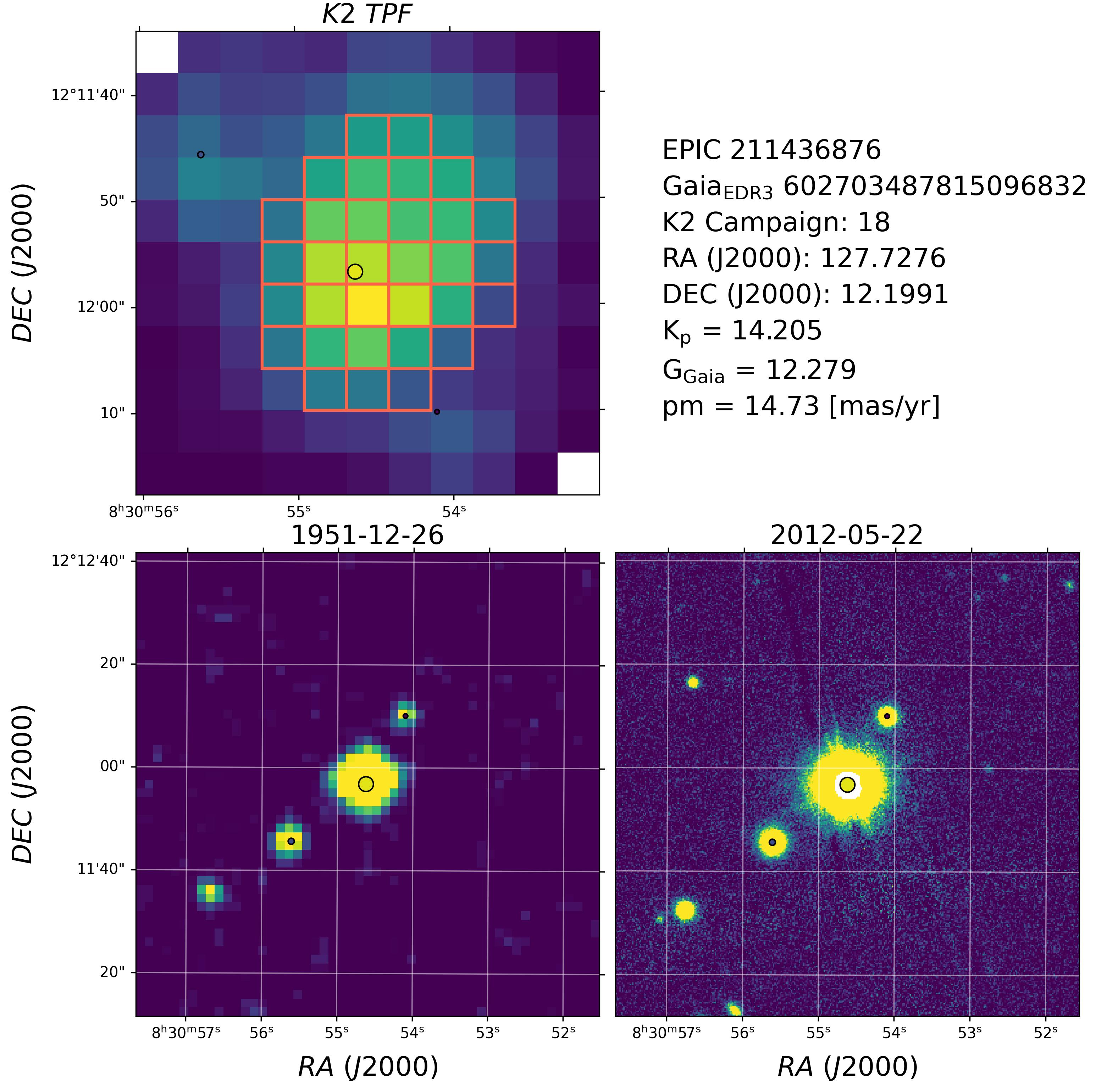}\\
                \includegraphics[width=0.6\columnwidth]{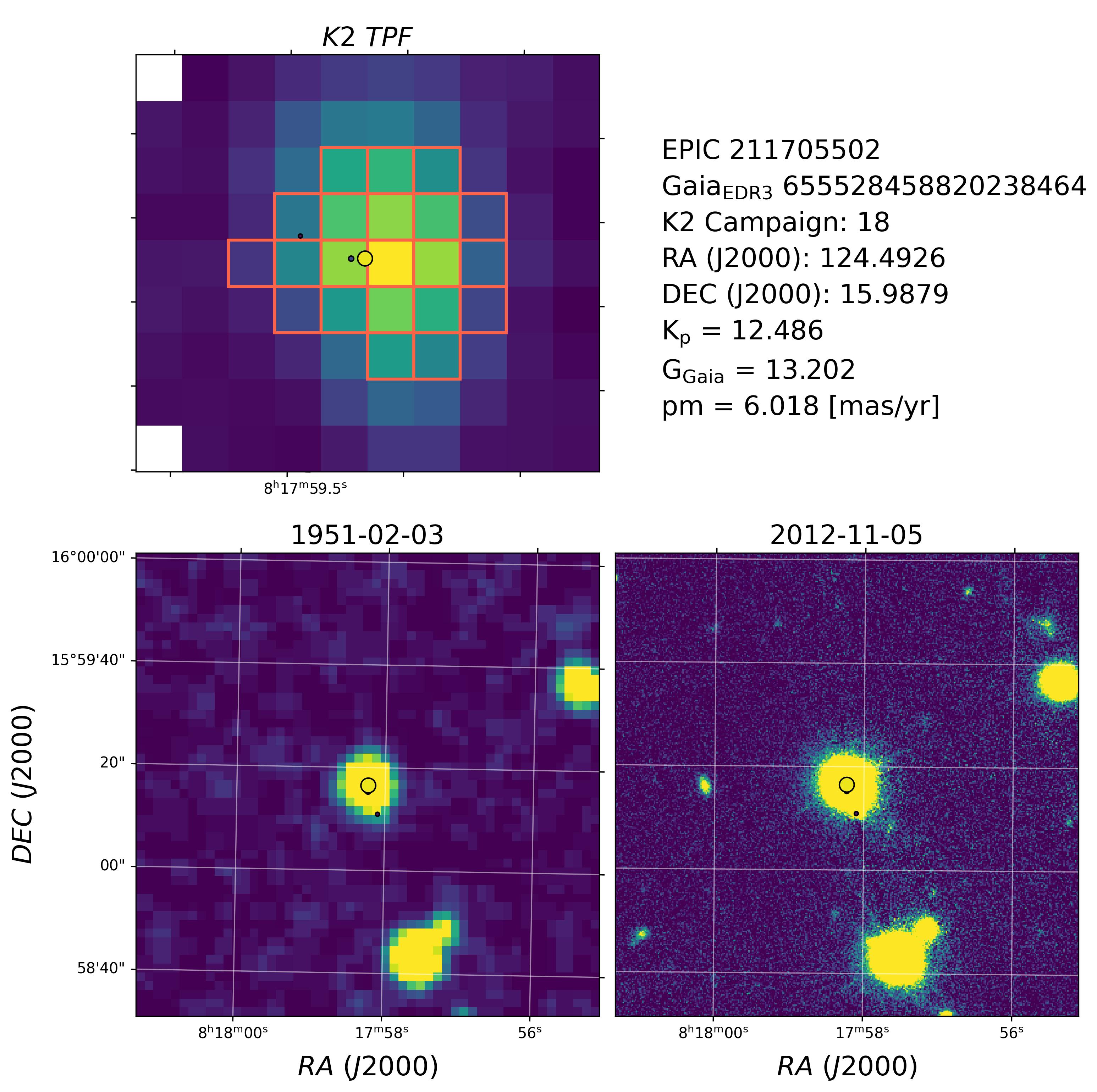}
                &
                \includegraphics[width=0.6\columnwidth]{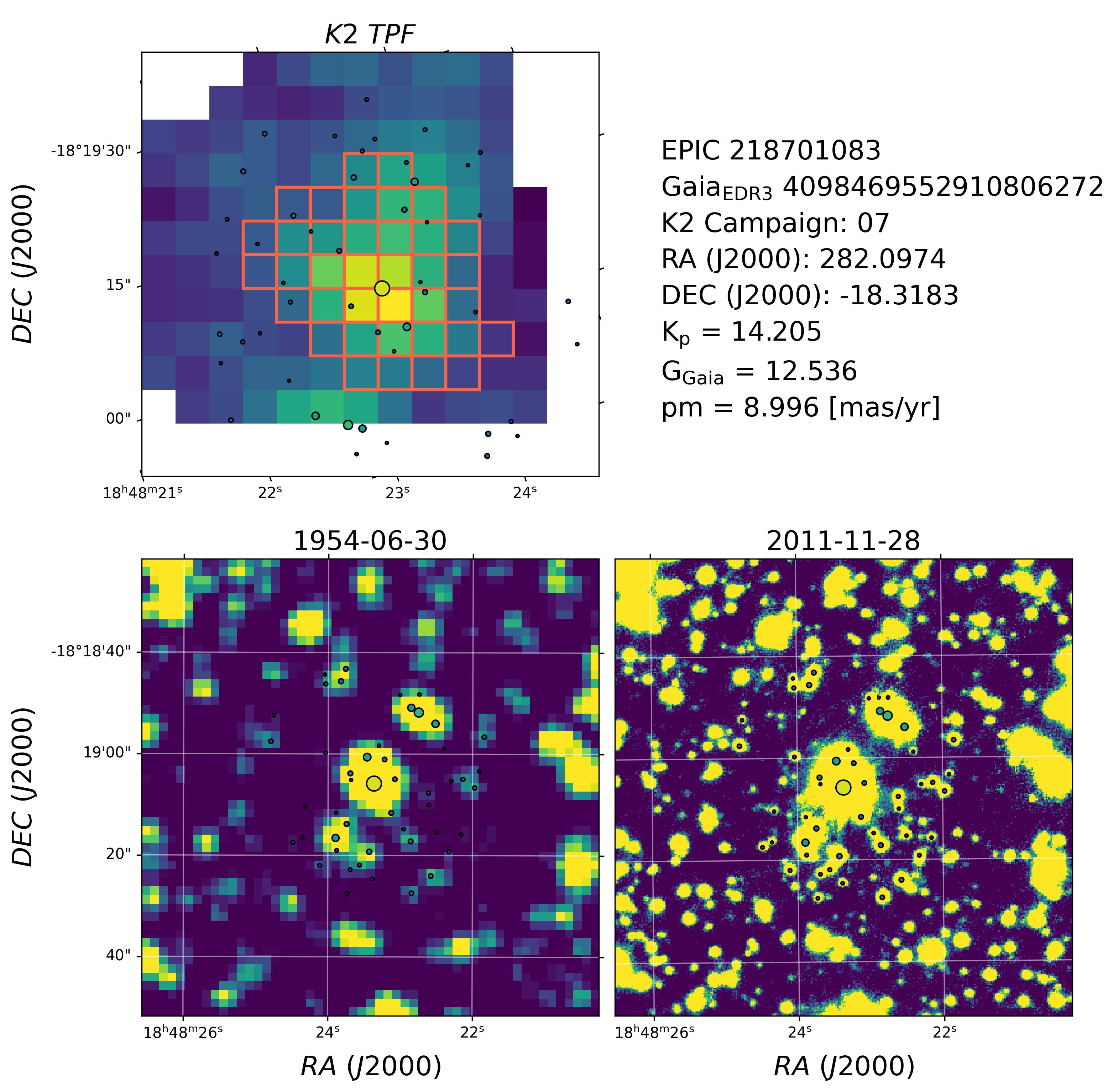}
                &
                \includegraphics[width=0.6\columnwidth]{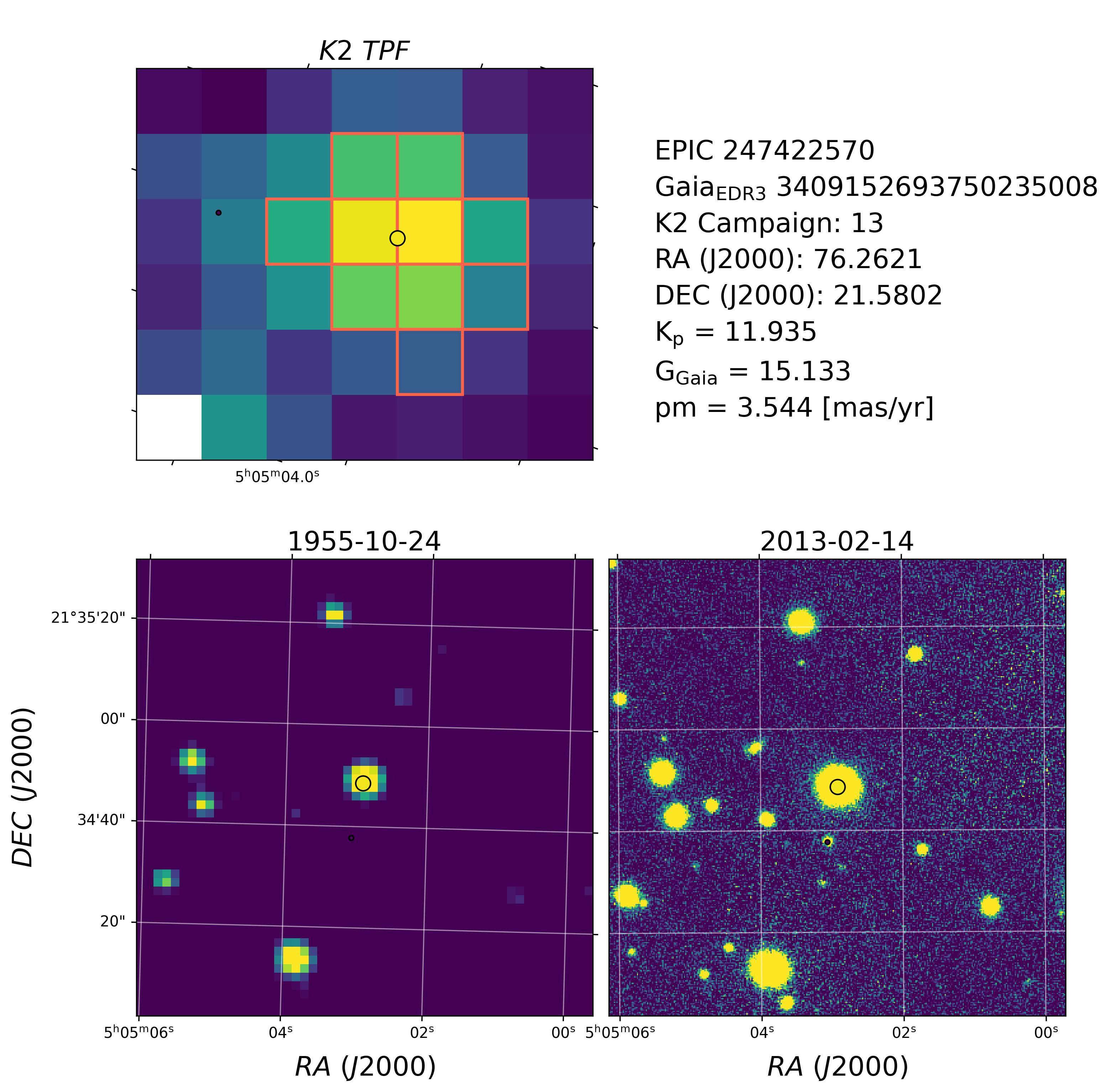}\\
                \includegraphics[width=0.6\columnwidth]{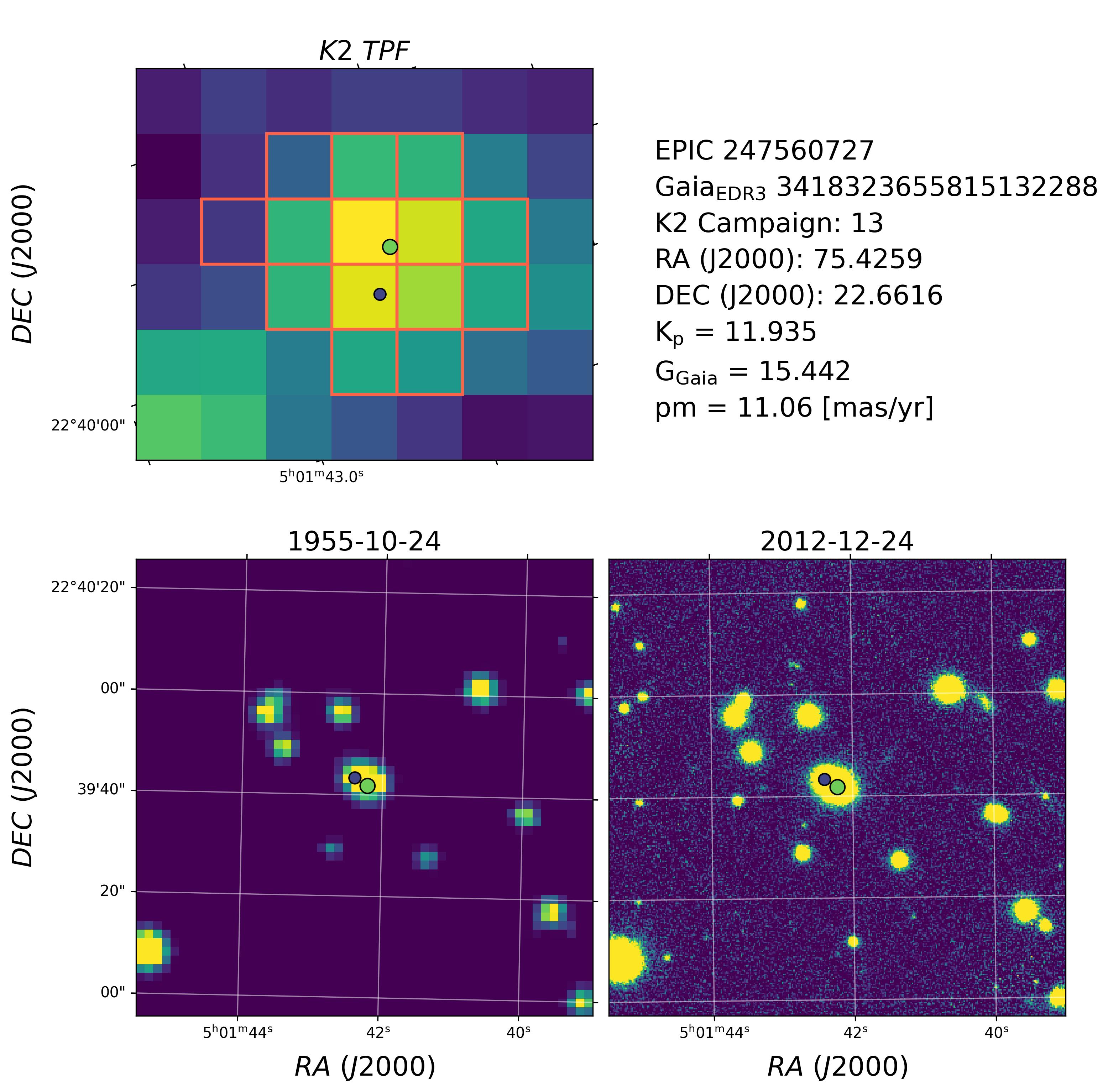}
                &
                \includegraphics[width=0.6\columnwidth]{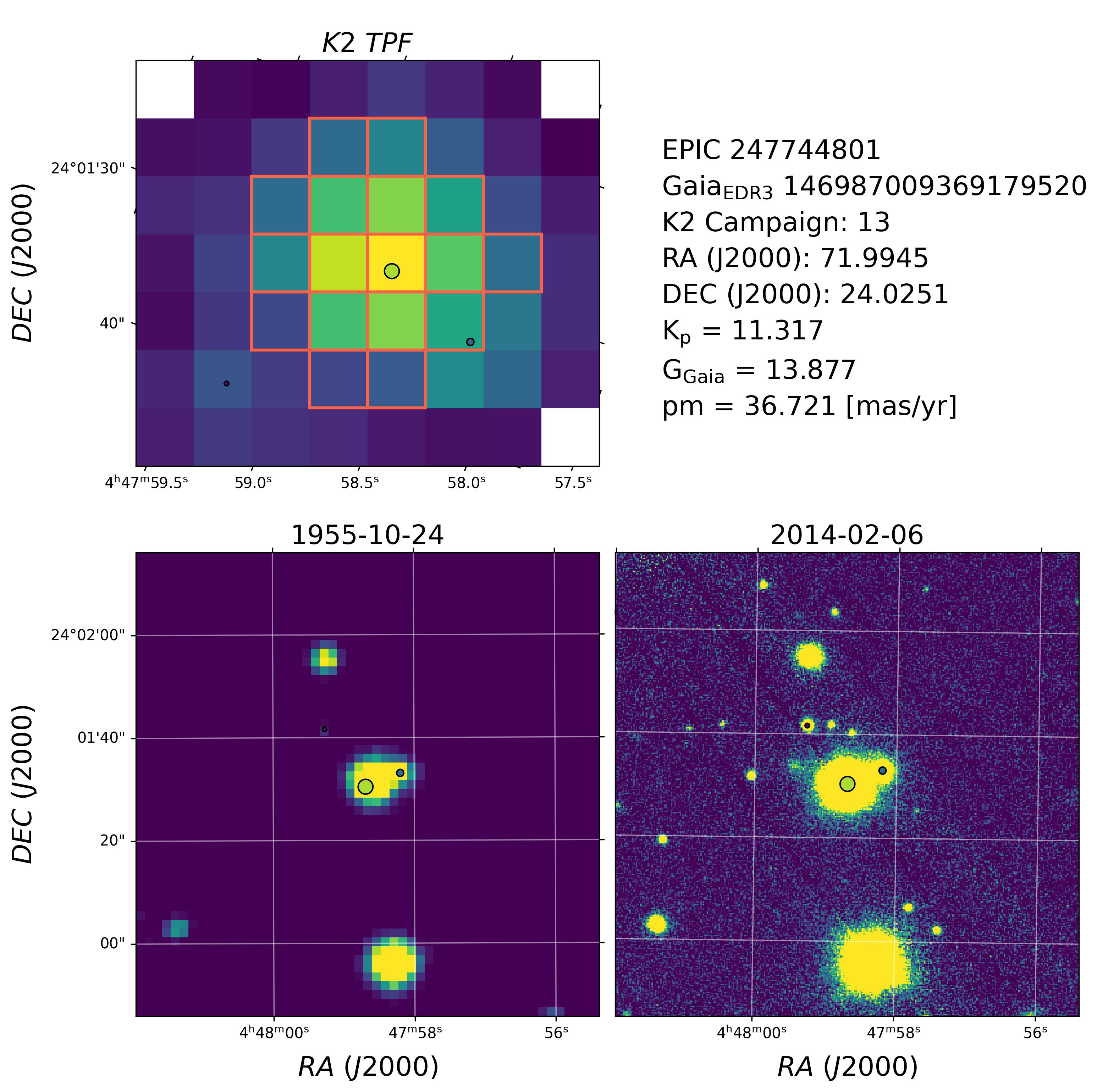}
                &
                \includegraphics[width=0.6\columnwidth]{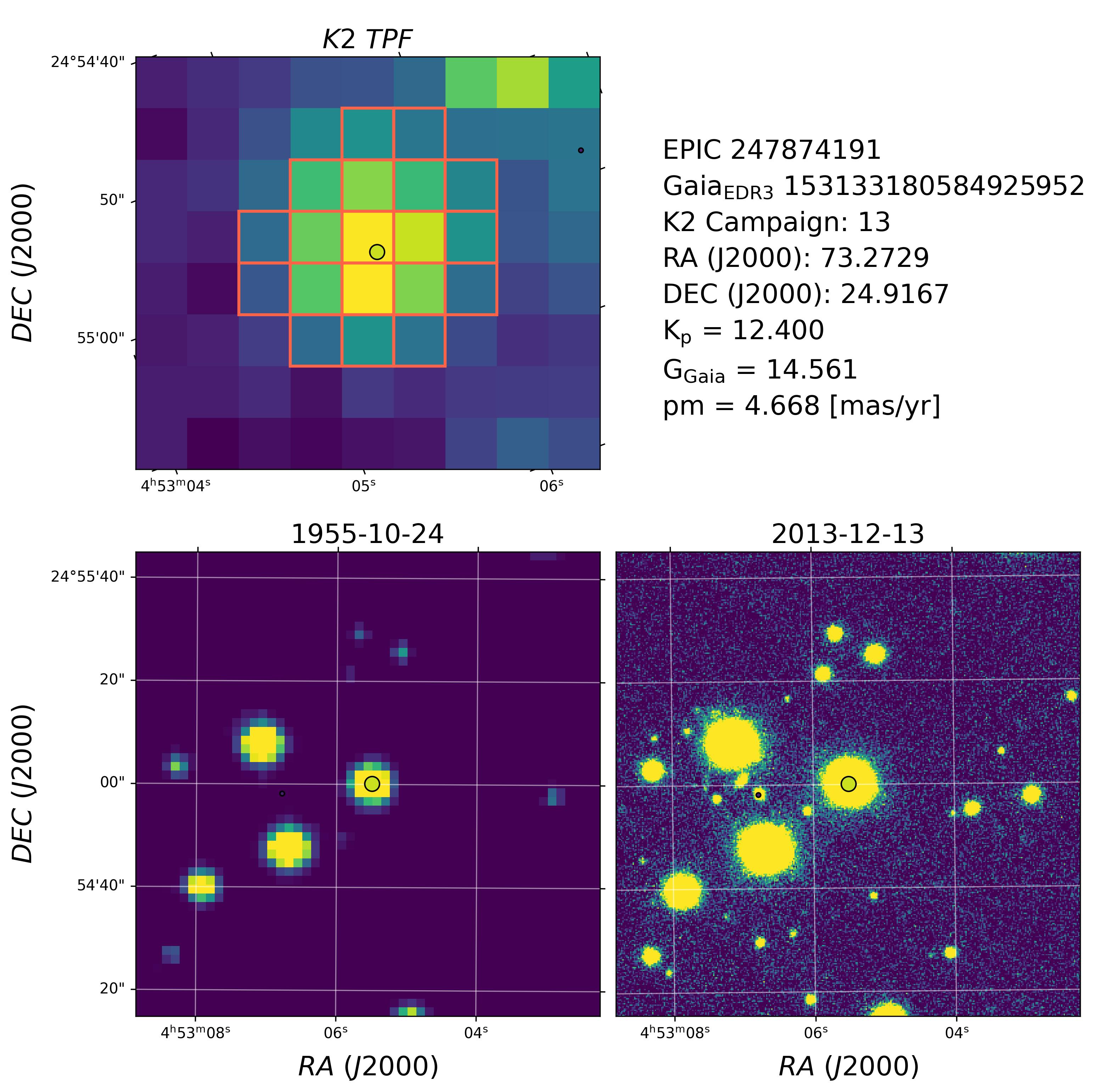}\\
            \end{tabular}
            \label{fig:gaia_cont}
            \caption{Seeing-limited imaging validation sheets for those targets in our sample with detected \gaia eDR3 companions within the \eve photometric aperture. In each validation sheet, \textbf{top left}: \ktwo Target Pixel File (TPF) with the \eve aperture (red squares) superposed. \textbf{Lower left}: Palomar Observatory Sky Survey (POSS-I) images. \textbf{Lower right}: Pan-STARRS DR2 images. \gaia eDR3 sources are represented with colored points.}
        \end{figure*}

The photoevaporation desert or Neptunian desert is a lack of planets between 2-4\rearth at very high insolations ($S/S_\oplus >650$ \citep{Lundkvist2016,West2019}. The mechanism, be it photoevaporation or core-powered mass loss, giving birth to the observed Neptunian Desert is currently unknown. Thus, planets found in and near the Neptune Desert boundaries are particularly valuable for the understanding of the origin of this phenomenon. Our planet candidate EPIC 218701083.01, with $R_p=2.1388^{+0.0898}_{-0.0924}$\rearth, and $S/S_\oplus$=1073.998 lies close to the edge of the Neptunian desert. We find a slightly smaller planetary radius than the one reported in \citet{Zink2021} ($R_p=2.594^{+0.173}_{-0.186}$\rearth). However, we cannot fully validate this candidate due to the presence of several fainter stars within the \eve photometric aperture (see Figure \ref{fig:gaia_cont}). Using \texttt{lightkurve}, we have tried to minimize the effects of the neighboring stars by modifying the photometric aperture and recomputing the light curve. In addition, by checking the \gaia astrometric parameters (see Section \ref{subsec:gaia}) of those stars still within the photometric aperture, we can rule out, up to a certain limit, the possibility of them being background eclipsing binaries. A comparison of the \gaia astrometric parameters for EPIC 218701083 and the nearest background contaminating stars is listed in Table \ref{tab:epic218701083}. Given the dilution in the transit depth due to the presence of these contaminating stars (especially from the brightest one, EPIC 218701831), if EPIC 218701083 is the transiting star, the real radius of the planet would be larger than the reported one (taking only EPIC 218701831 as secondary source, then $\gamma_{pri}\sim$1.06, and $R_p\sim 2.2$\rearth). This could put it inside the Neptunian desert region (see Figure \ref{fig:pl_rad}). However, the background eclipsing binary scenario cannot be fully discarded without further observations.

\begin{table*}
    \centering
    \caption{Comparison of the \gaia eDR3 astrometric properties for EPIC 218701083 and contaminating background stars.}
    \label{tab:epic218701083}
    \begin{tabular}{llllll}
\hline
EPIC      & \gaia eDR3 & $G$ [mag] & \verb|GOF_AL| & \verb|D| & $RUWE$ \\ 
\hline
218701083 & 4098469552910806272       & 12.54        & -1.43                       & 0.85                   & 0.93  \\ 
\hline
218701831 & 4098469557255647616       & 15.92        & -0.52                       & 0                       & 0.97  \\
218700307 & 4098469522895908480       & 18.25        & 0.55                        & 0                       & 1.03  \\
--        & 4098470313134530944       & 19.90        & 0.58                        & 1.11                    & 1.03  \\
--        & 4098470313133032576       & 18.39        & 2.37                        & 0.58                    & 1.13  \\
--        & 4098469557218787456       & 18.44        & 1.58                        & 1.07                    & 1.09  \\
--        & 4098469557255646720       & 18.08        & 0.23                        & 0.29                   & 1.01  \\
--        & 4098469557255646592       & 20.47        & -1.06                       & 0                       & --     \\ 
\hline
\end{tabular}
\end{table*}

\subsubsection{Habitability analysis}

\begin{figure*}
            \centering
            \includegraphics[clip,width=\columnwidth]{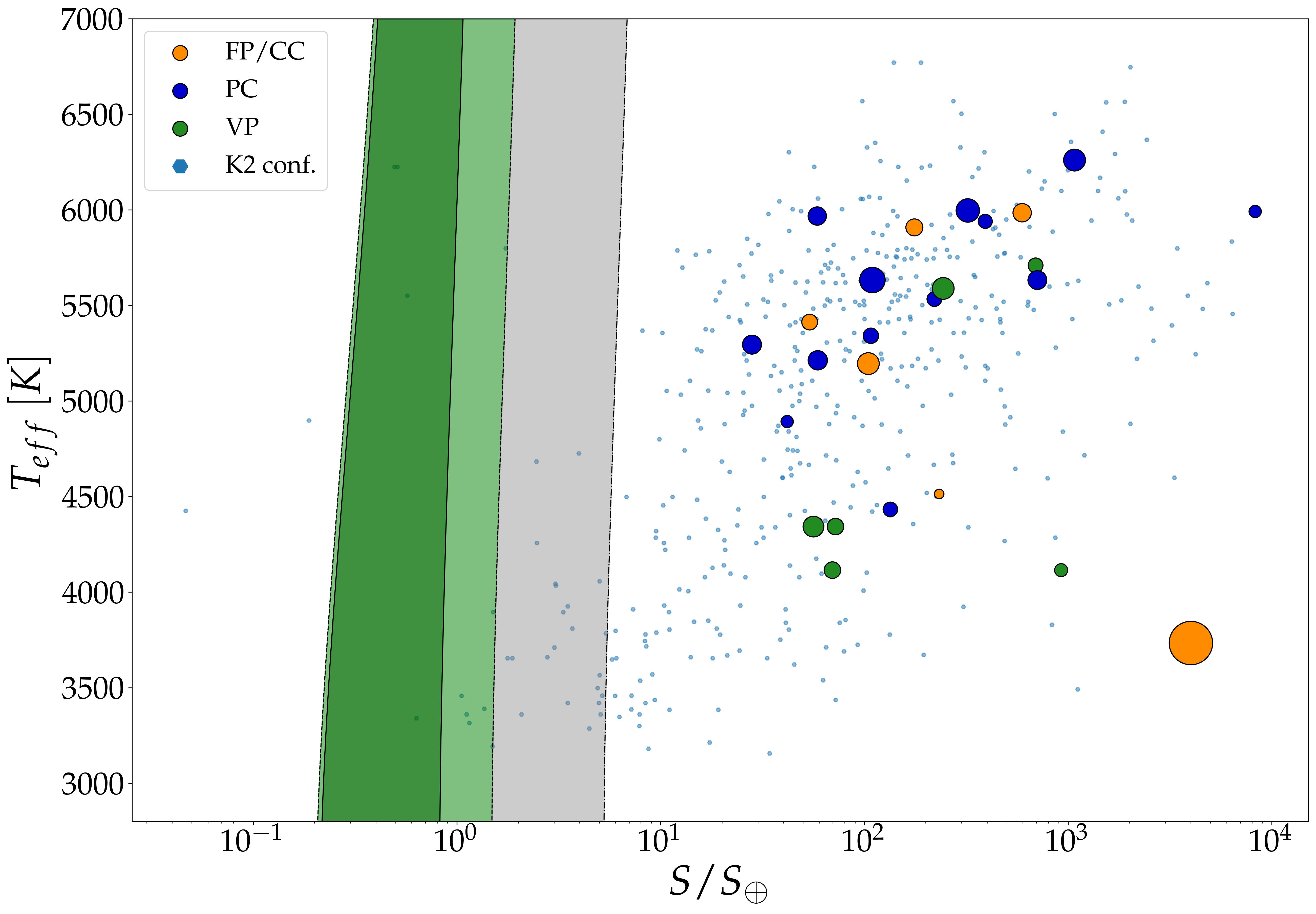}
            \caption{Stellar effective temperature as a function of the insolation fluxes received by our validated (green dots), candidate (dark blue dots), and false-positive (orange dots) sample versus the \ktwo confirmed sample from \textit{NASA} Exoplanet Archive (blue hexagons). Dot sizes from our sample candidates are scaled to their MCMC best-fit planetary radii. The conservative HZ (dark green region) is limited by the two solid lines corresponding to the moist greenhouse inner edge and the maximum greenhouse outer edge. The optimistic HZ (light green region) is bounded by two dashed lines
corresponding to the recent Venus inner limit and early Mars outer limit. The \citet{Zsom2013} hot desert world HZ (grey region) is limited by the dotted line.}
            \label{fig:habitability}
        \end{figure*}

In order to assess whether any of our validated or candidate planets could be in the Habitable Zone (HZ) of their host stars, we used the polynomial equations from \citet{Kopparapu2013} to
determine the limits of the HZ. The conservative HZ is delimited by the 'moist greenhouse' limit ($S/S_\oplus$=1.01; i.e. where the stratosphere becomes saturated by water and hydrogen begins to escape into space) and the 'maximum greenhouse' limit ($S/S_\oplus$=0.35; i.e. where the greenhouse effect fails as CO$_2$ begins to condensate from the atmosphere and the surface becomes too cold to hold liquid water). The optimistic HZ is delimited empirically by the recent Venus and early Mars limits (i.e. set by the last time that liquid surface water could have existed on Venus and Mars: $S/S_\oplus$=1.78 and $S/S_\oplus$=0.32, respectively \citep{Kasting1988}. We also include the more optimistic HZ set by \citet{Zsom2013}. It takes into account that the HZ for hot desert worlds (1\% relative humidity and terrestrial albedo, $A_B$=0.8, and assuming a surface pressure of 1 bar and a 10$^{-4}$ CO$_2$ mixing ratio) could be much closer to the star (as close as 0.38 AU around a solar-like star). Given the short orbital periods (typical of most of the \ktwo confirmed planets) of our candidate sample, and the effective temperatures of our host stars, none of the planets shown in this work are within the HZs discussed above (see Figure \ref{fig:habitability}). 
        
        \begin{figure*}
            \centering
            \begin{tabular}{ccc}
                \includegraphics[clip,width=0.5\columnwidth]{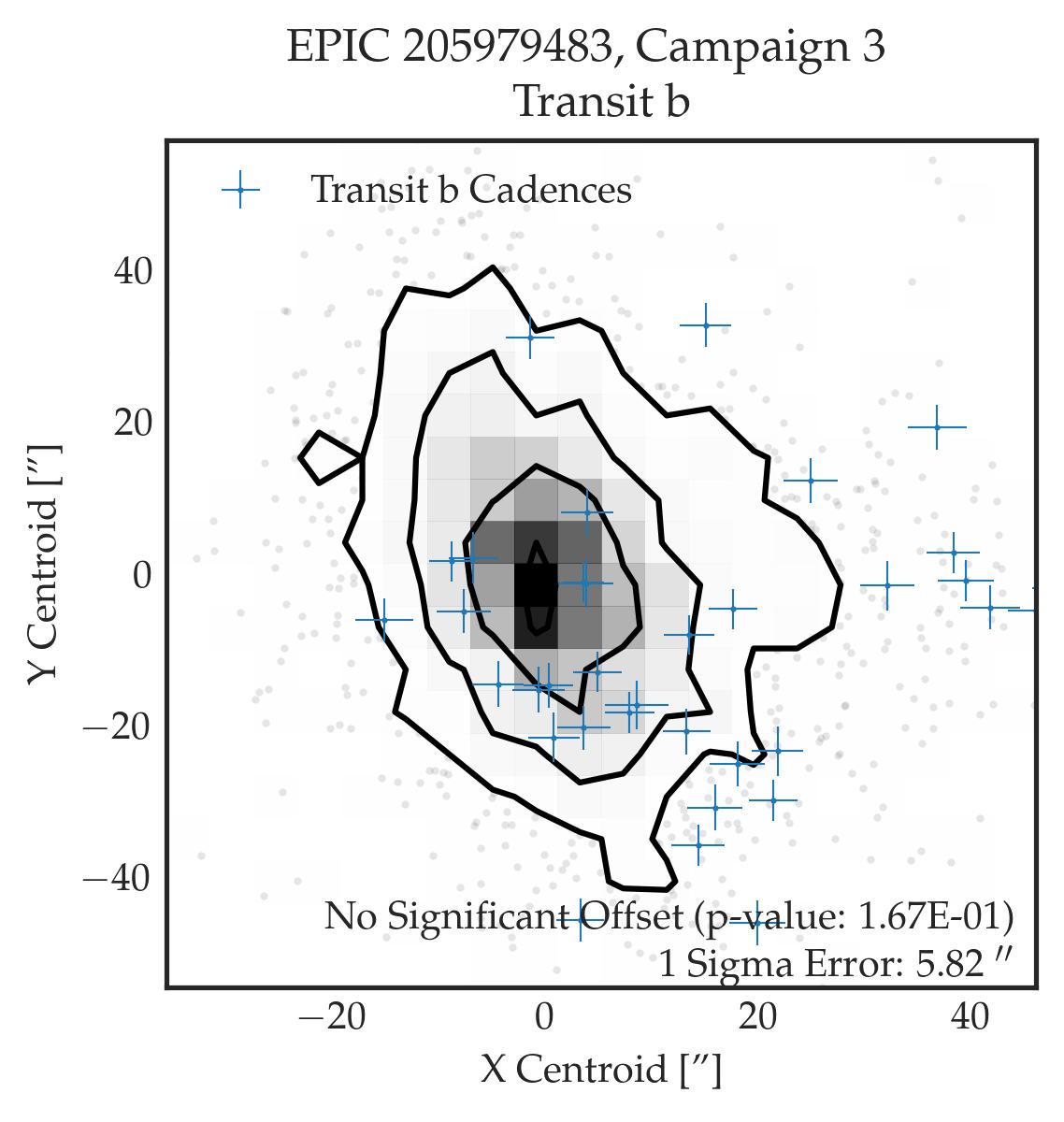}
                &
                \includegraphics[clip,width=0.5\columnwidth]{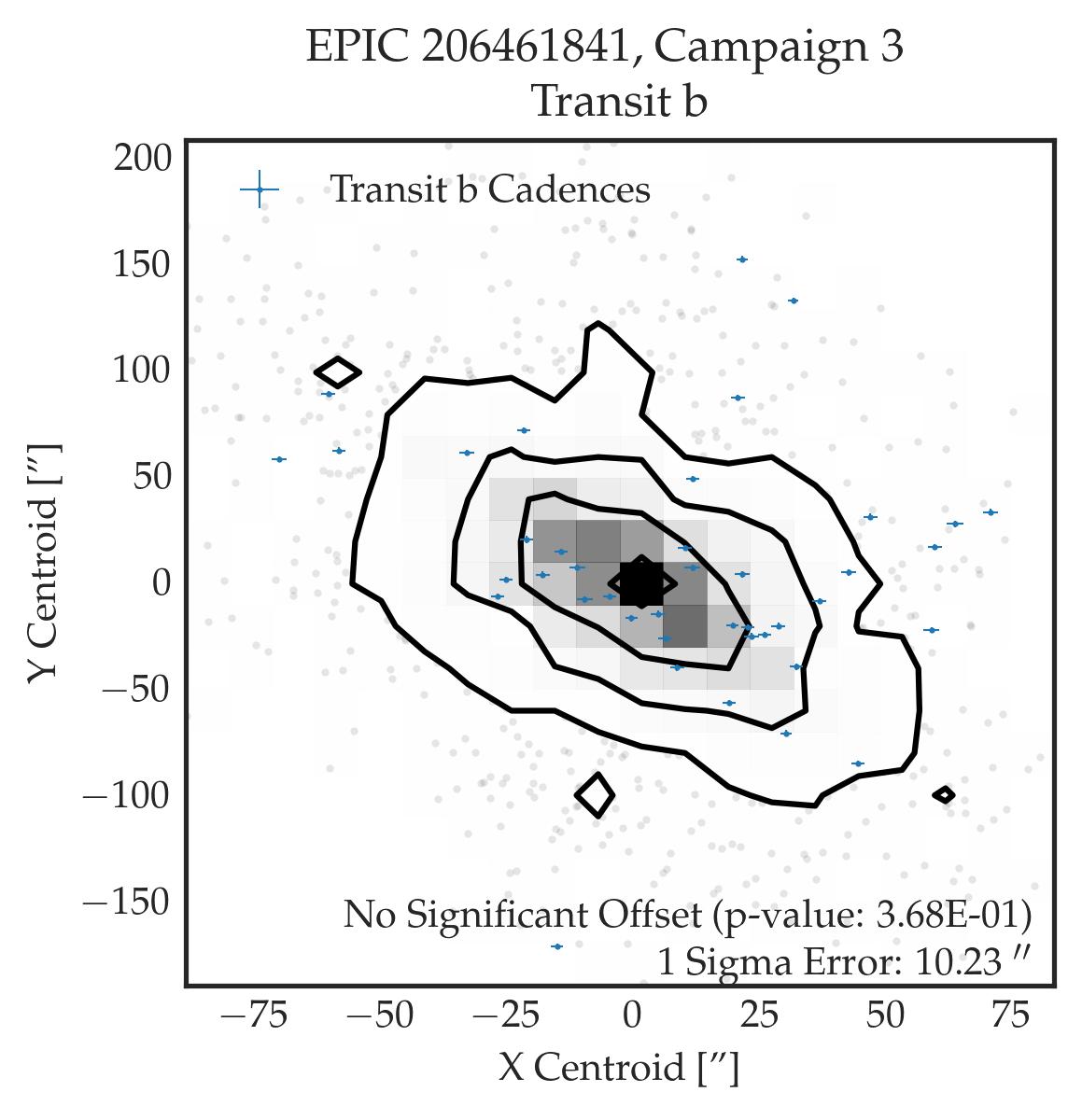}
                &
                \includegraphics[clip,width=0.5\columnwidth]{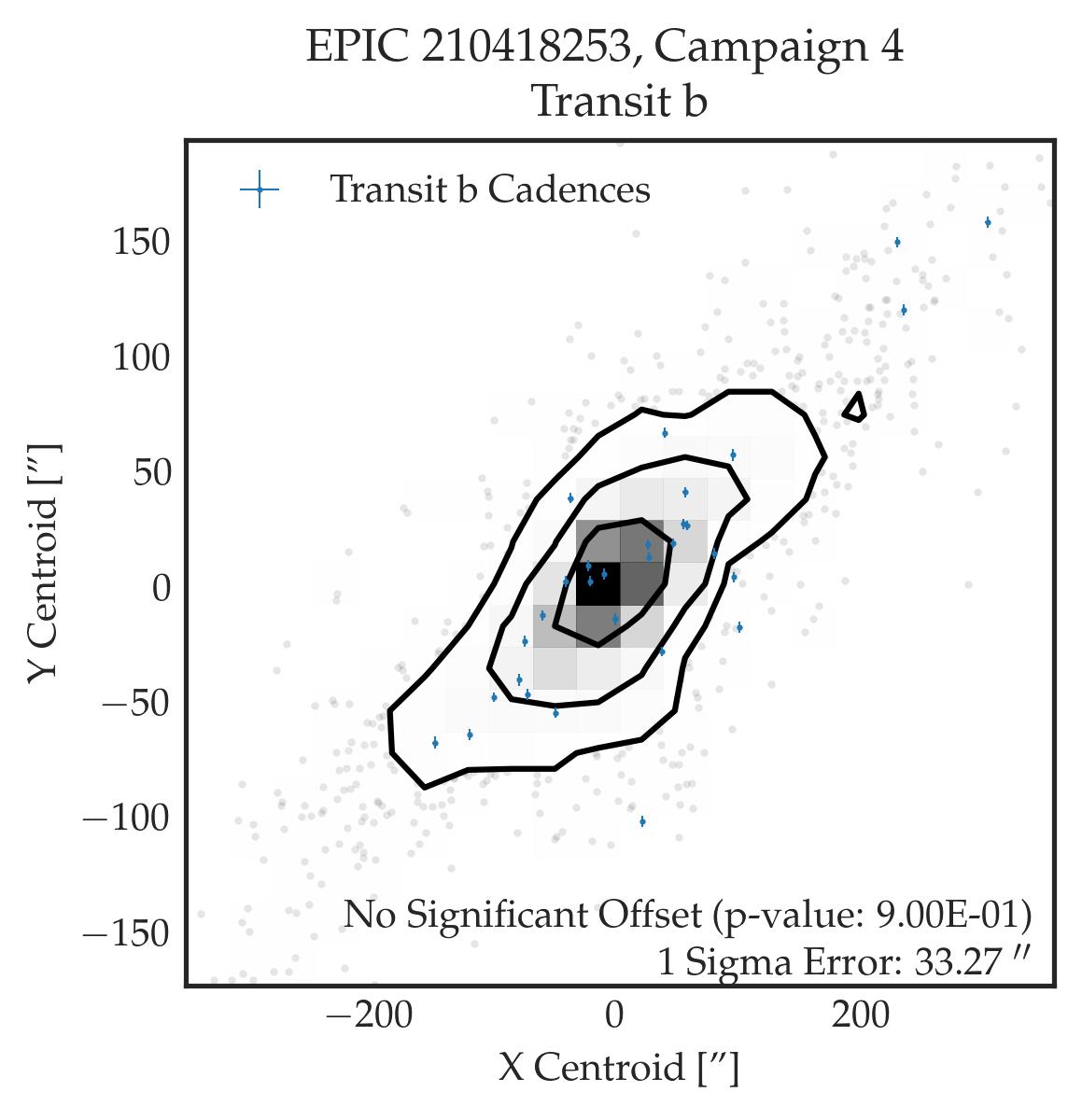}\\
                \includegraphics[clip,width=0.5\columnwidth]{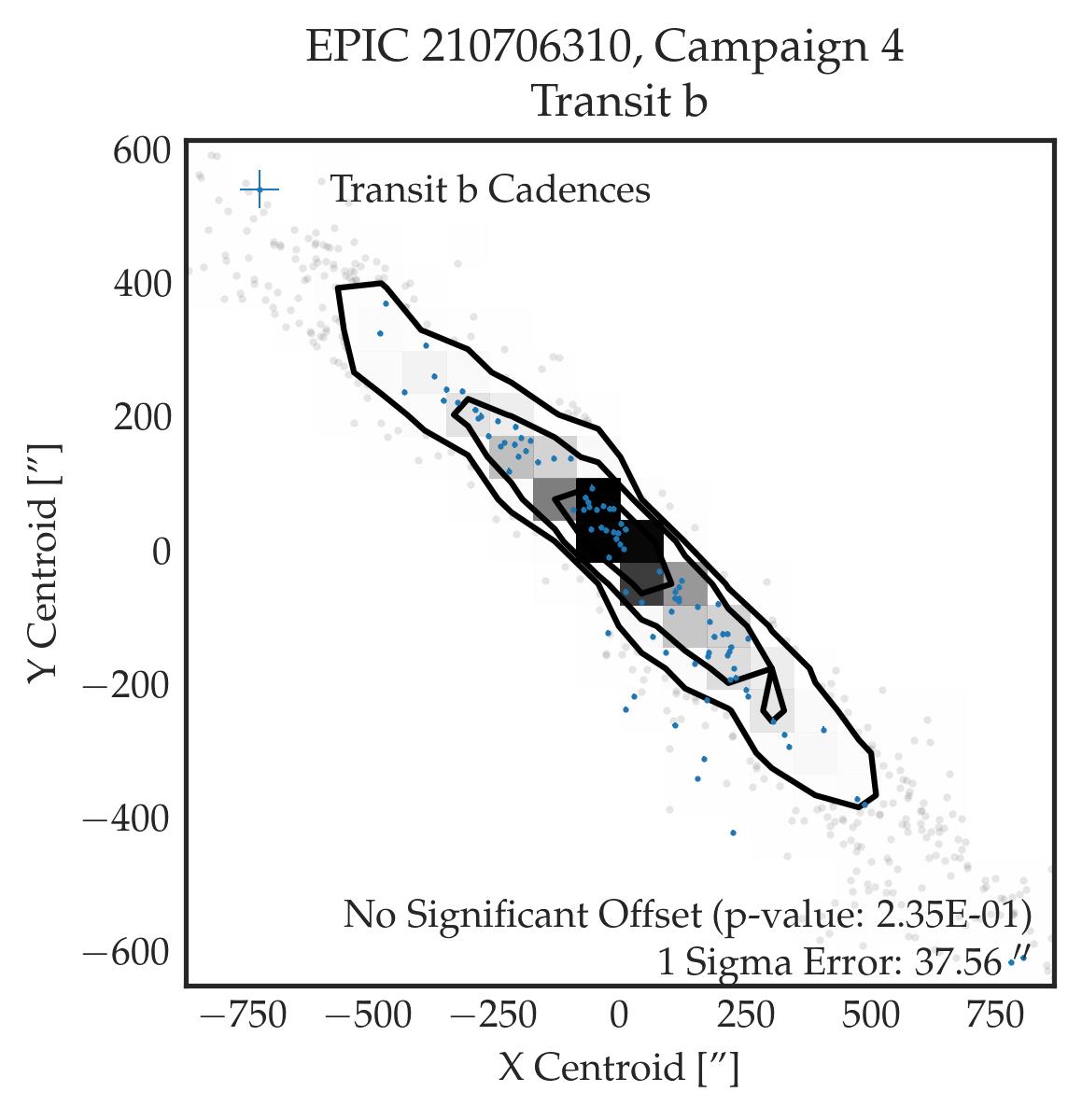}
                &
                \includegraphics[clip,width=0.5\columnwidth]{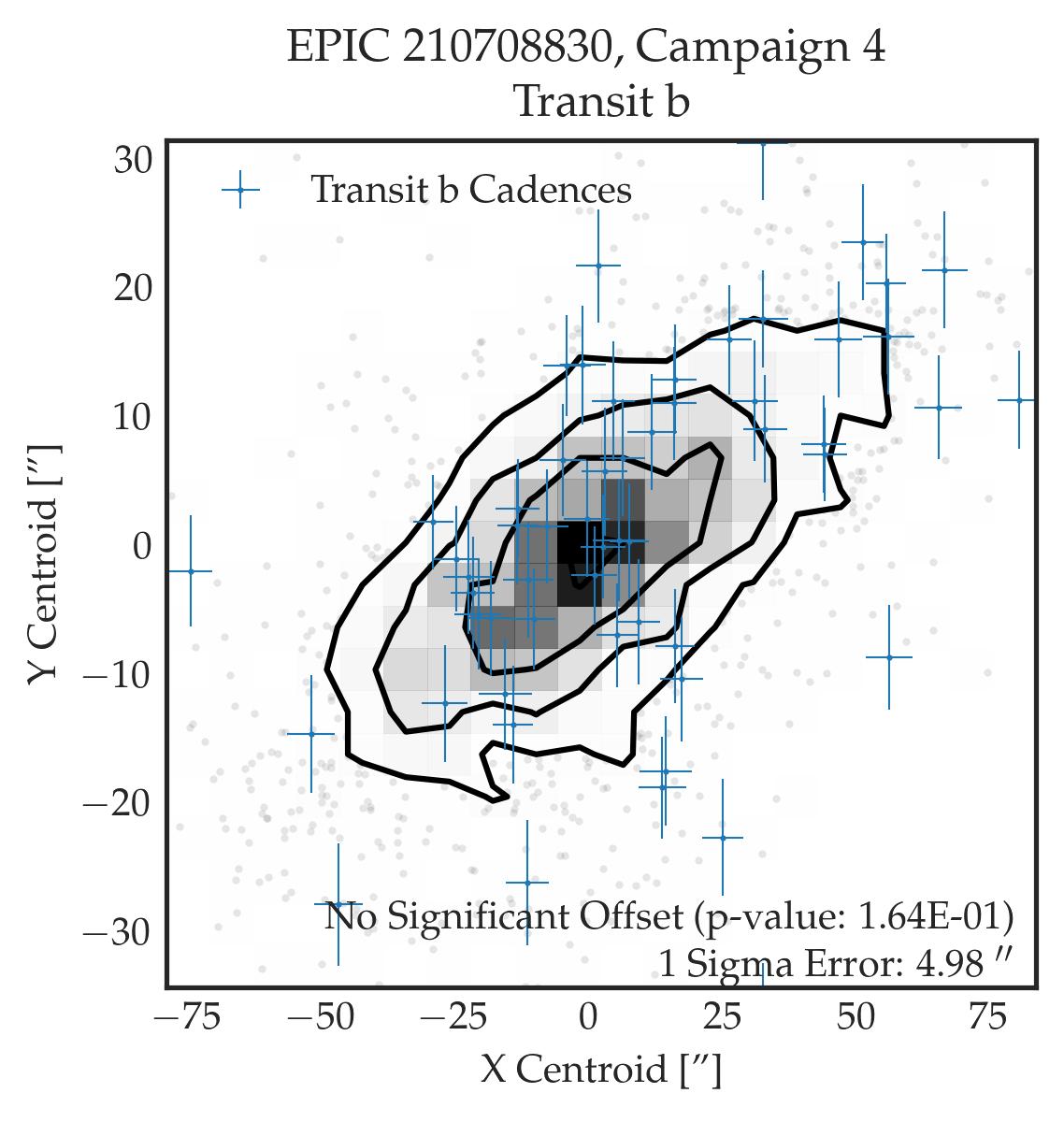}
                &
                \includegraphics[clip,width=0.5\columnwidth]{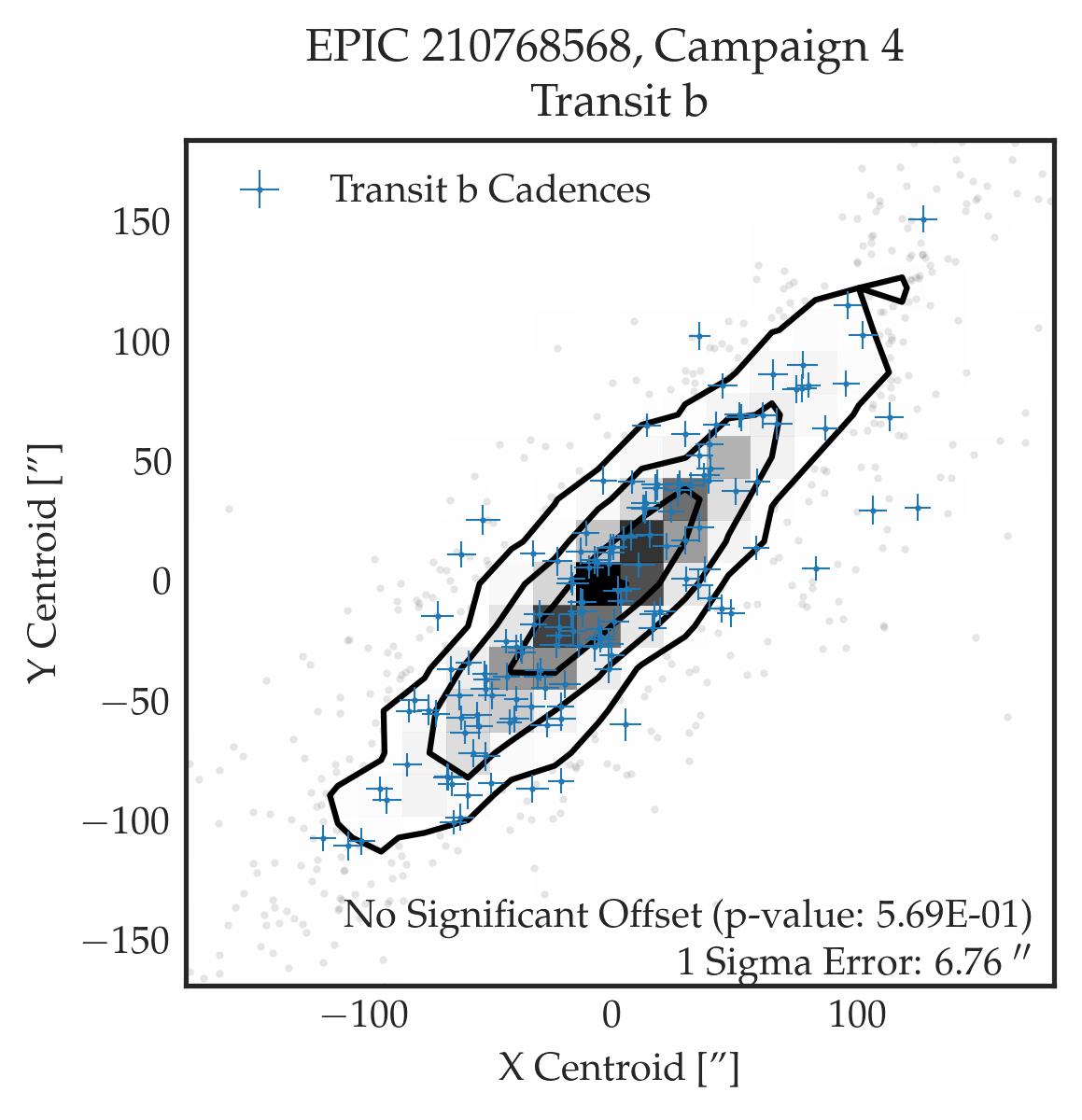}\\
                \includegraphics[clip,width=0.5\columnwidth]{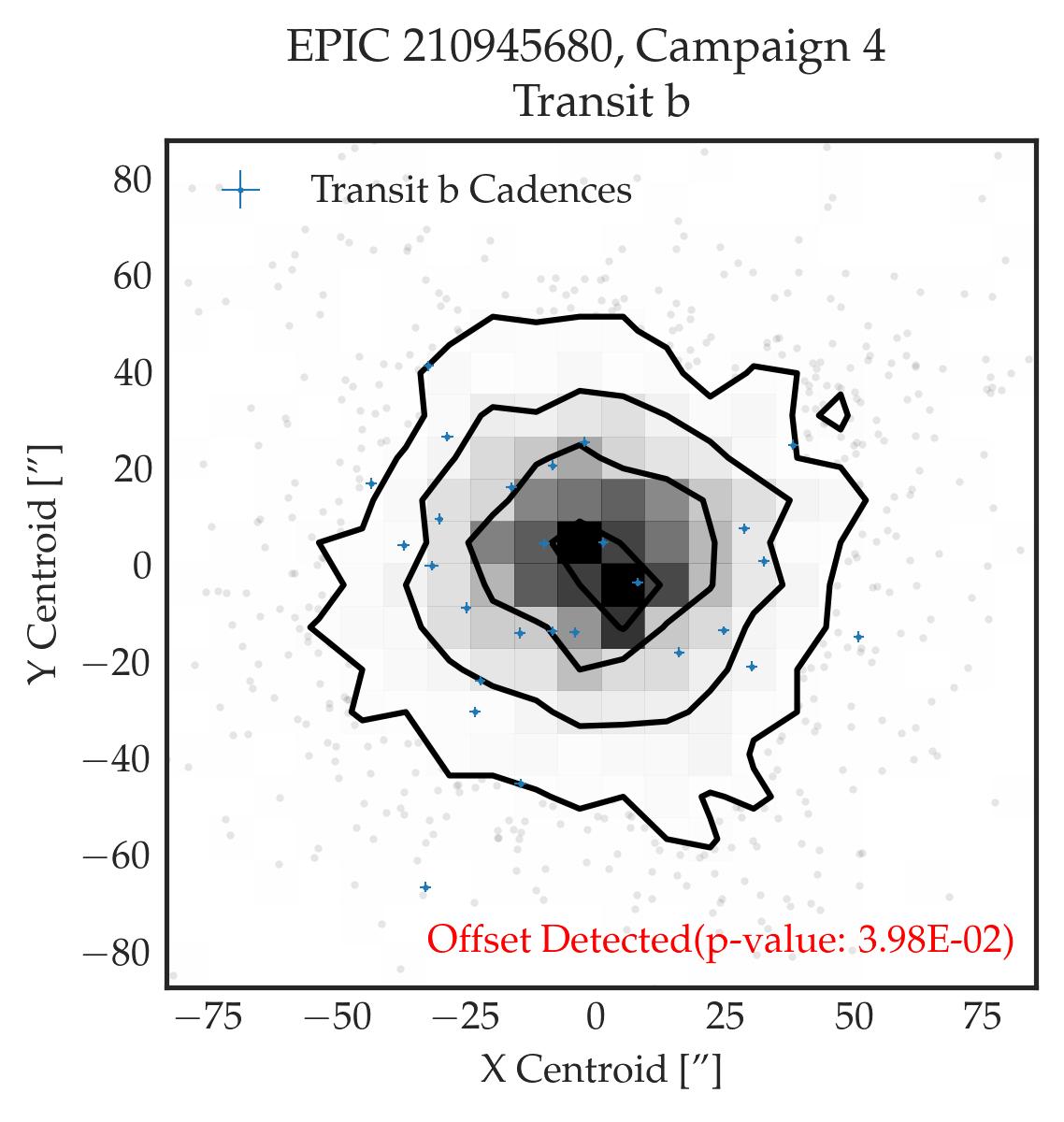}
                &
                \includegraphics[clip,width=0.5\columnwidth]{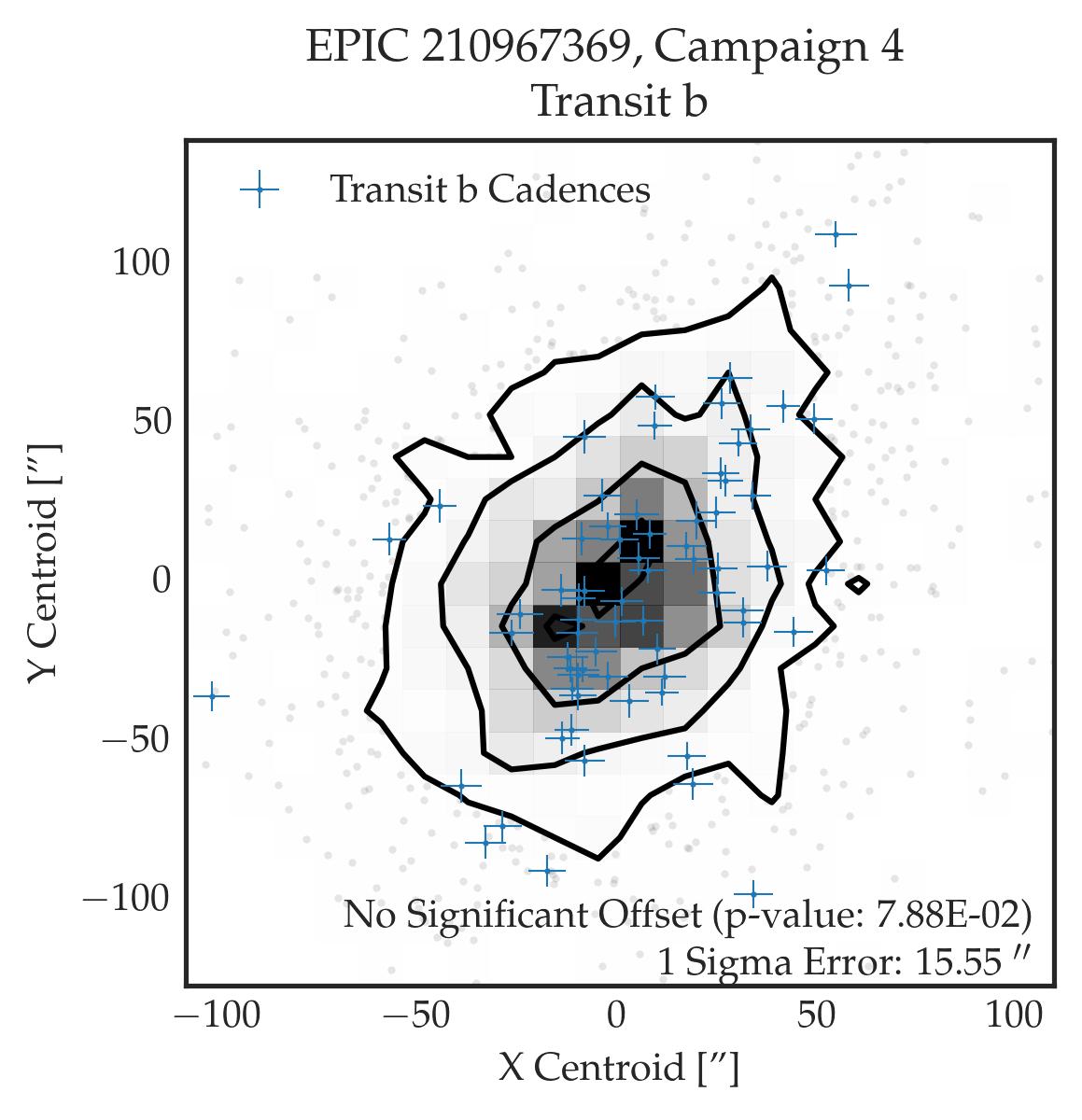}
                &
                \includegraphics[clip,width=0.5\columnwidth]{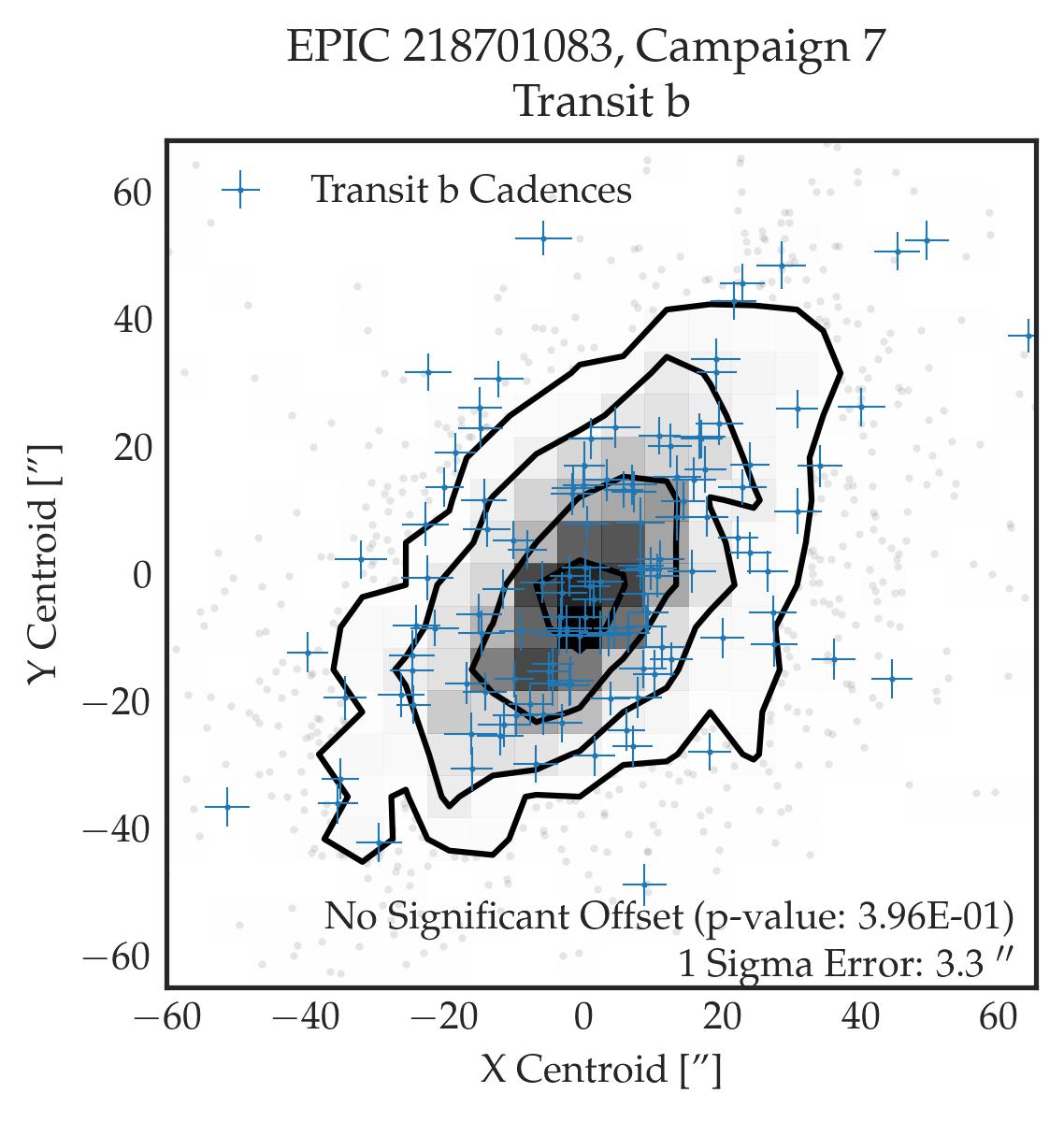}\\
                \includegraphics[clip,width=0.5\columnwidth]{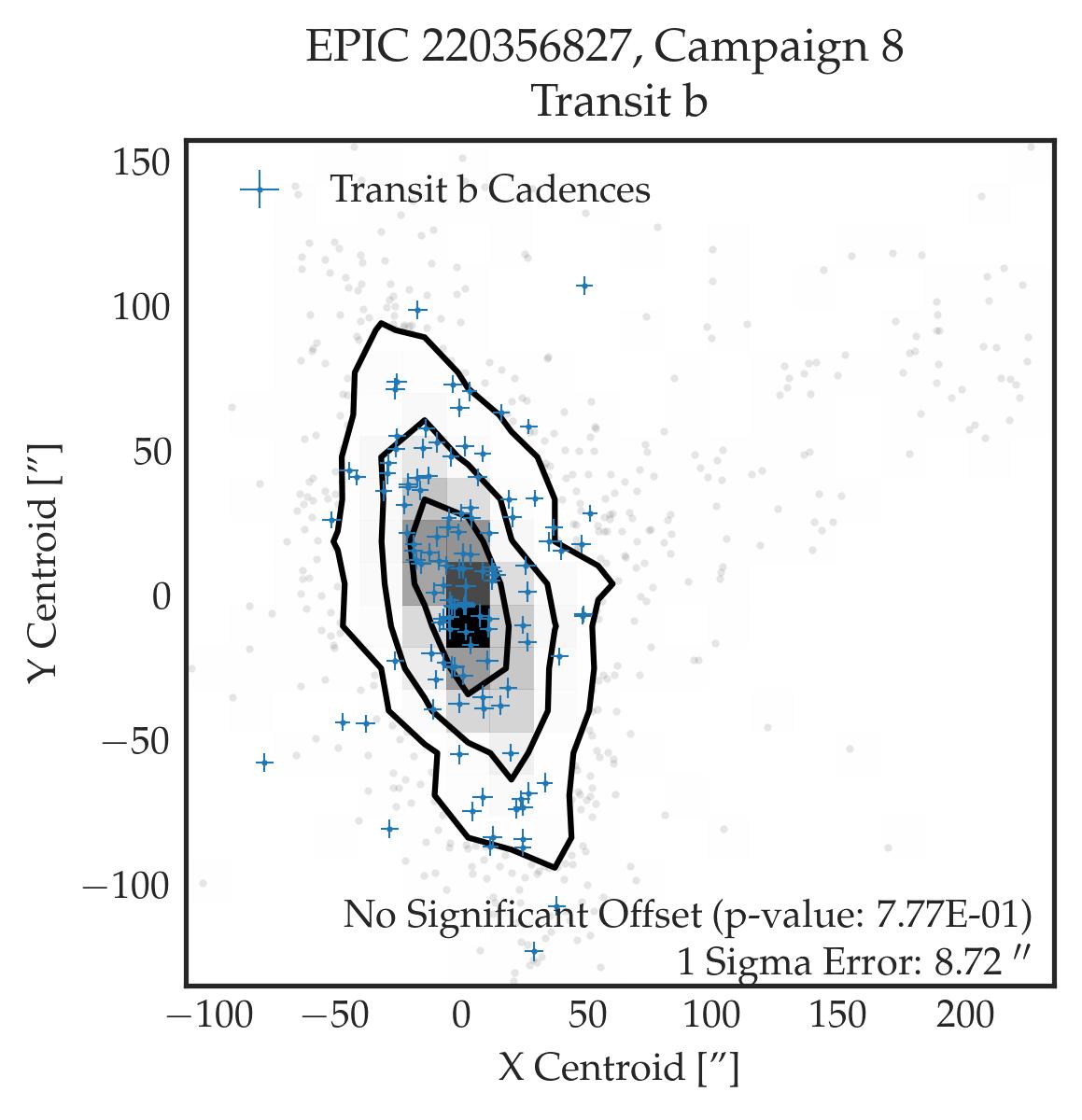}
                &
                \includegraphics[clip,width=0.5\columnwidth]{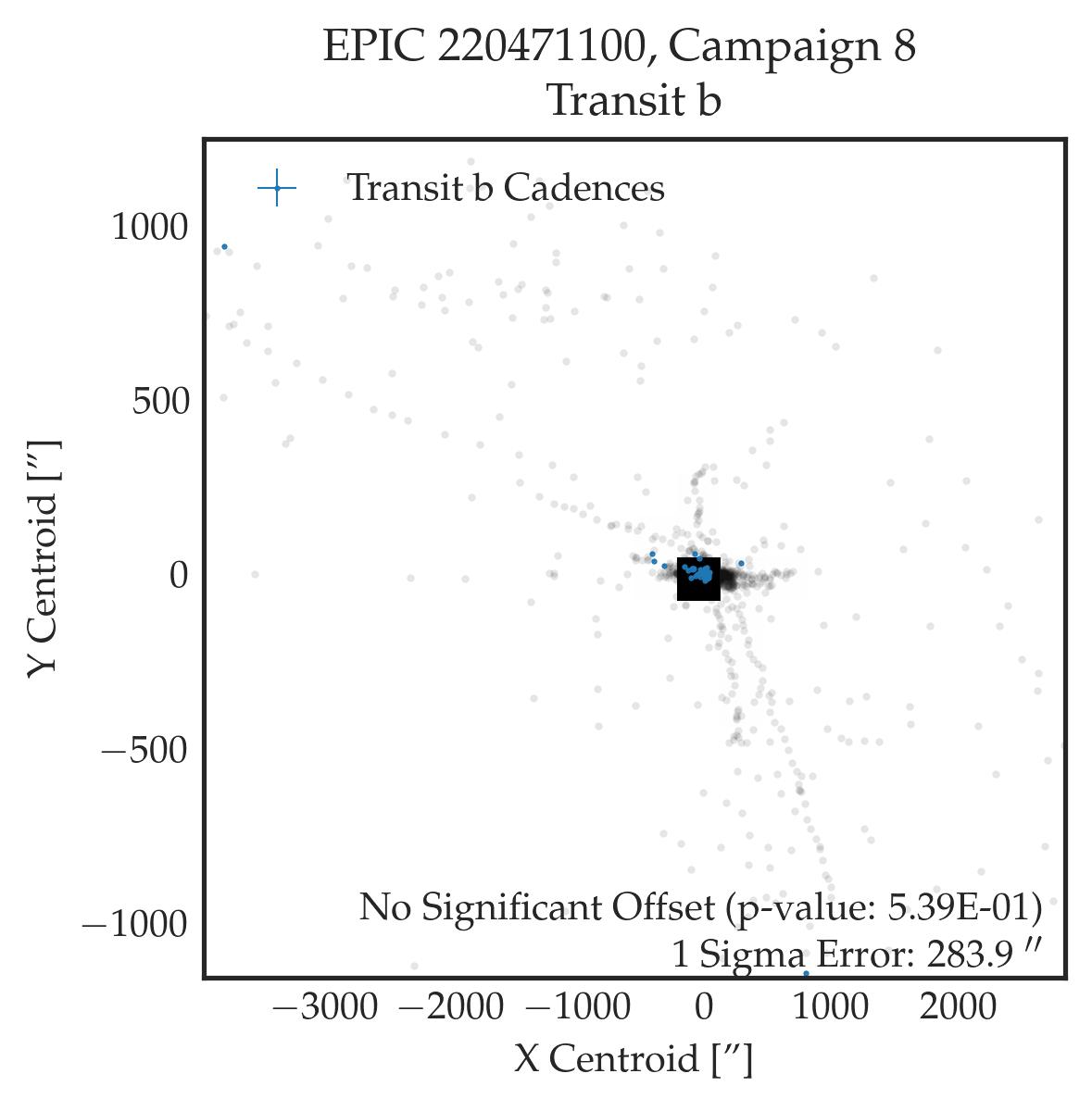}
                &
                \includegraphics[clip,width=0.5\columnwidth]{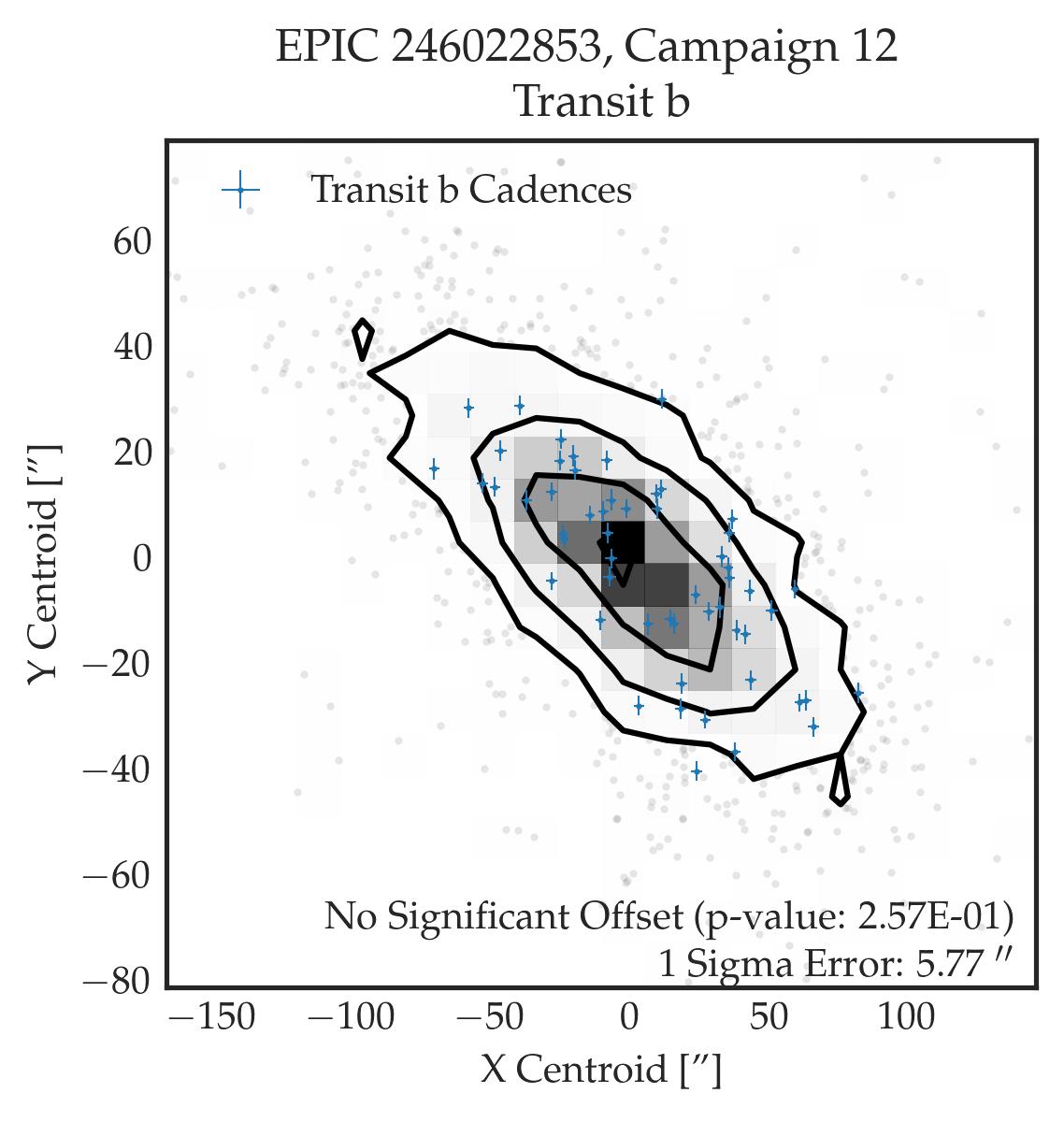}\\
            \end{tabular}
            \caption{Centroid plots for all single planet candidates listed in Table \ref{tab:planet}. The in-transit cadences centroid locations are denoted in blue while the out-of-transit centroid locations are denoted in grey. The 1$\sigma$, 2$\sigma$, and 3$\sigma$ contours of the centroids of the out-of-transit cadences are also represented. Candidates with significant centroid offsets (p-value<0.05) are denoted in red.}
            \label{fig:centsingle}
        \end{figure*}
        
        \begin{figure*}
            \centering
            \begin{tabular}{ccc}
                \includegraphics[clip,width=0.5\columnwidth]{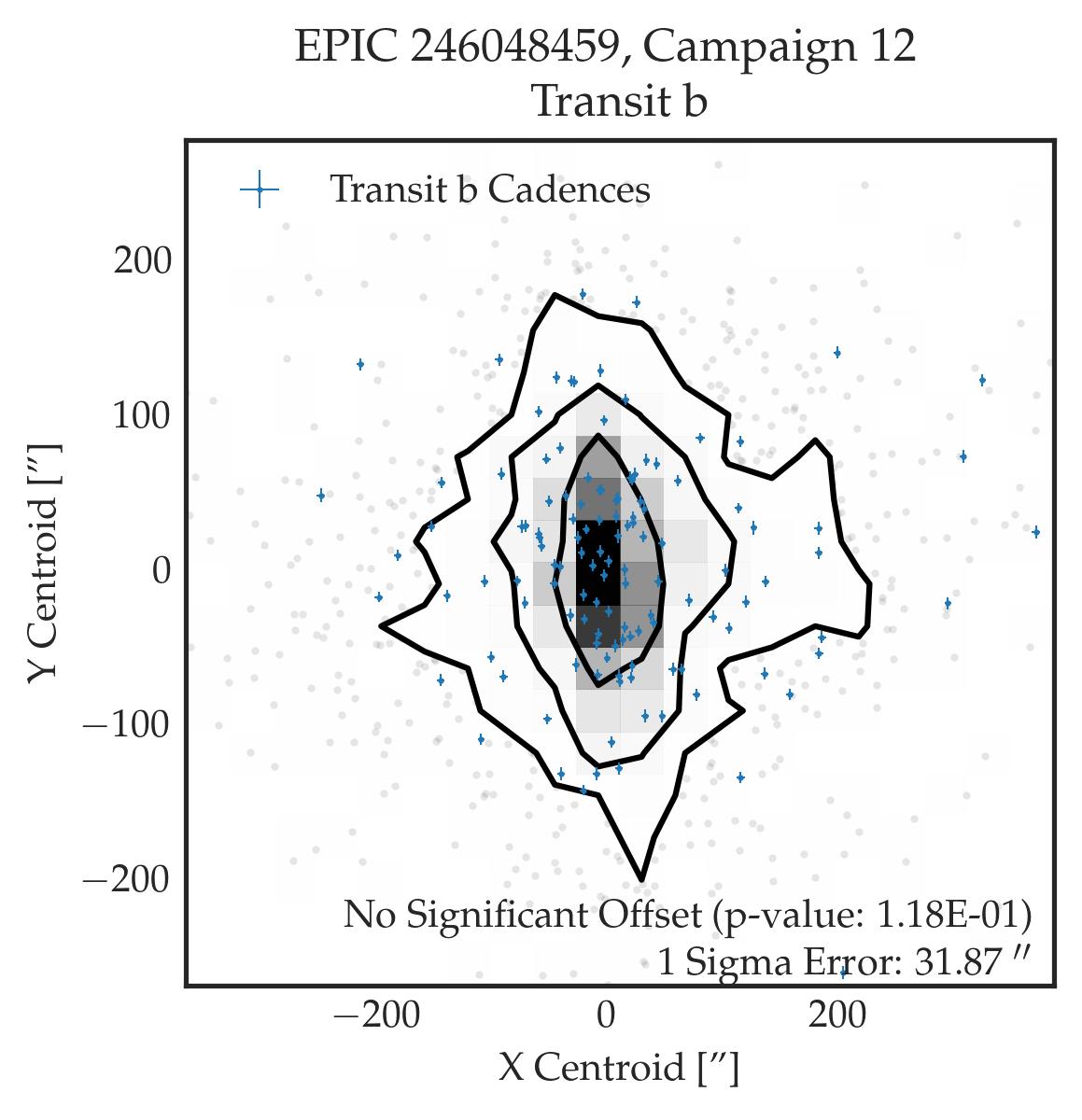}
                &
                \includegraphics[clip,width=0.5\columnwidth]{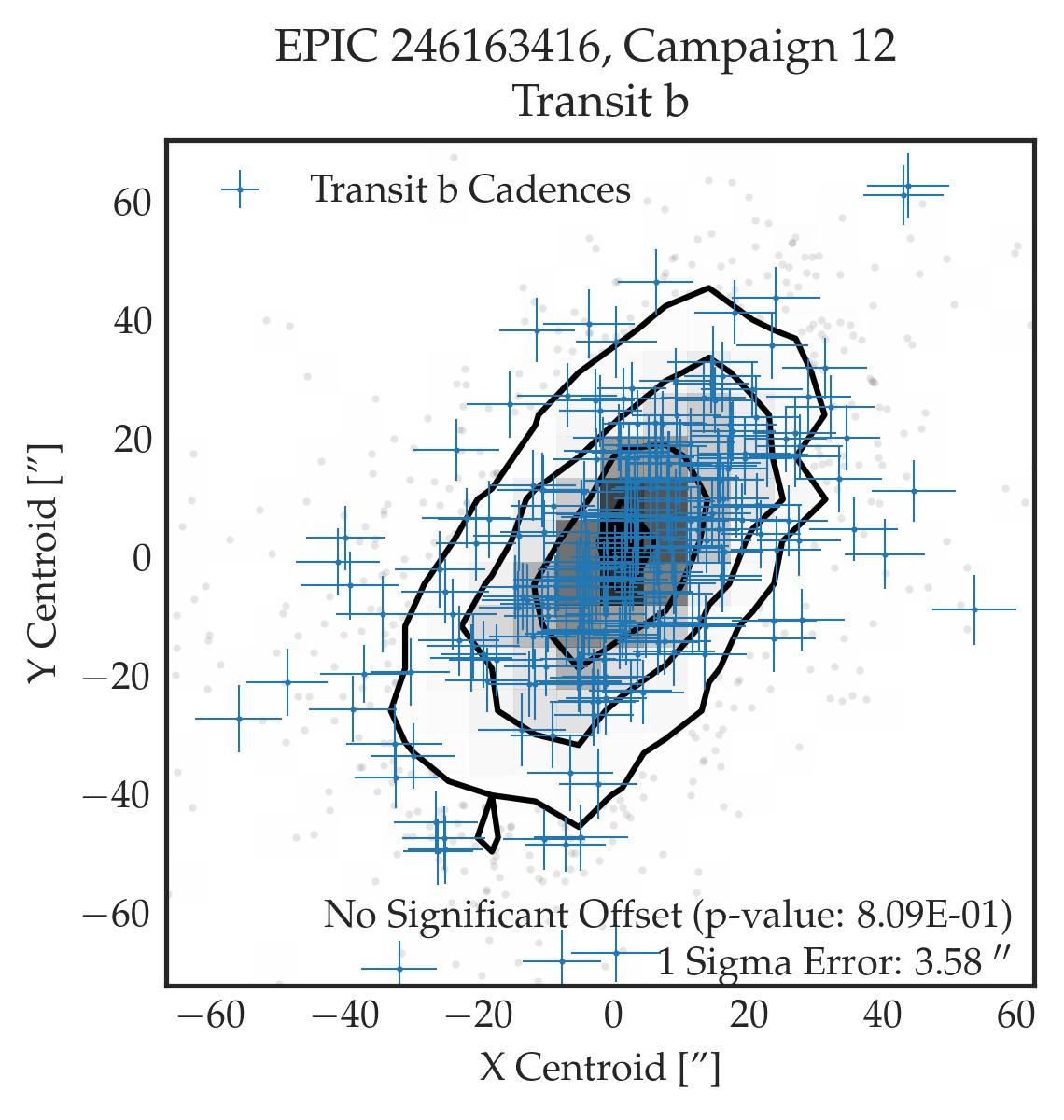}
                &
                \includegraphics[clip,width=0.5\columnwidth]{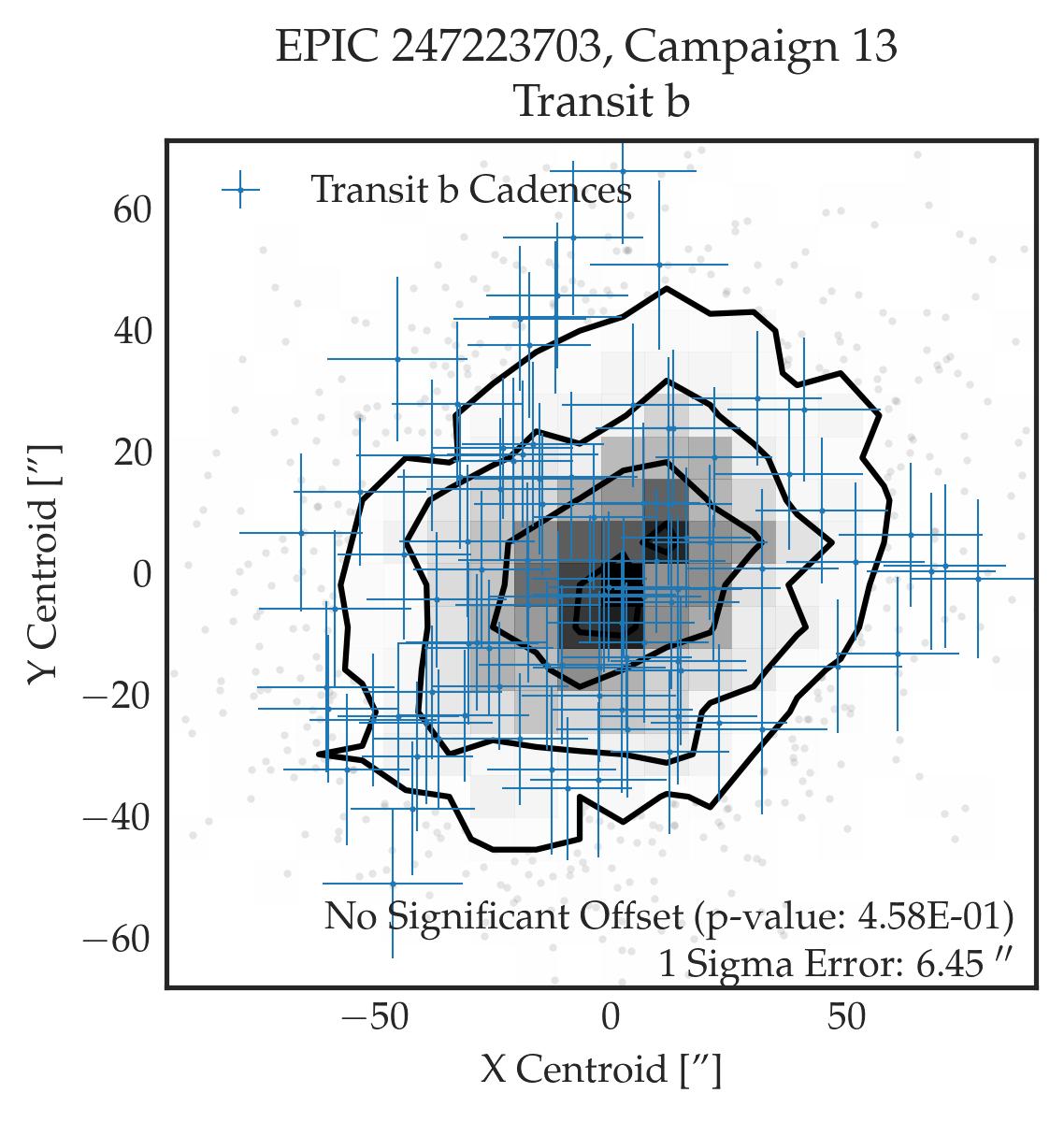}\\
                \includegraphics[clip,width=0.5\columnwidth]{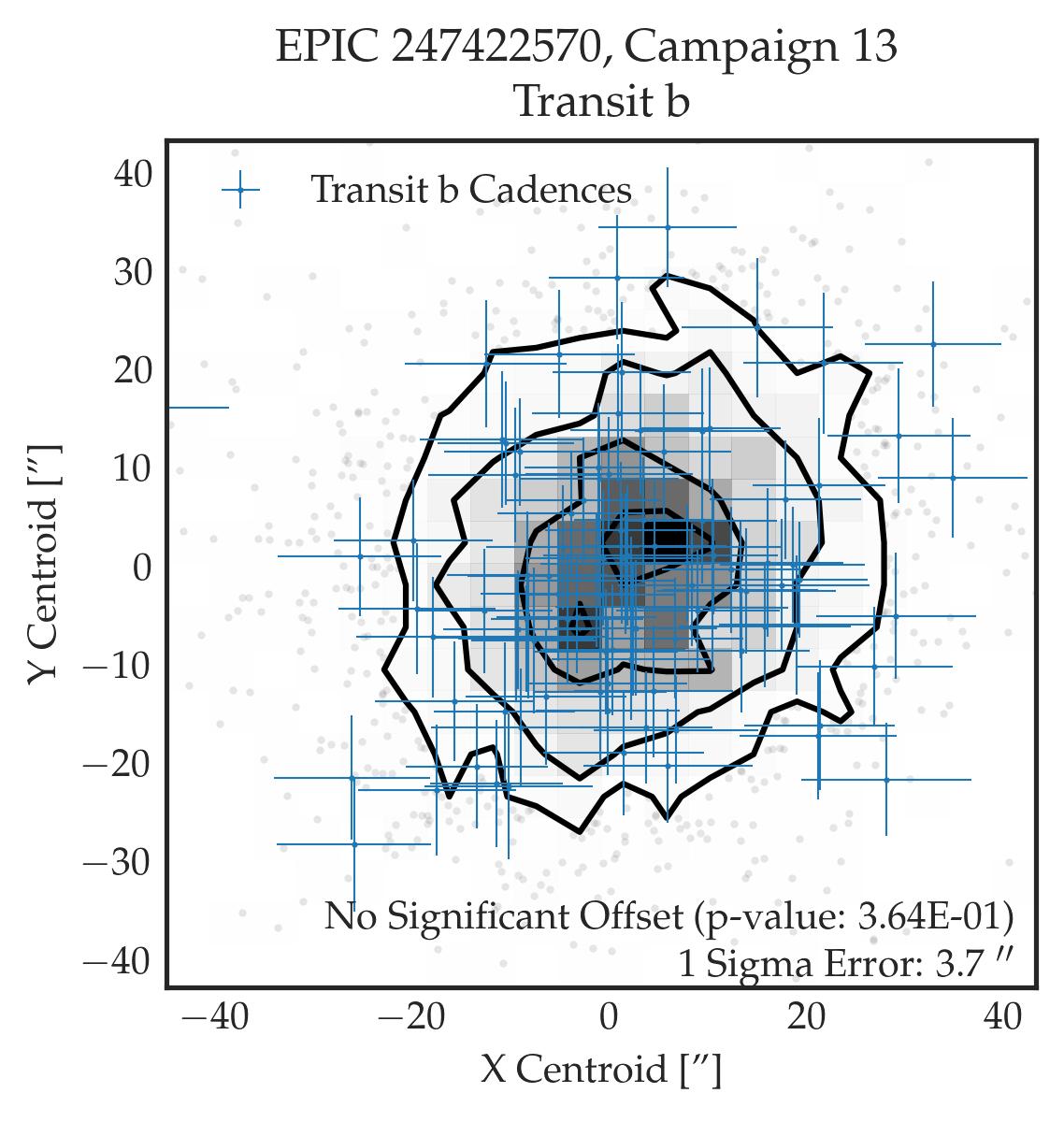}
                &
                \includegraphics[clip,width=0.5\columnwidth]{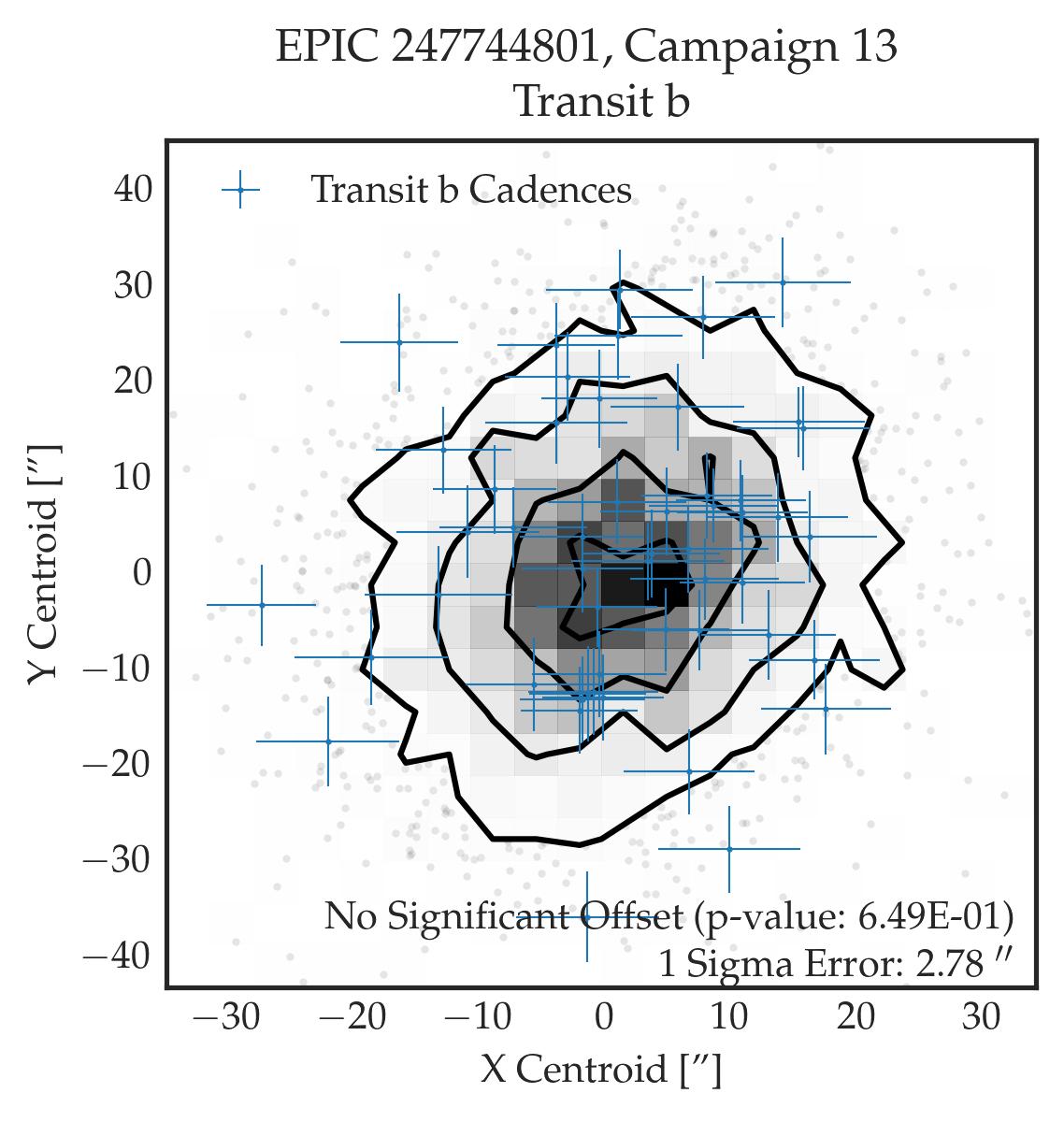}
                &
                \includegraphics[clip,width=0.5\columnwidth]{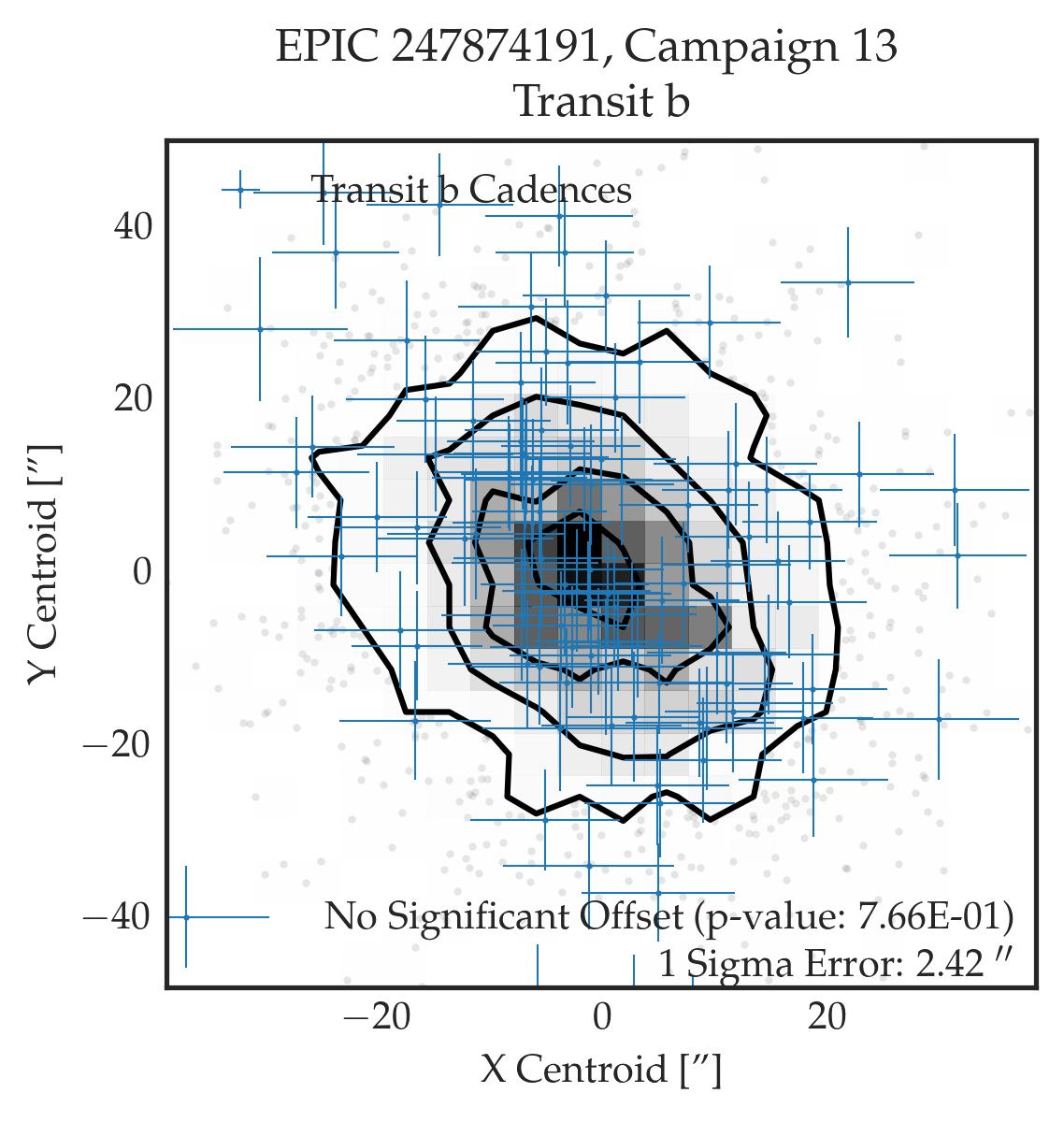}\\
                \includegraphics[clip,width=0.5\columnwidth]{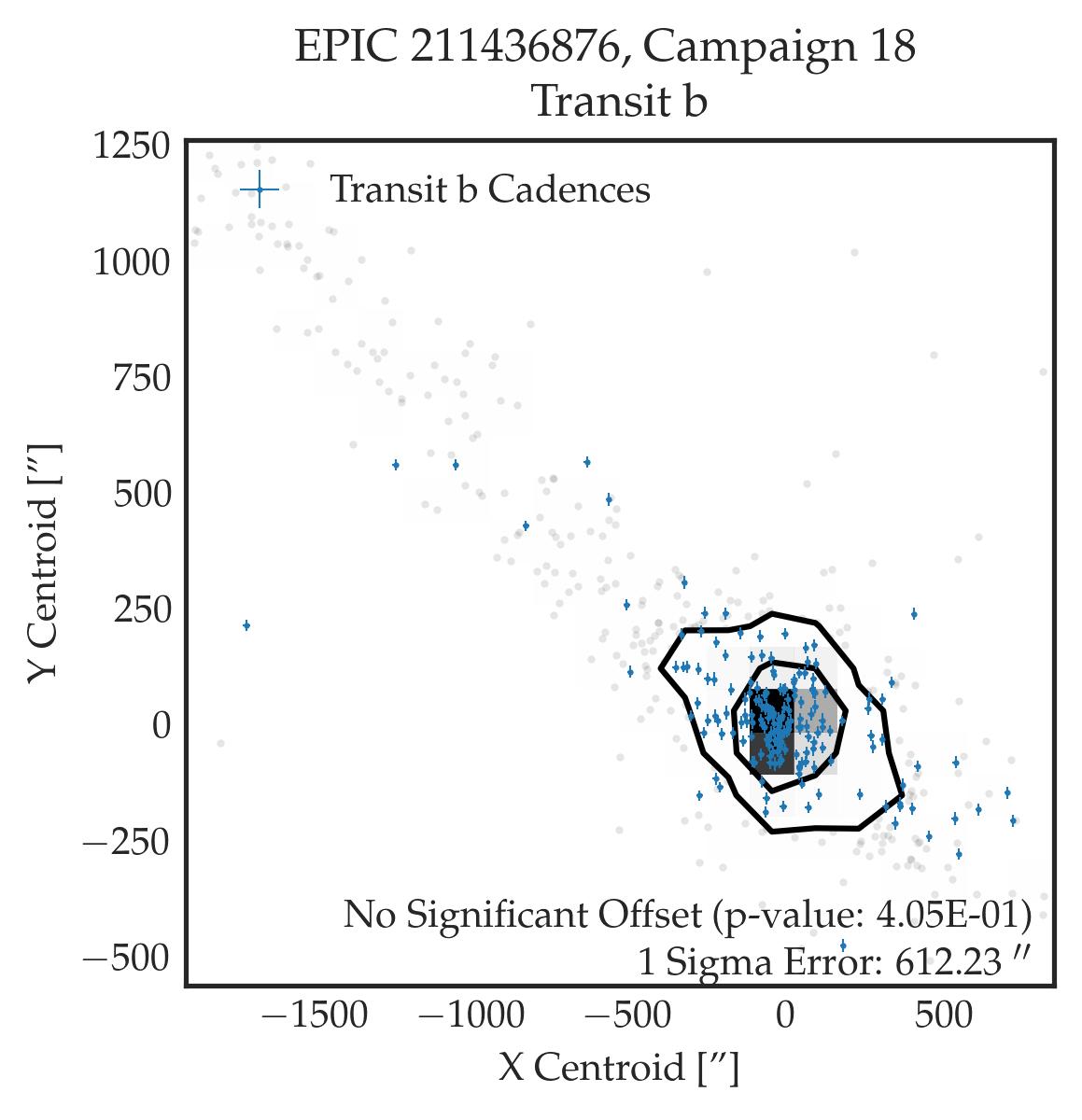}
                &
                \includegraphics[clip,width=0.5\columnwidth]{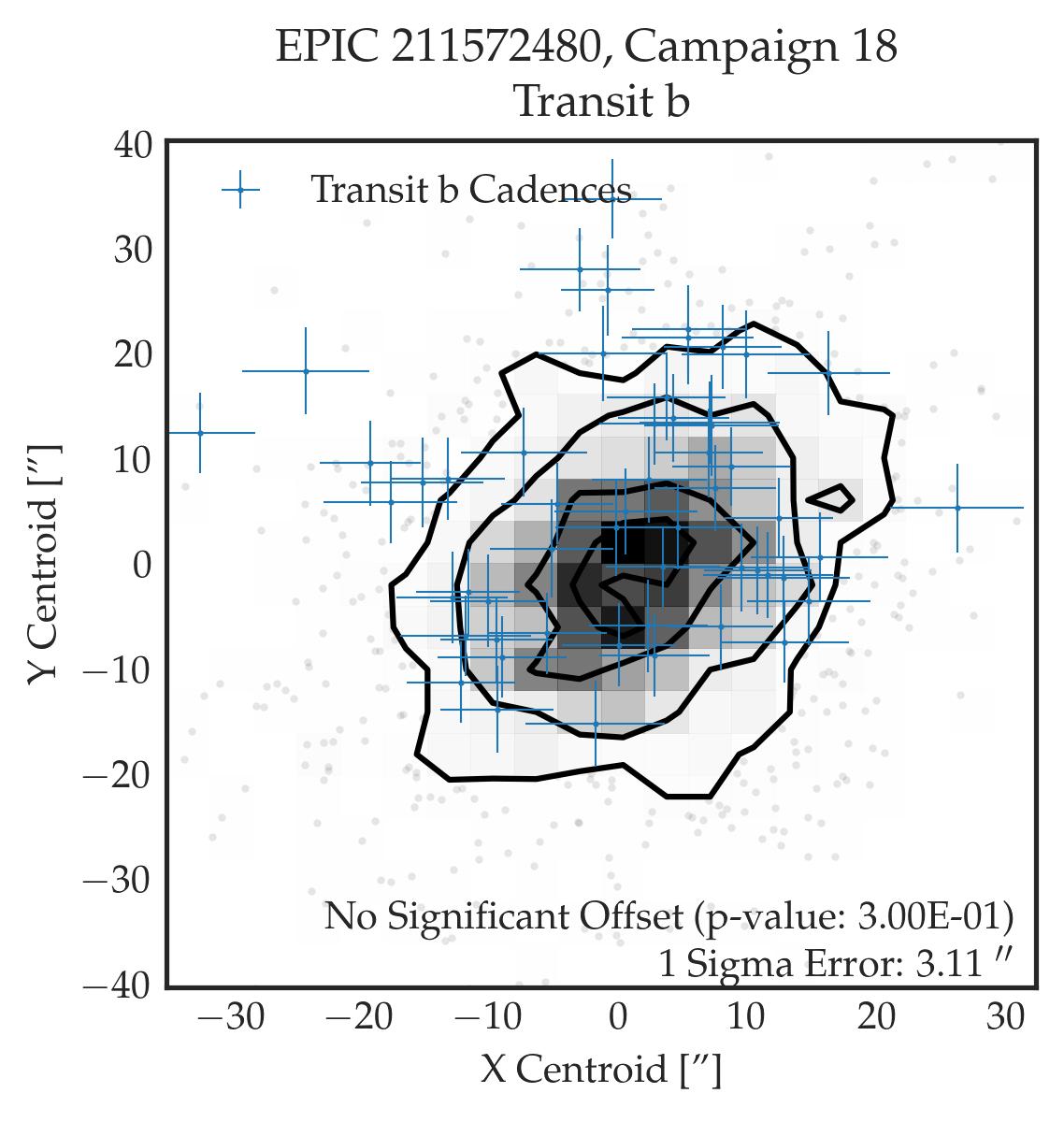}
                &
                \includegraphics[clip,width=0.5\columnwidth]{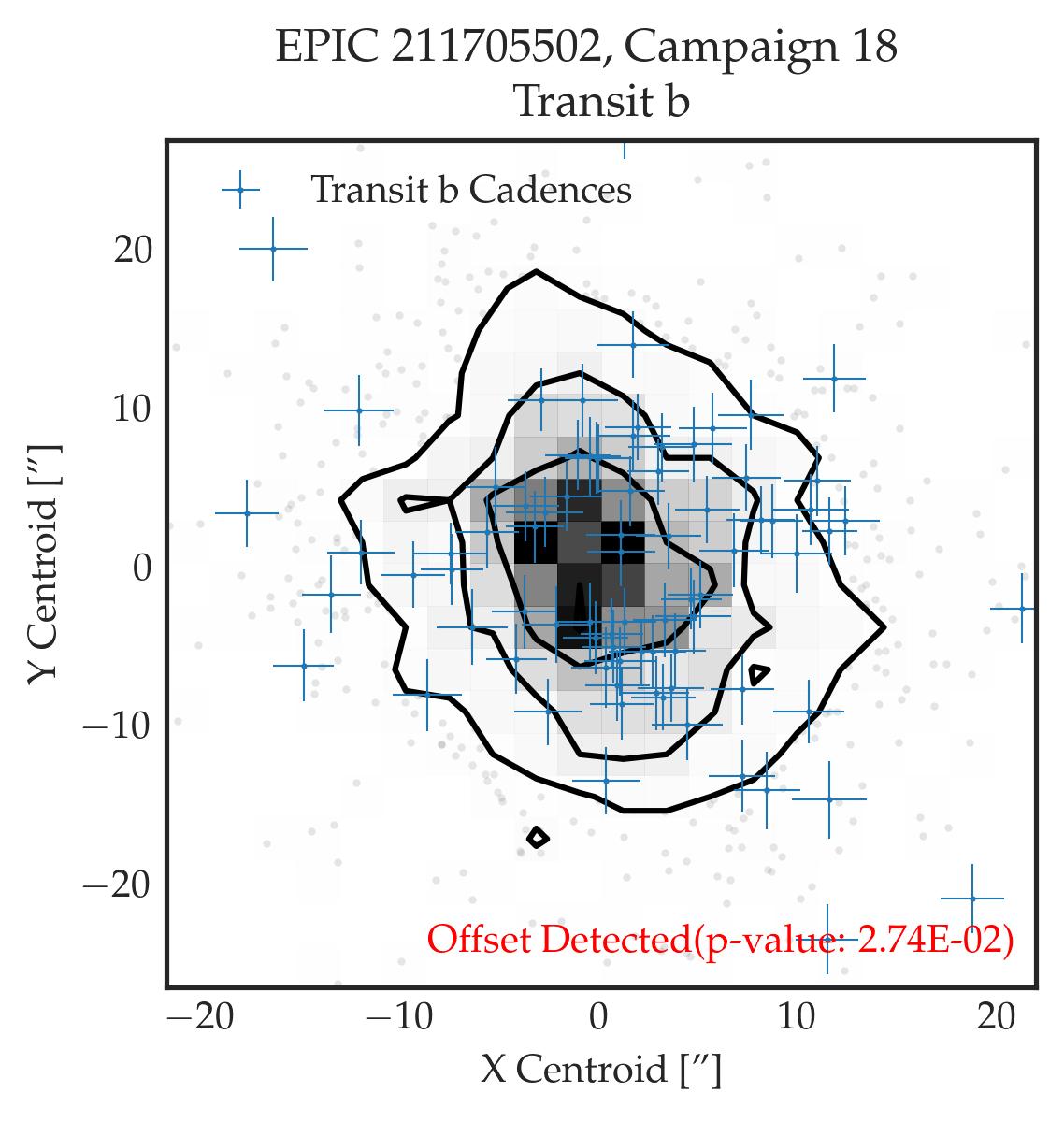}\\
            \end{tabular}
            \contcaption{Centroid plots for all single planet candidates listed in Table \ref{tab:planet}. The in-transit cadences centroid locations are denoted in blue while the out-of-transit centroid locations are denoted in grey. The 1$\sigma$, 2$\sigma$, and 3$\sigma$ contours of the centroids of the out-of-transit cadences are also represented. Candidates with significant centroid offsets (p-value<0.05) are denoted in red.}
        \end{figure*}
        
        \begin{figure*}
            \centering
            \begin{tabular}{l}
                \includegraphics[clip,width=0.5\textwidth]{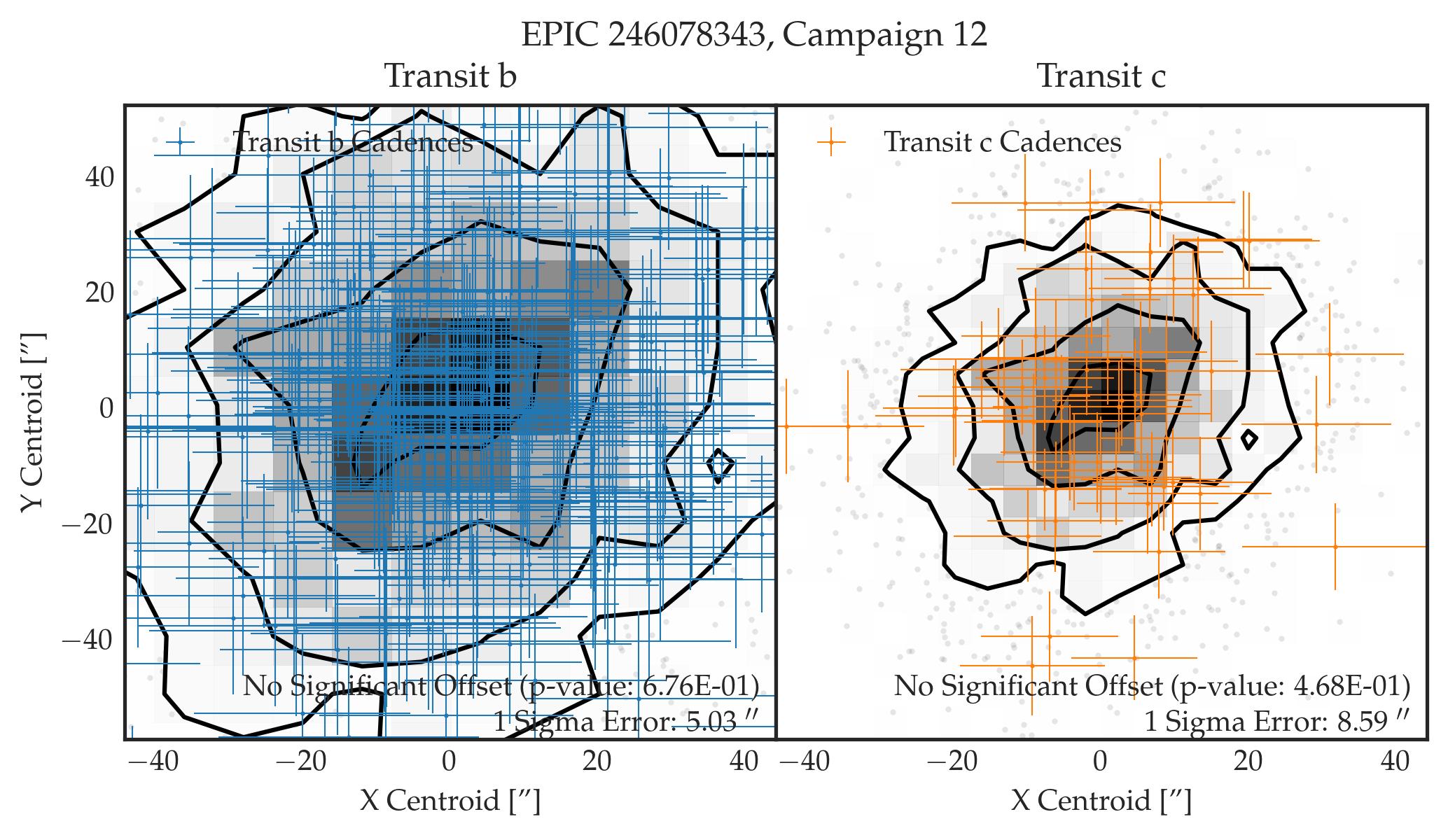}\\
                \includegraphics[clip,width=0.5\textwidth]{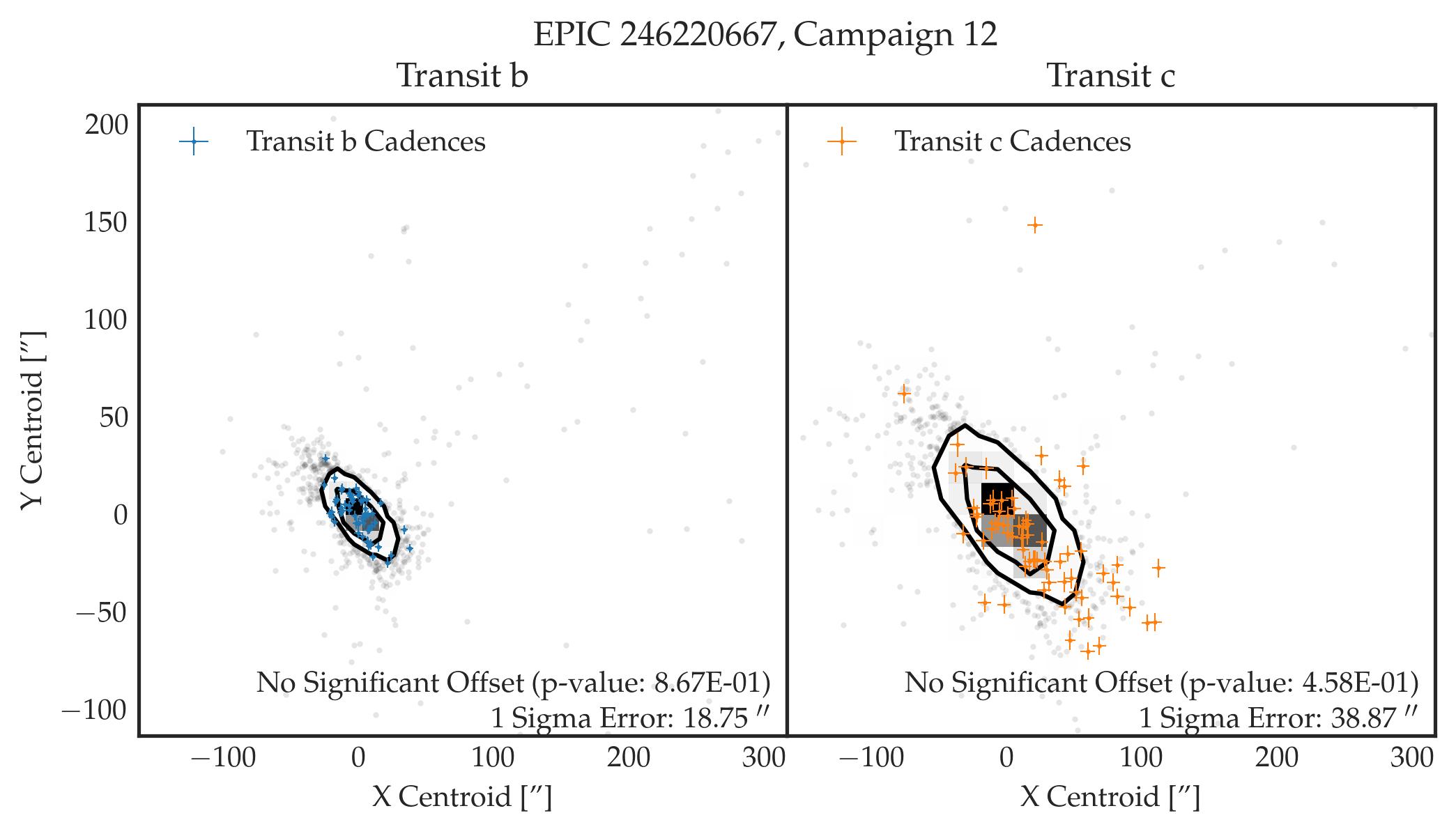}\\
                \includegraphics[clip,width=0.5\textwidth]{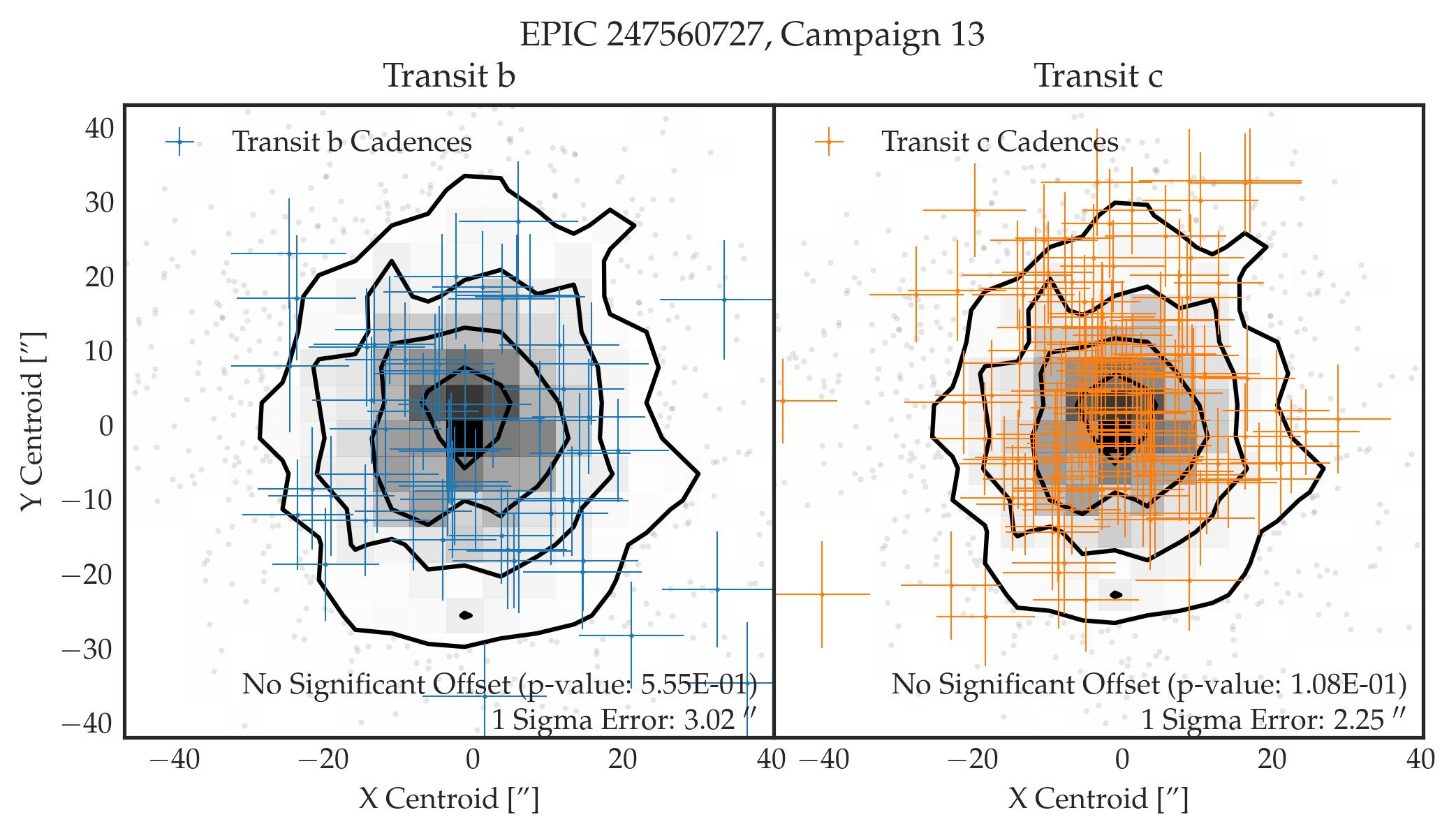}\\
            \end{tabular}
            \caption{Centroid plots for all multi planetary candidates listed in Table \ref{tab:planet}. The in-transit cadences centroid locations are denoted in blue while the out-of-transit centroid locations are denoted in grey. The 1$\sigma$, 2$\sigma$, and 3$\sigma$ contours of the centroids of the out-of-transit cadences are also represented.}
            \label{fig:centmulti}
        \end{figure*}

\subsection{Validated planets}
\label{subsec:validated}
\subsubsection{EPIC 210768568.01}
\label{subsubsec:210768568}

EPIC 210768568.01 is an Earth-sized planet ($R_{p}=0.9898^{+0.0498}_{-0.0486}$\rearth) orbiting around a relatively bright ($K_p$=11.935 mag, $G$=11.979 mag, $J$=10.704 mag) star (1.375$\pm0.068$\rsun, 1.018$\pm0.131$\msun) \citep{Paegert2021} observed by the \ktwo mission during the C4 campaign. Its coordinates are $(\alpha,~\delta)$(J2000) = (03:52:00.83,19:23:28.26), and its located at a distance of 296 pc \citep{Bailer-Jones2021}. EPIC 210768568.01 has an orbital semi-major axis of $0.0511^{+0.0021}_{-0.0028}~AU$, with a period of $3.2141\pm0.0002$ days, receiving a stellar insolation of $\sim$914$S/S_{\oplus}$. There is a nearby ($\sim$20.4$\arcsec$) fainter ($K_p$=16.689 mag) star partially affecting the \eve aperture. Following the vetting procedure explained in Section \ref{subsec:vetting}, we do not detect changes in the transit depth while modifying the aperture size to diminish the flux from the neighboring star. The \vespa and \tri FPP values are 0.0016, and 0.015, respectively. The centroid \texttt{p-value} for this target is 0.569 (see Figure \ref{fig:centsingle}), which is consistent with the target star being the source of the transiting signal. Also, the maximum computed separation that a background eclipsing binary could be at is 6.76$\arcsec$. We do not detect any companion star at closer separations using Speckle imaging data from SOAR and BTA. Using the \textit{Kepler} sample mass-radius relationship from \citet{Kanodia2019}, we predict a planetary mass of $\sim$2.34\mearth, which results in a RV semi-amplitude of $K\sim1m/s$ that is close to the detection limits of spectrographs like CARMENES \citep{Quirrenbach2010} and ESPRESSO \citep{Pepe2010}.

\subsubsection{EPIC 247422570.01}
\label{subsubsec:247422570}

EPIC 247422570.01 is a sub-Neptune planet ($R_{p}=2.1160^{+0.1050}_{-0.1052}$\rearth) orbiting a faint ($K_p$=15.160 mag, $G$= 15.133 mag, $J$=13.154 mag) G4 star (0.977$^{+0.066}_{-0.063}$\rsun, 0.893$^{+0.406}_{-0.275}$\msun) \citep{Hardegree2020} observed during \ktwo campaign C13. It is located at $(\alpha,~\delta)$(J2000) = (05:05:02.92,21:34:48.55) at a distance of $\sim$669 pc \citep{Bailer-Jones2021}. EPIC 247422570.01 orbits its star at a distance of 0.0586$^{+0.0032}_{-0.0033}~AU$ with a period of 5.9382$^{+0.0006}_{-0.0004}$ day, receiving a stellar insolation of $\sim$243$S/S_{\oplus}$. There are two nearby ($\sim$11\arcsec, and $\sim$16\arcsec) fainter ($G$=20.885 mag, and $G$=19.088 mag) stars partially within the \eve aperture. Following our vetting procedure, we have modified the photometric aperture to minimize the contamination from these two neighboring stars. In this case, changing the aperture we did not detect any significant changes in the transit depth. Also, the light curve obtained using \verb|lightkurve|, and centering a smaller aperture at the position of the fainter neighboring stars, does not produce a transiting feature at the listed period. In addition, the centroid \texttt{p-value}  for this target is 0.364, and the maximum computed separation for a background eclipsing binary is 3.7\arcsec (see Figure \ref{fig:centsingle}). Given that this distance is smaller than the angular separation of the neighboring stars, and the fact that we do not detect any other source with BTA speckle data, we consider EPIC 247422570 to be the host star of this transiting exoplanet. \vespa returns a FPP=0, and \tri returns a FPP of $\sim4\times10^{-3}$. Using the \textit{Kepler} sample from \citet{Kanodia2019}, we predict a planetary mass of $\sim$5.58\mearth.

\begin{table*}
    \centering
    \caption{Comparison of \gaia properties for EPIC 247422570 and contaminating nearby stars.}
    \label{tab:epic247422570}
    \begin{tabular}{llllll}
\hline
EPIC      & \gaia eDR3 & $G$ [mag] & \verb|GOF_AL| & \verb|D| & $RUWE$ \\ 
\hline
247422570 &  3409152693750235008      &   15.13      &    1.21    &   0.00    &  1.05 \\ 
\hline
 -- &   3409152629326319744     &    20.88     &   1.33    &    1.26     &    --  \\
 -- &   3409152625030757888     &    19.09     &   -0.65   &    0.00     & 0.97   \\ 
\hline
\end{tabular}
\end{table*}

\subsubsection{EPIC 246078343.01 \& EPIC 246078343.02}
\label{subsubsec:246078343}

\begin{figure*}
    \includegraphics[clip,trim={0 0 0 0},height=10cm, keepaspectratio]{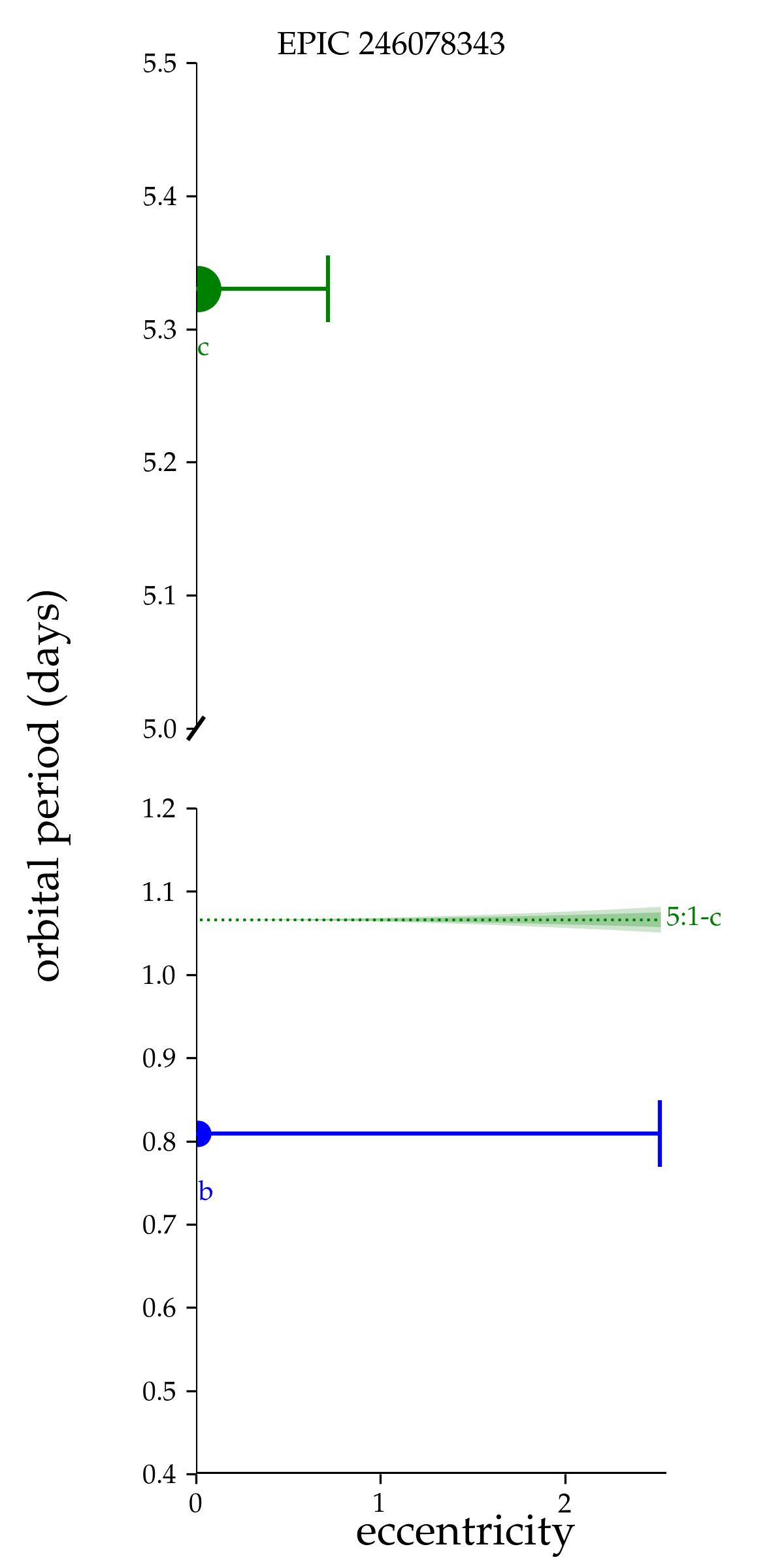}
    \caption{Resonance locations in the EPIC 246078343 system. The horizontal axis denotes the eccentricity while the vertical axis shows the orbital period (note that the y axis is discontinuous and not to scale). The location of the planets EPIC 246078343.01, and EPIC 246078343.02 are represented in blue and green circles respectively, and the solid horizontal lines extend to the eccentricity at which each planet would cross the next planet’s orbit. The orbital periods for these two planets are too separated to obtain resonant orbits.}
    \label{fig:res_246078343}
\end{figure*}

EPIC 246078343 is a faint ($K_{p}$=14.557 mag, $G$=14.565 mag, $J$=12.644 mag) K7 star (0.700$^{+0.055}_{-0.048}$\rsun,0.808$^{+0.364}_{-0.256}$\msun) \citep{Hardegree2020} observed during \ktwo campaigns C12 and C19. It is located at $(\alpha,~\delta)$(J2000) = (23:33:40.22,-07:36:42.98) at a distance of $\sim$253 pc \citep{Bailer-Jones2021}.
It is orbited by two planets: EPIC 246078343.01 is a sub-Earth USP planet (0.7599$^{+0.0758}_{-0.0496}$\rearth) with an orbital semi-major axis of 0.0117$^{+0.0009}_{-0.0012}~AU$, and a period of $0.8094\pm0.00003$ days. It has a \vespa FPP value of $6\times10^{-4}$ and a \tri FPP of 0.009 after applying the multiplicity boost. Using the mass-radius estimation from \citet{Chen2016}, we predict a planetary mass of $\sim$0.36\mearth. With these planetary parameters, EPIC 246078343.01 would be a planet with a similar structure to Mercury’s interior, as GJ 367 b \citep{Lam2021}. Also, the presence of a second planet in the system is to be expected given that USP planets are typically accompanied by other planets with orbital periods between 1-50 days \citep{SanchisOjeda2014}. EPIC 246078343.02 is a 1.2327$^{+0.0565}_{-0.0593}$\rearth super-Earth planet orbiting at a distance of 0.0426$^{+0.0016}_{-0.0019}~AU$, with a period of $5.3301\pm0.0003$ days. It was first reported by \citet{Dattilo2019}, as a candidate planet in a 5.3288-day orbit, in their \ktwo planet candidate training/test set. We detect this planet in \ktwo C12 \eve and \tfaw light curves, and also, as part of our vetting procedure, in the \verb|K2SFF| one. Given the shorter length ($\sim$6 days) of the good quality data points for the C19 campaign, we are not able to detect the planet using the available light curves. It has a \vespa FPP value of $2\times10^{-4}$ and a \tri FPP value of 0.007, after applying the multiplicity boost, and the available LDT contrast curves, as explained in Section \ref{subsec:validation}. The centroid \texttt{p-values} for both planets are 0.468 and 0.676, and the nearest background source would be at distances 8.59\arcsec, and 5.03\arcsec, respectively (see Figure \ref{fig:centmulti}). In both cases, they are consistent with the target star being the source of the transiting signals. Using the mass-radius estimation from \citet{Chen2016}, we calculate a planetary mass of $\sim$1.83\mearth. Given the orbital periods of these two planets we do not obtain resonant orbits in this system (see Figure \ref{fig:res_246078343}). 

\subsubsection{EPIC 246220667.01 \& EPIC 246220667.02}
\label{subsubsec:246220667}

\begin{figure*}
    \includegraphics[clip,trim={0 0 0 0}, height=10cm, keepaspectratio]{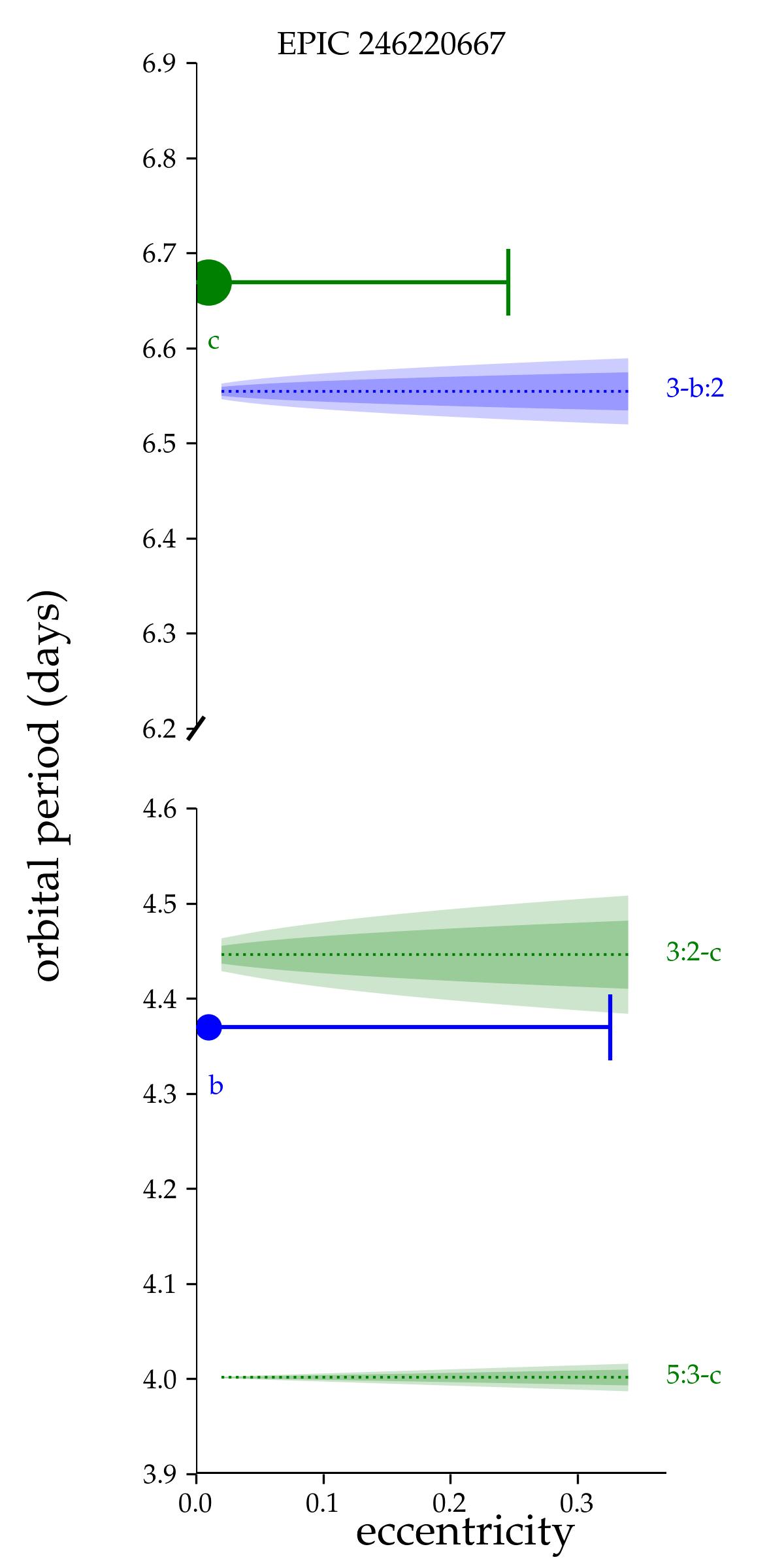}
    \caption{Resonance locations in the EPIC 246220667 system. Same notation as Figure \ref{fig:res_246078343}. The dashed lines show the location of the estimated 3:2 and 5:3 mean motion resonances for planets EPIC 246220667.01 and EPIC 246220667.02. The shaded regions around each resonance are the widths corresponding to the lower (dark shading) and upper (light shading) planet mass limits, propagated from the radii uncertainties, and estimated using the \textit{Kepler} mass-radius relationship from \citet{Kanodia2019}. The label “5:3-c” indicates, that a test particle at that location would complete 3 orbits in the same amount of time that planet c takes to complete 5 orbits.}
    \label{fig:res_246220667}
\end{figure*}

EPIC 246330667 is a faint ($K_{p}$=13.977 mag, $G$=13.929 mag, $J$=12.184 mag) K5 star (0.732$^{+ 0.055}_{- 0.052}$\rsun, 0.814$^{+ 0.363}_{- 0.251}$\msun) \citep{Hardegree2020} observed during \ktwo campaigns C12 and C19. It is located at $(\alpha,~\delta)$(J2000) = (23:26:32.7,-04:36:23.69) at a distance of $\sim$256 pc \citep{Bailer-Jones2021}. It is a multi-planetary system consisting of two planets: EPIC 246330667.01 with a period of 4.3606$\pm$0.0001, and EPIC 246330667.02 with 6.6690$\pm$0.0002 days. With the reported periods, they seem to be close to their 3:2 resonance (see Figure \ref{fig:res_246220667}). Although campaign C19 was not considered in our \tfaw survey (as there is no \eve light curve available for this campaign), during our vetting procedure we searched for these two planets in the available C19 light curves for this system. We detect one transit of EPIC 246330667.02 in the \verb|K2SFF| light curve as well as two transits from the TPF light curve obtained using the \verb|lightkurve| package. We also detect a transit-like feature in the phase-folded \verb|K2SFF| light curve for EPIC 246330667.01.
EPIC 246220667.01 is a validated super-Earth planet (1.2191$^{+0.0993}_{-0.0731}$\rearth) orbiting its host star at a distance of 0.0487$^{+0.0027}_{-0.0031}~AU$. It has \vespa and \tri FPP values of 0.97\% and 0.67\%, respectively, after the multiplicity boost is applied. Using the mass-radius estimation from \citet{Chen2016} we compute a planetary mass of $\sim$1.22\mearth.

EPIC 246220667.02 is a validated sub-Neptune planet (1.9288$^{+0.0621}_{-0.0719}$\rearth) orbiting its host star at a distance of 0.0552$^{+0.0017}_{-0.0023}~AU$. In this case, given the transit depth ($\sim$1ppt) of EPIC 246330667.02, the original \eve light curve presented trimmed transits. We recomputed the \eve light curve, by masking the transit and re-running the PLD analysis, to ensure unbiased results of the planetary radius. The \vespa and \tri FPP values for this planet are $10^{-3}$ and $5\times10^{-3}$ with the multiplicity boost applied. Using the \textit{Kepler} sample from \citet{Kanodia2019}, the estimated planetary mass is $\sim$5.03\mearth. With an incident flux of $\sim$56.13$S/S_{\oplus}$, EPIC 246220667.02 lies at the upper edge of the Radius Gap for a K-type star \citep{Fulton2017,Zeng2017,Petigura2022}.

The centroid \texttt{p-values} are 0.867 and 0.458, and the distances to the nearest background sources are 18.75\arcsec and 38.87\arcsec, respectively (see Figure \ref{fig:centmulti}). We do not detect any contaminating source within these distances neither with \gaia eDR3 data nor with our LDT speckle imaging observations.

\subsection{Highlights of our planet candidate sample}
\label{subsec:candidates}

\subsubsection{EPIC 247560727.01 \& EPIC 247560727.02}
\label{subsubsec:247560727}

\begin{figure*}
    \includegraphics[clip,trim={0 0 0 0}, height=10cm, keepaspectratio]{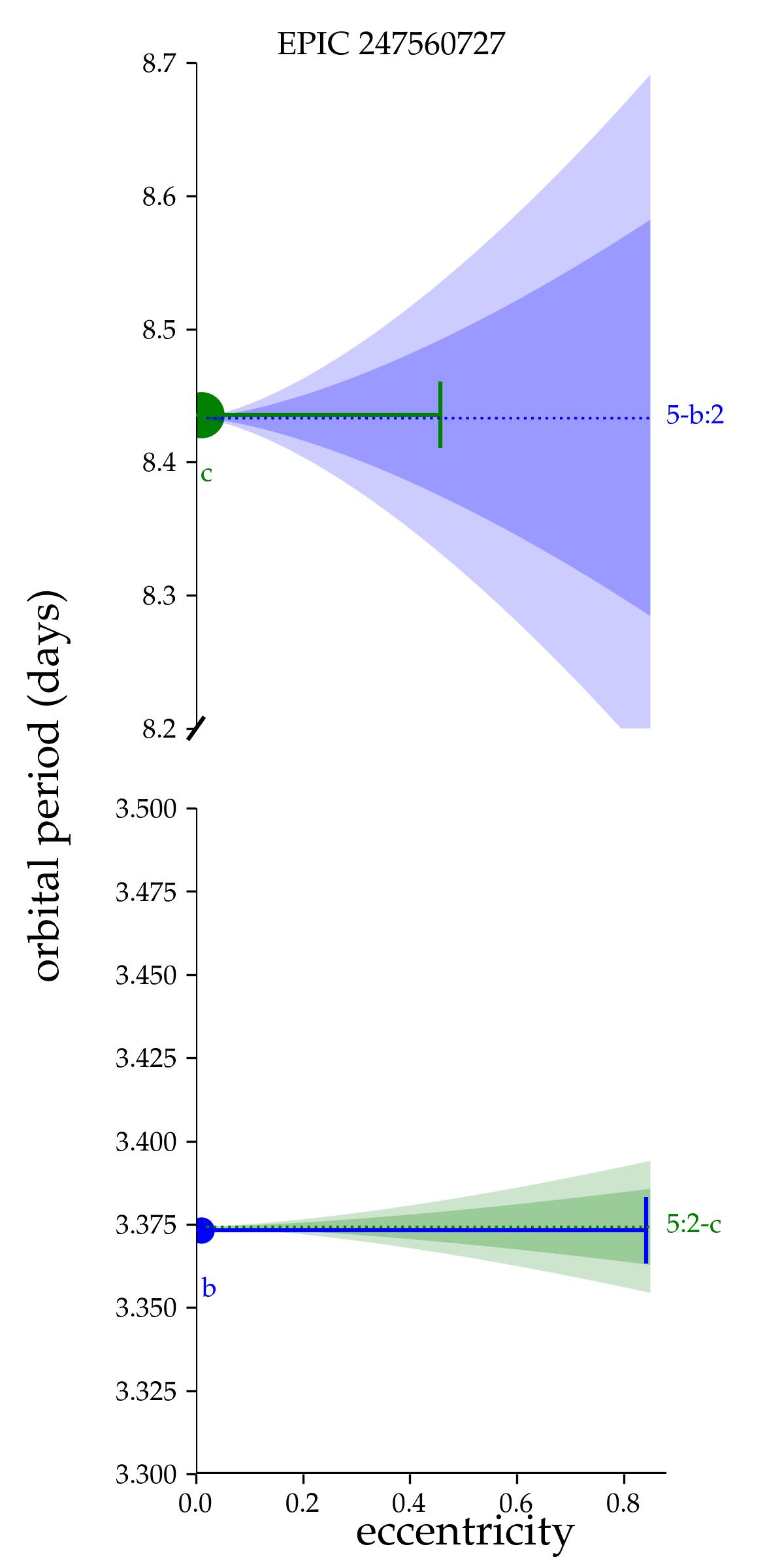}
    \caption{Resonance locations in the EPIC 247560727 system. Same notation as Figure \ref{fig:res_246078343}. The dashed lines show the location of the estimated 5:2 mean motion resonance for planets EPIC 247560727.01 and EPIC 247560727.02. The shaded regions around each resonance are the widths corresponding to the lower (dark shading) and upper (light shading) planet mass limits, propagated from the radii uncertainties and estimated using the \textit{Kepler} mass-radius relationship from \citet{Kanodia2019}. Both planet candidates are in a 5:2 resonance, similar to Jupiter and Saturn in the Solar System.}
    \label{fig:res_247560727}
\end{figure*}

EPIC 247560727 is a faint ($K_p$=15.164 mag, $G$=15.442 mag , $J$=13.577 mag) G8 star (0.779$^{+ 0.059}_{- 0.054}$\rsun, 0.693$^{+ 0.301}_{- 0.212}$\msun) \citep{Hardegree2020} observed during \ktwo campaign C13. It is located at $(\alpha,~\delta)$ = (05:01:42.22,22:39:41.81) at a distance of $\sim$681 pc \citep{Bailer-Jones2021}. It is a multi-planetary candidate system consisting of two planets with periods 3.3733$\pm$0.0002 days, and 8.4356$^{+0.0001}_{-0.0006}$ days. The \vespa and \tri FPP values are $10^{-3}$, and $0.013$ for EPIC 247560727.01, and $0.03$, and $0.028$ for EPIC 247560727.01. We do not validate this system due to the presence of a nearby, slightly fainter ($G$=16.584 mag) star at $\sim$3\arcsec~from EPIC 247560727. Given that this distance is of the order of the \textit{Kepler} pixel size, we can not differentiate the host star using the \ktwo photometry alone. Table \ref{tab:epic247560727} shows a comparison of the stellar properties for EPIC 247560727 and the neighboring star TIC 674662900. The latter seems to be a background star not bound to EPIC 247560727 given the differences in the proper motions and the parallaxes obtained from \gaia eDR3. In addition, the astrometric information from \gaia eDR3 (i.e. \verb|GOF_AL|, \verb|D|, and $RUWE$) for both targets initially rules out the possibility of both stars being binaries on their own. The centroid computed distances to the nearest neighboring star (see the last row of Figure \ref{fig:centmulti}) also cannot discard the possibility of TIC 674662900 being the host star, though the 2.25$\arcsec$ distance for EPIC 247560727.02 seems to favor the brightest star as the transiting one.
Assuming that EPIC 247560727 is the host star of these planet candidates, EPIC 247560727.01 is a super-Earth (1.5838$\pm$0.0781\rearth) orbiting at a distance of 0.0279$^{+0.0013}_{-0.0016}~AU$, and EPIC 247560727.02 is a sub-Neptune (2.9192$^{+0.4080}_{-0.3350}$\rearth) orbiting at a distance of 0.0708$^{+0.0051}_{-0.0056}~AU$. Using the \textit{Kepler} mass-radius relationship from \citet{Kanodia2019}, we estimate planetary masses of $\sim$4.24\mearth and $\sim$6.78\mearth. Using these planetary masses, the detected periods, and the resonance analysis explained in Section \ref{subsec:resonance}, we find that both planets are in a 5:2 resonant orbit (see Figure \ref{fig:res_247560727}), similar to Jupiter and Saturn in the Solar System. This fact suggests that both planet candidates are orbiting the same star rather than each one of them orbiting a different host star. Using the dilution factor (see Section \ref{subsec:blending}), and assuming that the depths in the \gaia bandpass are of the same order as in the \textit{Kepler} one (both filters are centered approximately at the same wavelength, and have similar bandwidths), the planetary radii would be a factor $\sim$1.16$\times$ larger if the planets are orbiting EPIC 247560727, and $\sim$1.96$\times$ larger if the host star is TIC 674662900. This would still put both candidates well below the $R_p<8$\rearth brown-dwarf/stellar limit.

\begin{table*}
    \centering
    \caption{Comparison of stellar properties for EPIC 247560727 and contaminating background star TIC 67462900.}
    \label{tab:epic247560727}
    \begin{tabular}{lllllllllll}
\hline
EPIC     & TIC  & \rsun        & \msun        & \teff & \logg                              & \verb|GOF_AL| & \verb|D| & $RUWE$ & pm [mas/yr] & $\Pi$ [mas] \\ \hline
27560727 & 69054629  &  0.779$^{+ 0.059}_{- 0.054}$ & 0.693$^{+ 0.301}_{- 0.212}$ & 5634$\pm$138         & 4.494$^{+ 0.150}_{- 0.150}$ & -0.83                         & 0.0                     & 0.96   & 4.92                           & 1.431                        \\
-        & 674662900 &  1.288$^{\dagger}$                       & 1.210$^{\dagger}$                       & 6252$\pm$128$^{\dagger}$         & 4.301$^{\dagger}$                                             & -1.24                         & 0.0                     & 0.94   & 1.90                           & 0.338     \\                  
\hline
\multicolumn{11}{l}{$\dagger$: data from TIC catalogue \citep{Paegert2021}}
\end{tabular}
\end{table*}

\subsubsection{EPIC 211436876.01}
\label{subsubsec:211436876}

EPIC 211436876 is a relatively bright ($K_p$=12.302 mag, $G$=12.279 mag , $J$= 11.335 mag) G2 star (1.057$^{+0.022}_{-0.020}$\rsun, 0.992$^{+0.068}_{-0.063}$\msun) \citep{Hardegree2020} observed by \ktwo during campaigns C5 and C18. It is located at $(\alpha,~\delta)$ = (08:30:54.63, 12:11:56.77), at a distance of $\sim$370 pc \citep{Bailer-Jones2021}.
We detect a significant period of 1.1524$^{+0.0003}_{-0.0004}$ days in the \eve and \tfaw light curves for both sectors (and in the combined C5+C18 light curves), and a harmonic of the period in the \texttt{K2SFF} light curves (although they have $\sim$1.6$\times$ worse photometric precision than the \eve ones). EPIC 211436876.01 is a candidate sub-Earth (0.6746$^{+0.0454}_{-0.0392}$\rearth) orbiting at a distance of 0.0125$^{+0.0009}_{-0.0011}~AU$, and receiving a stellar insolation of $\sim$8270 $S/S_{\oplus}$. The \vespa and \tri FPP values for this target are 0.4624 and 0.1054, respectively. The centroid \texttt{p-value} is 0.405, and the maximum computed separation for a background eclipsing binary is 612.23\arcsec (see Figure \ref{fig:centsingle}). There are two nearby ($\sim$15.5\arcsec, and $\sim$18\arcsec), fainter ($G$=17.983 mag, and $G$=16.591 mag) stars partially affecting the \eve aperture. Following our vetting procedure (see Section \ref{subsec:vetting}), we recomputed the light curves for both campaigns, changing the aperture size in order to minimize the flux contribution from these neighboring stars. Also, we could not recover the transiting signal when creating custom apertures centered in the neighboring stars using the \texttt{lightkurve} pipeline. The \gaia astrometric parameters (see Table \ref{tab:epic211436876}) for these two stars seem to rule-out the chances of them being background binary stars.
Using the mass-radius estimation from \citet{Chen2016}, we compute an estimated mass of $\sim$0.24\mearth, this results in a very small RV semi-amplitude of $K\sim0.15m/s$. Also, although orbiting a relatively bright star, the photometric follow-up of this target is challenging given the small transit depth (<0.1 ppt). However, if confirmed, it would be one of the few very short-period (<1.5 days) sub-Earths (with $R_p<0.7$\rearth) to be detected (LHS 1678 b \citep{Silverstein2022}; Kepler-1351 b, and Kepler-1087 b \citep{Morton2016}), the second in the \ktwo mission (after K2-89 b \citep{Crossfield2016}), and also, the second around a G-type star (after Kepler-1087 b).

\begin{table*}
    \centering
    \caption{Comparison of \gaia properties for EPIC 211436876 and nearby stars.}
    \label{tab:epic211436876}
    \begin{tabular}{llllll}
\hline
EPIC      & \gaia eDR3 & $G$ [mag] & \verb|GOF_AL| & \verb|D| & $RUWE$ \\ 
\hline
 211436876 &    602703487815096832    &  12.28 & -2.72  &  2.36  &  0.88 \\ 
\hline
 211437101 &    602703487814896384    &  17.98 &  1.07  &  0.70  &  1.04 \\
 211436674 &    602702731900653568    &  16.59 &  0.39  &  0.00  &  1.02 \\ 
\hline
\end{tabular}
\end{table*}

\subsubsection{EPIC 210706310.01}
\label{subsubsec:210706310}

EPIC 210706310 is a relatively bright ($K_p$=12.294 mag, $G$=12.296 mag, $J$=11.083 mag), metal-poor (\feh=-0.402$\pm$0.235$[dex]$, \citet{Hardegree2020}; \feh=-0.463428$^{+0.35536}_{-0.230723}[dex]$, \citet{Anders2022}; \feh=-0.252370$\pm$0.081465$[dex]$, \citet{Buder2021}), F7 star (0.954$^{+ 0.055}_{- 0.052}$\rsun, 0.709$^{+ 0.304}_{- 0.219}$\msun) \citep{Hardegree2020} observed by \ktwo in campaign C4. It is located at $(\alpha,~\delta)$ = (03:57:30.02, 18:27:13.13), at a distance of $\sim$275 pc \citep{Bailer-Jones2021}. We detect a significant period of 5.1718$\pm$0.0002 days in the \ktwo pipeline, \eve, and \texttt{K2SFF} light curves. EPIC 210706310.01 is a candidate sub-Earth (0.8891$^{+0.0529}_{-0.0443}$\rearth) orbiting at a distance of 0.0510$^{+0.0024}_{-0.0029}$ AU, and receiving a stellar insolation of $\sim$391 $S/S_{\oplus}$. Even though its \vespa FPP is below the 1\% threshold, we do not validate this target due to the presence of a very faint ($G$=20.247 mag) background star at a distance of $\sim$6.4$\arcsec$. Also, the \tri results point as the most probable scenarios either the transiting planet around the target (57\%), the unresolved bound companion, with the transiting planet around the primary star (16\%), or the secondary star (22\%). The \gaia eDR3 astrometric parameters (\verb|GOF_AL|=1.32, \verb|D|=7.07, $RUWE$=1.052) seem to disfavour the binary scenario for this target. Also, the astrometric values (\verb|GOF_AL|=-1.23, \verb|D|=1.27$\times$10$^{-15}$, $RUWE$=0.944) for the faint background star seem to discard it from being a background eclipsing binary. Data from future \gaia releases might help to improve the characterization of this system. Candidates orbiting metal-poor stars like this one can help planet formation theories by setting limits to the lowest metallicity that protoplanetary disks can have to form planets \citep{Matsuo2007,Gaspar2016,Petigura2018}.

\subsection{False positives}
\label{subsec:false}

Out of our sample of 27 planetary candidates, 8 of them have either not passed the vetting procedure in Section \ref{subsec:vetting} or have FPPs exceeding the thresholds defined in Section \ref{subsec:validation}. EPIC 220356827.01 (with FPP$_{\tri}=0.4558$ and FPP$_{\vespa}$=0.9834), and EPIC 246048459.01 (with FPP$_{\tri}=0.9836$ and FPP$_{\vespa}$=0.3407) have failed the validation process. In the case of the former, the transit shape is v-shaped and different during the egress. In addition, there is some excess flux during the ingress that could point towards a binary nature of the system. In the case of the latter, although the \gaia astrometric parameters, and the SOAR speckle imaging, seem to rule out the presence of contaminating stars, the FPP values make us mark this candidate as a false positive.  

\subsubsection{False positives by \gaia eDR3}
\label{subsubsec:falsegaia}

\begin{figure*}
            \begin{tabular}{cccc}
                \includegraphics[clip,width=0.45\columnwidth]{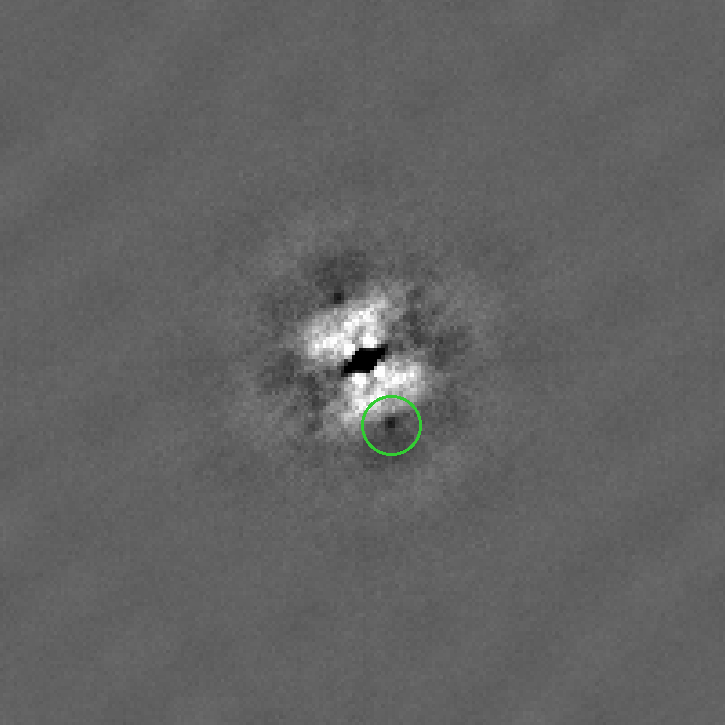}
                &
                \includegraphics[clip,width=0.45\columnwidth]{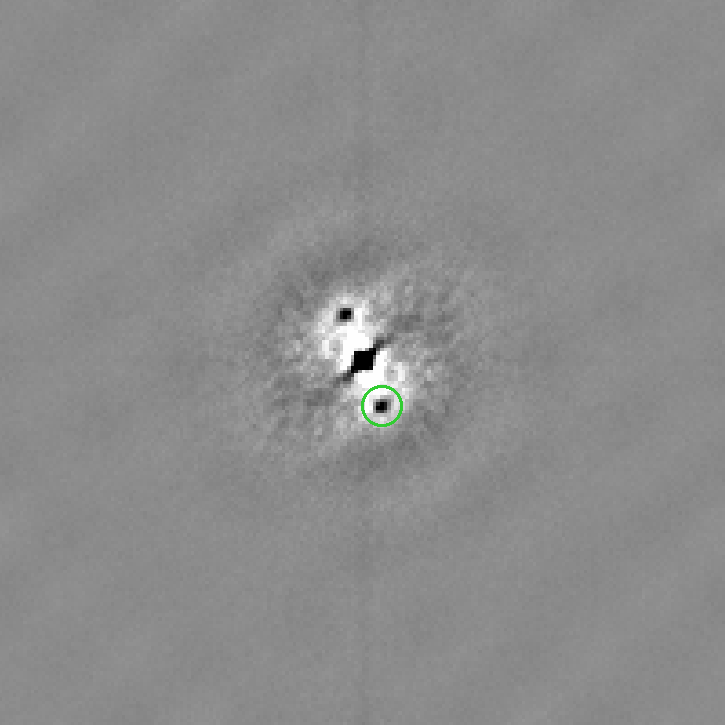}
                &
                \includegraphics[clip,width=0.45\columnwidth]{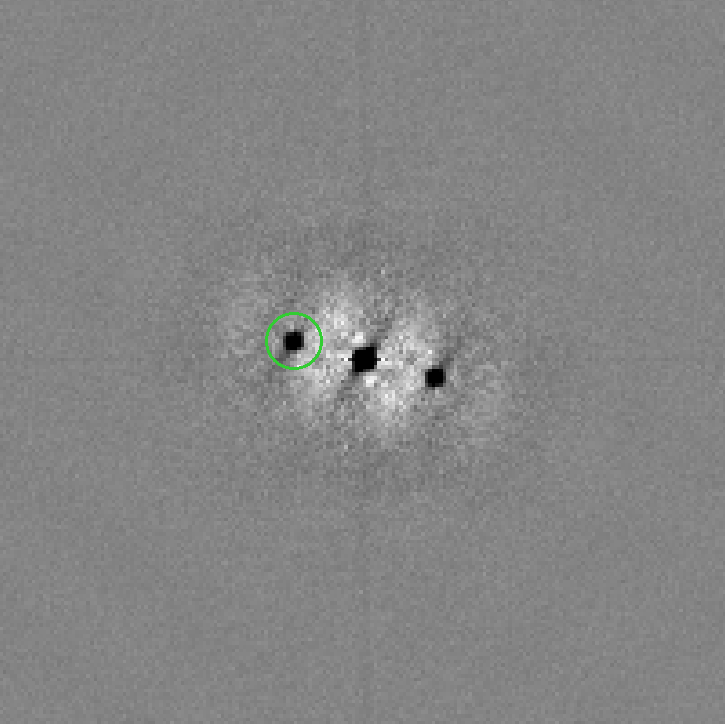}
                &
                \includegraphics[clip,width=0.47\columnwidth]{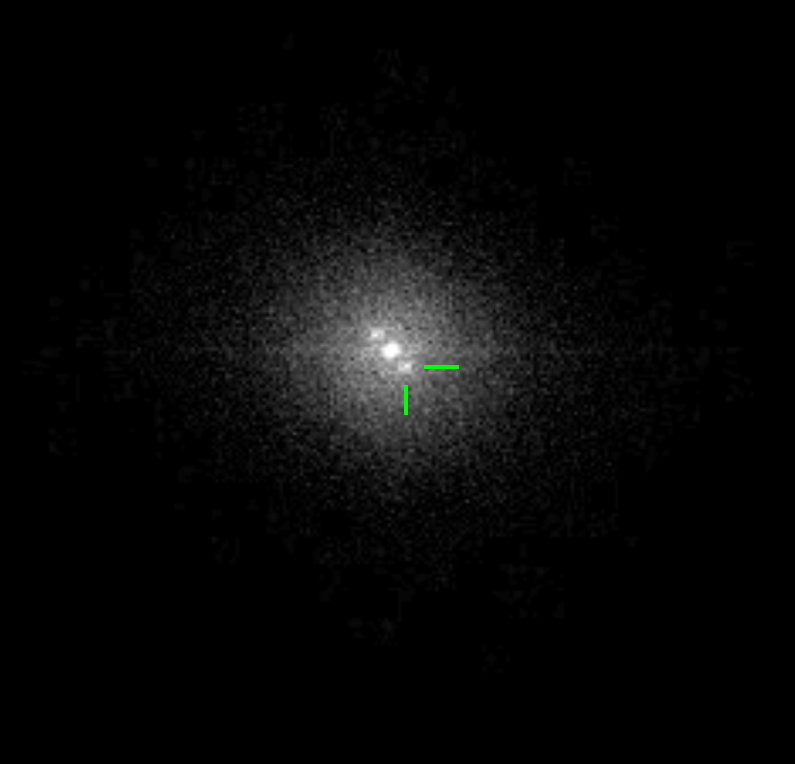}\\
            \end{tabular}
            \caption{SOAR speckle auto-correlations for EPIC 205979843 (left), EPIC 246022853 (middle left), and EPIC 246163416 (middle right) with detected nearby companions marked in green. The field size is 3.15 arcsec, north up and east left. BTA observations for EPIC 211572480 (right) with the detected companion marked in green. North up and east left }
            \label{fig:speckle_cont}
\end{figure*}

Of the remaining six false positives systems, five of them (EPIC 211572480, EPIC 211705502, EPIC 220471100, EPIC 246022853, and EPIC 246163416) have been discarded following the criteria in Section \ref{subsec:gaia} for \gaia eDR3 \verb|GOF_AL|, \verb|D|, and $RUWE$ values. EPIC 211572480, and EPIC 211705502 have missing stellar properties both from the EPIC catalogue \citep{Huber2017} and from \citet{Hardegree2020} data. 
EPIC 211572480 has very large \verb|GOF_AL|, \verb|D|, and $RUWE$ values (see Table \ref{tab:stars}) that point towards the binary nature of the system. We detect a companion at $\sim$0.1$\arcsec$ and $\Delta$mag$\sim$0 using BTA speckle imaging using the 550/50 filter (see Figure \ref{fig:speckle_cont}).
EPIC 211705502 was first reported to have a transiting object of $R_p$=10.29\rearth in a $P$=2.58 day orbit by \citet{Castro-Gonzalez2021}. They used \textit{isochrones}-derived stellar parameters to derive the planetary parameters, they took into account the presence of two fainter and nearby stars (separated $\sim$1.17$\arcsec$, and $\sim$5.88$\arcsec$), and with \gaia DR2 \verb|GOF_AL|=0.57, and \verb|D|=0.00, FPP$_{\vespa}$=0.99 values, they cataloged EPIC 211705502.01 as a candidate transiting exoplanet. Using updated \gaia eDR3 data (\verb|GOF_AL|=30.94, \verb|D|=57.8, and $RUWE$=2.42), we denote this candidate as a false positive. The two other nearby stars, with \verb|GOF_AL|=3.87, \verb|D|=4.54, and \verb|GOF_AL|=0.83, \verb|D|=0.48, and $RUWE$=1.031, respectively do not seem to be the source of the transiting signal. Given the low SNR BTA observations for this target, we could not obtain conclusive results for the presence of companions.
EPIC 246022853, and EPIC 246163416 have resolved companions from SOAR speckle imaging data (see Figure \ref{fig:speckle_cont}, and Table \ref{tab:speckle_cont}), as well as large values for the \gaia parameters. Except for the case of EPIC 220471100, where we do not detect any companion star using BTA, there seems to be a good agreement between speckle imaging and \gaia eDR3 astrometric parameters. These seem to confirm that the use of these parameters could be a good way of determining probable false positive scenarios during the vetting stage of future planet candidates.

\begin{table*}
    \centering
    \caption{Systems with detected companions in SOAR speckle imaging data (2021.75-2021.80.}
    \label{tab:speckle_cont}
    \begin{tabular}{llll}
\hline
EPIC      & $\rho$ [$\arcsec$] & $\Theta [^{\circ}]$ & $\Delta$I [mag] \\
\hline
205979483 & 0.5751            & 203.2    & 4.0       \\
246022853 & 0.4305            & 201.2    & 2.7        \\
246163416 & 0.6289            & 75.9     & 0.7         \\ 
211572480 & 0.1000            & 225.0    & 0.1          \\
\hline
\end{tabular}
\end{table*}

The remaining false positive system, EPIC 205979483, has a very faint ($\Delta$I=4 mag) companion detected through SOAR speckle imaging at a distance of 0.5751$\arcsec$ (see Figure \ref{fig:speckle_cont}). Interestingly, both \verb|D|=15.1, and $RUWE$=1.41, exceed the threshold values defined in Section \ref{subsec:gaia}, and although the \verb|GOF_AL|=7.19 is smaller than the defined limit, it is the highest value of all the systems that have passed these vetting criteria. This result seems to indicate that caution has to be taken for planetary candidates whose \gaia parameters are close to the theoretical values, and also, reinforces the fact that high-resolution imaging through speckle and/or adaptive optics are needed in order to better characterize these systems. Regarding this, one consideration has to be done for our planet candidate EPIC 210967369.01. Even though it has a slightly large $RUWE$ value of 1.28, a slightly smaller value of \verb|GOF_AL| (5.44) than EPIC 205979483, and relatively large FPP values (FPP$_{\vespa}$=0.8165, and FPP$_{\tri}$=0.266) due to the presence of a nearby ($\sim$19.7$\arcsec$), fainter ($G$=19.727) background star, we classify it as a planet candidate; but taking into consideration that it might benefit from new astrometric values from future \gaia data releases.

In the case of EPIC 246163416 we detect a slightly fainter ($\Delta$I=0.7 mag) companion at a distance of 0.6289$\arcsec$ (see Figure \ref{fig:speckle_cont}) through SOAR speckle imaging. The \gaia eDR3 parameters for this target (see Table \ref{tab:stars}) point towards the binary nature of the system. However, the angular separation of the SOAR companion is not compatible with the detected transiting period of $P$=0.8768 days. Thus, a third object is present as either part of a trinary system (more observations would be needed in order to determine whether the SOAR companion is gravitationally bound to the EPIC target) or transiting one of the stars in a binary configuration. The MCMC best-fit planetary radius ($R_p$=8.4683$^{+4.3149}_{-2.8121}$\rearth), and the retrograde, and high orbit inclination angle ($i$=140.52$^{\circ}$ $^{+2.2073}_{-2.1373}$) seem to point towards a grazing binary scenario as the most probable one for this candidate.

\section{Conclusions}
\label{sec:conclusions}

The \ktwo sub-Neptune-sized planetary legacy is a niche that still remains to be fully exploited. Algorithms able to increase the photometric precision of the \ktwo light curves can help increase the number of detected exoplanets orbiting fainter stars or of different spectral types. In this sense, \tfaw denoising, together with \tls improved detection capabilities, offer a new way of detecting and characterizing planetary transit candidates missed by previous works. In this work, we have presented the results from a first sample of 27 planetary candidates from the \tfaw survey. Combining \vespa and \tri FPPs, we statistically validate six planets in four different stellar systems and present 12 planetary candidates, of which 11 are new detections. Our sample of validated and candidate planets is comprised of three sub-Earth planets, seven Earth-sized planets, four super-Earths, and four sub-Neptunes. With respect to individual systems, we highlight the following: a validated highly-irradiated Earth-sized planet (EPIC 210768568.01), and a validated sub-Neptune planet (EPIC 247422570.01) orbiting a G4 star. Two validated multi-planetary systems, EPIC 246078343 and EPIC 246220667; the latter near its 3:2 mean motion resonance. A candidate multi-planetary system EPIC 247560727 consists of a super-Earth and sub-Neptune in a 5:2 resonant orbit. And EPIC 21436876.01 is a very-short period sub-Earth candidate, and one of the few detected orbiting around a G2 star.
In addition, one of our validated planets, and one candidate are USP planets. Given their estimated escape velocities, and effective temperatures, four of our planet candidates, and one validated planet are close to the He atmospheric escape threshold and within the Radius Gap. With its estimated density, candidate planet EPIC 247560727.02 would probably be a water world. Although affected by the presence of contaminating background stars, planet candidate EPIC 218701083.01 could be one of the few planets within the Neptunian desert. Given the improvements obtained with \tfaw, eight listed planets have radii below the Radius Gap.
Finally, we classify eight candidates as false positives. We find from combining speckle imaging and \gaia eDR3 photometric and astrometric information that \gaia data can be a powerful tool that can benefit the vetting process of future planet candidates.

By increasing the number of statistically validated, and candidate planets, \tfaw aims to expand the statistical information of the population of planets. This can have an impact on improving the planet occurrence rates, affect the current and future planet formation and evolution theories and their role on habitability conditions, and improve our understanding of star-planet interactions, atmospheric erosion, and other phenomena.

\section*{Acknowledgements}

This research has made use of the NASA Exoplanet Archive, which is operated by the California Institute of Technology, under contract with the National Aeronautics and Space Administration under the Exoplanet Exploration Program. This work made use of NASA ADS Bibliographic Services. This research has made use of \texttt{Aladin} sky atlas developed at CDS, Strasbourg Observatory, France. This work has made use of data from the European Space Agency (ESA) mission \gaia. DdS acknowledges funding support from RACAB. DdS and OF acknowledge the support by the Spanish Ministerio de Ciencia e Innovaci\'{o}n (MICINN) under grant PID2019-105510GB-C31 and through the ``Center of Excellence Mar\'{i}a de Maeztu 2020-2023'' award to the ICCUB (CEX2019-000918-M). MdA acknowledges financial support from the Universitat de Barcelona-Reial Acad{\`e}mia de Ci{\`e}ncies i Arts de Barcelona (RACAB) collaboration grant 2020.2.RACAB.1.

\section*{Data Availability}

The data presented in this article will be shared on reasonable request to the corresponding author.



\bibliographystyle{mnras}
\bibliography{ref} 



\bsp	
\label{lastpage}
\end{document}